\documentclass[aps,superscriptaddress,twocolumn,showpacs]{revtex4}
\usepackage{epsfig}

\newcommand{\simle}
{\raisebox{-0.75ex}[-1.5ex]{$\;\stackrel{<}{\sim}\;$}}
\newcommand{\simge}
{\raisebox{-0.75ex}[-1.5ex]{$\;\stackrel{>}{\sim}\;$}}
\def\d{{\partial}}
\def\s{{\sigma}}
\def\e{{\epsilon}}
\def\k{{ {\bf k} }}
\def\p{{ {\bf p} }}
\def\q{{ {\bf q} }}
\def\Q{{ {\bf Q} }}

\def\w{{\omega}}
\def\a{{\alpha}}

\def\g{{\gamma}}

\begin{document}

\def\runtitle{
Theory of infrared conductivity and 
Hall conductivity Based on the Fermi Liquid Thoery
}
\def\runauthor
Hiroshi {\sc Kontani}

%\draft
\title{
Theory of infrared conductivity and 
Hall conductivity Based on the \\ Fermi Liquid Theory: 
analysis of high-$T_{\rm c}$ superconductors
}

\author{
Hiroshi {\sc Kontani}
}

\address{
Department of Physics, Nagoya University,
Furo-cho, Nagoya 464-8602, Japan.
}

\date{\today}

\begin{abstract}
We study optical conductivities for high-$T_{\rm c}$ superconductors
under the magnetic field on the basis of the microscopic Fermi liquid theory.
Current vertex corrections (CVC's) are correctly
taken into account to satisfy the conservation laws,
which has been performed for the first time for optical conductivities
based on the fluctuation-exchange (FLEX) approximation.
We find that the CVC emphasizes the $\w$-dependence of $\s_{xy}(\w)$
significantly when the antiferromagnetic (AF) fluctuations are strong.
By this reason, the relation $\s_{xy}(\w) \sim \{\s(\w)\}^2$,
which is satisfied in the extended-Drude model given by the relaxation
time approximation (RTA), is totally violated 
for a wide range of frequencies.
Consequently, the optical Hall coefficient $R_{\rm H}(\w)$ 
strongly depends on $\w$ below the infrared frequencies,
which is consistent with experimental observations.
We also study the mystery about a simple-Drude form of the 
optical Hall angle $\theta_{\rm H}(\w)$ observed by Drew et al.,
%follows a simple Drude-form in the infrared region, 
which is highly nontrivial in terms of the 
RTA since the strong $\w$-dependence of the relaxation time 
should modify the Drude-form.
%[J. Cerne et al., Phys. Rev. Lett. {\bf 84} (2000) 3418.]
We find that a simple Drude-form of $\theta_{\rm H}(\w)$
is realized because the $\w$-dependence of the CVC almost 
cancels that of the relaxation time.
In conclusion, anomalous optical transport phenomena
in high-$T_{\rm c}$ superconductors,
which had been frequently assumed as an 
evidence of the breakdown of the Fermi liquid state,
are well understood in terms of the nearly AF Fermi liquid
once the CVC is taken into account.
\end{abstract}


\sloppy

\maketitle

%\begin{multicols}{2}
%\narrowtext
%%%%%%%%%%%%%%%%%%%%%%%%%%%%%%%%%%%%%%%%%
\section{Introduction}
%%%%%%%%%%%%%%%%%%%%%%%%%%%%%%%%%%%%%%%%%

% Importance of transport phenomena in HTSC

In cuprate high-$T_{\rm c}$ superconductors (HTSC's),
various physical quantities in the normal state
deviate from the conventional 
Fermi liquid behaviors in usual metals,
which are called the non-Fermi liquid (NFL) behaviors.
These NFL behaviors has caused controversial
discussions on its ground state.
One of the most predominant candidates
is the Fermi liquid state with strong 
antiferromagnetic (AF) fluctuations.
 \cite{Yamada-rev,Moriya,Pines,Kontani-review}.
In fact, spin fluctuation theories like the
SCR theory
 \cite{Moriya} 
and the fluctuation-exchange (FLEX) approximation
 \cite{Bickers,Monthoux-Scalapino}
can reproduce the Curie-Weiss like behavior of $1/T_1T$ and the 
$T$-linear resistivity in HTSC's, as well as an appropriate 
optimum $T_{\rm c}$ of the order of 100K with the correct symmetry,
$d_{x^2\mbox{-}y^2}$.

Especially, anomalous
transport phenomena under the magnetic field 
in HTSC's have been long-standing problems,
as an strong objection against a simple Fermi 
liquid picture.
For example, 
the Hall coefficient $R_{\rm H}$ is positive 
in hole-doped systems like
YBa$_2$Cu$_3$O$_{7-\delta}$ (YBCO) and
La$_{2-\delta}$Sr$_\delta$CuO$_4$ (LSCO)
whereas it is negative
in Nd$_{2-\delta}$Ce$_\delta$CuO$_4$ (NCCO),
although they possess similar hole-like
Fermi surfaces (FS's)
 \cite{Satoh}.
In each compound,
$R_{\rm H}\propto T^{-1}$ is observed 
below $T_0\sim 700$K, and $|R_{\rm H}|\gg 1/ne$
($n$ being the electron filling number)
at lower temperatures.
Moreover, the magnetoresistance
$\Delta\rho/\rho_0$ is proportional to
$T^{-4}$ below $T_0$
 \cite{Kimura,Ando}.
As a result,
so called the modified Kohler's rule,
$\Delta\rho\cdot\rho_0\propto R_{\rm H}^2$,
is well satisfied in HTSC's.
They cannot be explained on the same footing
within the relaxation time approximation (RTA)
even if one assume an extreme momentum and 
energy dependences of $\tau_\k(\e)$:
If one assume a huge anisotropy of $\tau_\k(0)$
to explain the enhancement of $R_{\rm H}$ 
experimentally at lower temperatures, 
then $\Delta\rho/\rho_0$ should increase much faster 
than experiments ($\propto T^{-4}$) because 
$\Delta\rho/\rho_0$ is much sensitive to the 
anisotropy of $\tau_\k$ in terms of the RTA.
Thus, we cannot explain the modified Kohler's rule
on the basis of the RTA
 \cite{Ioffe,Kontani-MR}.

Resent theoretical works
have shown that the origin of these
anomalous DC-transport phenomena in HTSC's 
is the {\it vertex correction
for the total current ${\bf J}_\k$},
which is known as the {\it back-flow}
in Landau-Fermi liquid theory
 \cite{Kontani-review,Kontani-Hall}.
${\bf J}_\k$ becomes totally different
from the quasiparticle velocity ${\bf v}_\k$
when strong AF fluctuations exist.
Reflecting this fact,
the DC-conductivities $\s_{\mu\nu}$
($\mu,\nu=x,y$) behaves as
 \cite{Kontani-Hall}
\begin{eqnarray}
\s \propto \tau, \ \ \ 
\s_{xy} \propto \chi_Q\cdot\tau^2 ,
 \label{eqn:DC-s}
\end{eqnarray}
where $\tau$ represents the relaxation time
of quasiparticles (at the cold spot),
and $\chi_Q$ is the staggered susceptibility,
which follows the Curie-Weiss like behavior in HTSC.
As a result, $R_{\rm H}\propto \chi_Q \propto T^{-1}$ is concluded.
By taking account of the back-flow,
we can naturally explain anomalous behaviors of 
$R_{\rm H}$, the magnetoresistance ($\Delta\rho/\rho$), 
the thermoelectric power ($S$)
and the Nernst coefficient ($\nu$) {\it in a unified way}
 \cite{Kontani-review,Kontani-MR,Kontani-Hall,Kontani-S,Kontani-N}.

%punch lines
Dynamical transport phenomena in HTSC's are
furthermore mysterious.
For example, optical conductivities 
under the magnetic field 
shows striking deviation from the extended-Drude forms
in HTSC's.
%They put severe constraints on theories
%in the normal state of HTSC's
Previous theoretical works,
many of them were based on the RTA,
unable to give comprehensive understanding for them
 \cite{Drew04,Drew02,Drew00,Drew00-c,Drew96,Zimmers}.
In the present paper,
we study the role of the back-flow in the diagonal
optical conductivity $\s(\w)$ and the off-diagonal one
$\s_{xy}(\w)$ based on the Fermi liquid theory.
Here, we develop the method of calculating
$\s(\w)$ and $\s_{xy}(\w)$ using the FLEX approximation
by taking the current vertex correction (CVC),
which represent the back-flow contribution,
to satisfy the conservation laws.
We call it the CVC-FLEX approximation.
By this approximation,
both AC and DC transport phenomena in HTSC are explained
{\it on the same footing}.
\cite{Letter}.

%Here we explain in detail
%the optical conductivities observed by Drew et al.,
%which would eliminate any theory of DC transport 
%phenomena in HTSC based on the RTA.
In the spirit of the RTA,
optical conductivities are given by the
following extended-Drude (ED) forms when $\w\simle\tau^{-1}(\w)$:
\begin{eqnarray}
\s^{\rm RTA}(\w)&\approx& \frac{\Omega}{\tau^{-1}(\w)-i\w} ,
 \label{eqn:Drude-sig}\\
\s_{xy}^{\rm RTA}(\w) &\approx& 
 \frac{\Omega_{xy}}{(\tau^{-1}(\w)-i\w)^{2}} ,
 \label{eqn:Drude-sxy} 
\end{eqnarray}
where $\tau(\w)$ is the relaxation time of quasiparticles.
Here, $\w$-dependences of $\Omega$ and $\Omega_{xy}$ have been
dropped for simplicity.
[This simplification will not be allowed in heavy fermion systems
because of the strong $\w$-dependence of the renormalization factor.]
%whose $\w$-dependence is prominent in strongly correlated systems.
%In general $\tau(\w)$ is a monotonous decrease function of $\w$.
In usual Fermi liquids, $\tau\propto (\w^{2}+(\pi T)^2)^{-1}$.
In HTSC, $\w$-dependence of $\tau(\w)$ is much stronger;
according to a spin fluctuation theory
\cite{Pines-opt},
$\tau(\w) \propto  \langle {\rm Im}\Sigma_\k^{-1}(\w-i\delta)
 \rangle_{\rm FS} \propto (\w+\pi T)^{-1}$
for a wide range of ($\w$, $T$).
Actually, the relaxation time deduced from the experimental
optical conductivity, which is proportional to 
${\rm Re}\s^{-1}(\w)$, follows the above relation.

When the ED-form is satisfied,
$R_{\rm H}(\w)$ becomes real and $\w$-independent because
$\s_{xy}^{\rm RTA}(\w) \propto \{\s^{\rm RTA}(\w)\}^2$.
%One may optimistically guess that
This relation is approximately satisfied in Cu and Au;
for $\w\approx 1000{\rm cm}^{-1}$ 
where $\w\simge \tau^{-1}(\w)$ is satisfied,
the reduction of Re$R_{\rm H}(\w)$ from the DC-value
as well as the ratio of Im$R_{\rm H}(\w)$ to the real part
are about 10$\%$ for Cu, and are about 20$\%$ for Au,
respectively \cite{DrewCuAu}.
However, $R_{\rm H}(\w)$ shows strong $\w$-dependence in HTSC,
which means that the ED-forms are totally violated.
In fact, we show in the present work that 
$\s_{xy}(\w)$ strongly deviates from the extended-Drude form
due to the $\w$-dependence of the back-flow
in the presence of strong AF fluctuations.
According to experiments for the optimally-doped YBCO,
Im$R_{\rm H}(\w)$ takes the maximum value at 
$\w_{\rm RH}\sim50{\rm cm}^{-1}$, and
Im$R_{\rm H}(\w_{\rm RH})\sim{\rm Re}R_{\rm H}(\w_{\rm RH})$
 \cite{Drew96}
On the other hand, Im$\s(\w)$ takes the maximum value at 
$\w_{xx}\sim150{\rm cm}^{-1}$.
This relation $\w_{\rm RH} \ll \w_{xx}$,
which cannot be reproduced by the RTA
even if one assume strong ($\k,\w$)-dependences of $\tau_\k(\w)$,
is well reproduced in the present study.

We also discuss the optical Hall angle 
$\theta_{\rm H}(\w) \equiv \s_{xy}(\w)/\s(\w)$,
whose ED form is given by
$\theta^{\rm RTA}(\w)\propto (\tau^{-1}(\w)-i\w)^{-1}$.
Quite surprisingly,
$\theta_{\rm H}(\w)$ in HTSC's follows a simple Drude form
even in the infrared (IR) region ($\w\sim1000{\rm cm}^{-1}$)
 \cite{Drew00,Drew04}.
For instance, the real part of $\theta_{\rm H}^{-1}(\w)$
in HTSC is almost $\w$-independent.
This experimental fact cannot be understood in the framework 
of the RTA since the $\w$-dependence of $\tau(\w)$ is prominent
in HTSC as mentioned above.
Thus, the optical Hall angle in HTSC's have put very severe 
constraints on theories in the normal state of HTSC's.
In the present paper, we show that the 
simple Drude form of $\theta_{\rm H}(\w)$ in HTSC
is a natural consequence of the cancellation between
the $\w$-dependence of $\tau$ and that of the CVC.
This fact further confirms the importance of the CVC
for both DC and AC transport phenomena in HTSC's.
\cite{Letter}.

%In the present work,
%we explain these mysterious behaviors
%of the optical transport phenomena in HTSC's
%using the CVC-FLEX approximation,
%%in a comprehensive way,
%for enough wide range of frequencies and temperatures.
%%without assuming any fitting parameters.
%The main origin is that
%the $\w$-linear term of $\s_{xy}(\w)$,
%which is pure imaginary, is strongly enhanced by the CVC 
%when the AF fluctuations are dominant.
%This fact leads to the violation of the ED-form for $\s_{xy}(\w)$.
To clarify the reason
we derive the general expression for 
the $\w$-linear terms of $\s(\w)$ and $\s_{xy}(\w)$
from Kubo formula based on the microscopic Fermi liquid thoery.
They are exact up to the most divergent terms with respect to $\tau$.
By analysing the back-flow in the obtained expression,
we find that the relation
${\rm Im}\s_{xy}(\w)/i\w \propto T^{-2}\cdot \tau^3$ holds.
Because $\s_{xy}(0)\propto \chi_Q\tau^2 \propto T^{-1}\tau^2$
according to eq.(\ref{eqn:DC-s}),
the ED-form in eq.(\ref{eqn:Drude-sxy}) fails 
due to the back-flow in nearly AF metals.
[If we extend DC-$\s_{xy}$ in eq.(\ref{eqn:DC-s}) 
to finite frequencies in the spirit of the RTA,
we obtain eq. (\ref{eqn:Drude-sxy}) with 
$\Omega_{xy}\propto \chi_Q$.
However, such an easy extension is {\it not true} 
because it gives that
${\rm Im}\s_{xy}(\w)/i\w \propto \chi_Q \tau^3$.]
In summary, the enhancement of ${\rm Im}\s_{xy}(\w)/i\w$
due to the CVC is more prominent than that of $\s_{xy}(0)$,
which leads to the violation of the ED-form for $\s_{xy}(\w)$.
The present study confirms the significant role of the back-flow 
on the optical (as well as DC) conductivities in HTSC's.

We shortly mention the recent theoretical 
progress on the optical conductivity
in Fermi liquids.
%Until now, there is no theoretical work
%which can reproduce the functional form of 
%$\s_{xx}(\w)\equiv\s(\w)$ and $\s_{xy}(\w)$ 
%observed in HTSC in a unified way.
For example, 
studies by the dynamical-mean-field-theory (DMFT)
have been revealed important strong correlation 
effect on the optical conductivity
 \cite{Kotliar-rev,Vollhardt-rev}.
However, the effect of back-flow
is totally dropped in DMFT, which is known
to give the enhancement of $R_{\rm H}$
in nearly AF metals; see eq.(\ref{eqn:DC-s}).
We also comment that
the effect of back-flow on the Drude weight 
of $\s(\w)$ was studied based on the Fermi liquid theory
at zero temperature
 \cite{Okabe,Jujo}.
However, the overall behavior of $\s(\w)$
at finite temperatures in strongly correlated 
systems is highly unknown.

The contents of the present paper is the following:
In \S II,
we explain how to calculate the self-energy
and the conductivity by the FLEX approximation
In \S III, we derive the exact expression for
$\lim_{\w\rightarrow0}\s_{\mu\nu}(\w)/i\w$
based on the Kubo formula.
We discuss the deviation from the Fermi liquid like
behavior due to the back-flow
in the presence of the AF fluctuations.
In \S IV, we address the numerical results for $\s(\w)$,
$\s_{xy}(\w)$, $R_{\rm H}(\w)$ and $\theta_{\rm H}(\w)$,
and we compare them with experimental results.
We succeed in reproducing their characteristic behaviors
in a natural way 
at the same time.
This is the main part of the present study.
Summary of the present work is 
addressed in \S V.
A physical meaning of the back-flow is explained.

%%%%%%%%%%%%%%%%%%%%%%%%%%%%%%%%%%%%%%%%
\section{Numerical Calculation}
%%%%%%%%%%%%%%%%%%%%%%%%%%%%%%%%%%%%%%%%

%%%%%%%%%%%%%%%%%%
\subsection{FLEX approximation for HTSC}
%%%%%%%%%%%%%%%%%%

In this subsection,
we explain the fluctuation-exchange (FLEX) approximation, 
which is one of a self-consistent spin fluctuation theory
 \cite{Bickers}.
The FLEX approximation is  
classified as a conserving approximation whose framework 
was constructed by Baym and Kadanoff
 \cite{Baym-Kadanoff,Baym}.
In the conserving approximation,
correlation functions given by the solution of the
Bethe-Salpeter equations 
automatically satisfy the macroscopic 
conservation laws.
This is a great advantage of the FLEX approximation
in studying transport coefficients.
In fact, it is well known that
approximations which violate conservation laws,
like the relaxation time approximation (RTA),
frequently give unphysical transport phenomena.

Origin of anomalous behaviors in the normal state
in HTSC, which are frequently called the non-Fermi-liquid 
(NFL) like behaviors, have been studied intensively for 
almost 20 years.
Recently, many of them are consistently explained based on the
Fermi liquid picture with strong antiferromagnetic (AF)
fluctuations, using the FLEX approximation,
the perturbation theory with respect to $U$,
SCR theory, and so on
 \cite{Moriya,Yamada-rev,Pines}.
The range of applicability of the FLEX approximation
is wide; from the over-doped region
till the slightly under-doped region ($\delta\sim10$\%)
above the pseudo-gap temperature $T^\ast\sim200$K.
By taking the CVC into account, we can reproduce
various NFL-like behaviors in transport phenomena 
above $T^\ast$ by the FLEX approximation
 \cite{Kontani-Hall}, 
or even below $T^\ast$ by the FLEX+T-matrix approximation 
 \cite{Kontani-N}.
As for the organic superconductor $\kappa$-(BEDT-TTF),
the $d$-wave superconductivity 
 \cite{Kino-Kontani,Kondo,Schmalian},
as well its Curie-Weiss like behavior of $R_{\rm H}$
 \cite{Kontani-RH-kappa},
are also well reproduced by the FLEX approximation.

Here, we study the following square lattice Hubbard model:
\begin{eqnarray}
H=\sum_{\k\s}\e_\k c_{\k\s}^\dagger c_{\k\s}
 + U\sum_i n_{i\uparrow}n_{i\downarrow} ,
 \label{eqn:Hub}
\end{eqnarray}
where $U$ is the on-site Coulomb interaction, and
$\e_\k$ is the dispersion of a free electron.
In the tight-binding approximation,
$\e_\k= 2t(\cos(k_x)+\cos(k_y))
 + 4t'\cos(k_x)\cos(k_y)
 + 2t''(\cos(2k_x)+\cos(2k_y))$,
where $t$, $t'$, and $t''$ are the nearest,
the next nearest, and the third nearest
neighbor hopping integrals, respectively.
In the present study, we
use the following set of parameters:
(I) YBCO (hole-doped):
 $t_0=-1$, $t_1=1/6$, $t_2=-1/5$, $U=8$.
(II) NCCO (electron-doped):
 $t_0=-1$, $t_1=1/6$, $t_2=-1/5$, $U=5.5$.
(III) LSCO (hole-doped):
  $t_0=-1$, $t_1=1/10$, $t_2=-1/10$, $U=5$.
These parameters are equal to that
used in ref.\cite{Kontani-Hall},
except that $U$ for LSCO is changed from 6 to 5.
These hopping parameters were determined by
fitting to the Fermi surface (FS)
observed by ARPES or obtained by the LDA band calculations.
The shape of the FS's for YBCO, LSCO, NCCO
are shown in ref.\cite{Kontani-Hall}.
Because $t_0\sim4000$K in real systems,
$T=0.01$ corresponds to $\sim40$K.

First, we calculate the self-energy numerically
using the FLEX approximation.
The expression for the self-energy is given by
 \cite{Bickers}
\begin{eqnarray}
\Sigma_\k(\e_n)
&=& T\sum_{l,\p}V_{\p}^{\rm FLEX}(\w_l)
 G_{\k-\p}(\e_n-\w_l) ,
 \label{eqn:FLEX-S} \\
V_{\p}^{\rm FLEX}(\w_l)
&=& U^2\left\{ \frac32 \chi_\p^{\rm s}(\w_l)
 + \frac12 \chi_\p^{\rm c}(\w_l) - \chi_\p^0(\w_l)  \right\} ,
 \\
\chi_\p^{\rm s}&=& \frac{\chi_\p^0}{1- U\chi_\p^0} , \ \
\chi_\p^{\rm c}= \frac{\chi_\p^0}{1+ U\chi_\p^0} ,
 \\
\chi_\p^0(\w_l)
 &=& -T\sum_{n,\k}G_\k(\e_n) G_{\k+\p}(\e_n+\w_l) ,
 \label{eqn:FLEX-chi0}
\end{eqnarray}
where $G_\k(\e_n)=(i\e_n+\mu-\e_\k-\Sigma_\k(\e_n))^{-1}$
is the thermal Green function, and $\mu$ is the chemical
potential.
$\e_n\equiv \pi(2n+1)T$, $\w_l\equiv 2\pi lT$
are the Matsubara frequencies for fermion and boson,
respectively.
$\chi^{\rm s}$ and $\chi^{\rm c}$ represent the 
spin and charge susceptibilities.
We solve the above eqs. (\ref{eqn:FLEX-S})
and (\ref{eqn:FLEX-chi0}) self-consistently,
under the constraint of constant electron density $n$
by choosing the chemical potential.
Hereafter, we study mainly the case for $n=0.9$ for LSCO and YBCO,
and for $n=1.15$ for NCCO.
In the present numerical study,
$64\times64$ $\k$-meshes and
512 Matsubara frequencies are used.

%%%%%%%%%%%%%%%%%%%%%%%%%%%%%%%%%%%%%%%%%%%%%%%%%%%%%%
\begin{figure}
%\vspace{10mm}
\begin{center}
\epsfig{file=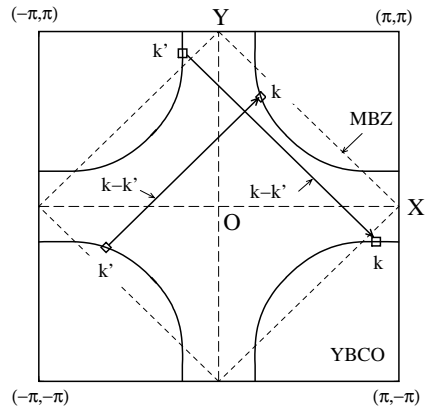,width=7.5cm}
\epsfig{file=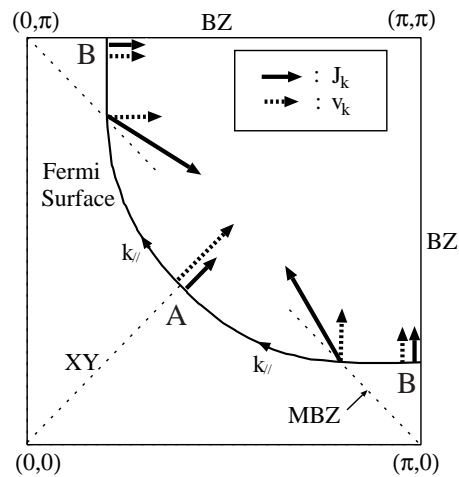,width=6cm}
\end{center}
%\vspace{10mm}
\caption{
Schematic $\k$-dependences of 
${\bf J}_\k$ and ${\bf v}_\k$ when the 
AF fluctuations with ${\bf Q}\approx(\pi,pi)$ are strong.
The cold-spot on the Fermi surface for hole-doped 
(electron-doped) systems exist around A (B).
}
  \label{fig:FS}
\end{figure}
%%%%%%%%%%%%%%%%%%%%%%%%%%%%%%%%%%%%%%%%%%%%%%%%%%%%%

In the present study,
the Stoner factor $\a_{\rm S}\equiv U\chi_{\bf Q}^0(0)$
at $T=0.02$ exceeds 0.99 both for $n=0.9$ (LSCO, YBCO) and
$n=1.15$ (NCCO).
(Note that $\a_{\rm S}<1$ at finite temperatures
in two dimensional systems because Marmin-Wagner
theorem is satisfied in the FLEX approximation.)
In this case,
$\chi_{\bf Q}^s \gg \chi_{\bf Q}^c $ is realized, and
$V^{\rm FLEX} \approx (3U^2/2)\chi^{\rm s}$.
Then,
by reflecting the strong $\q$-dependence of $\chi_\q^{\rm s}$,
$\gamma_\k\equiv {\rm Im}\Sigma_\k(-i\delta)$ on the FS
becomes very anisotropic.
($\gamma_\k$ represents the quasiparticle damping rate.)
$\gamma_\k$ takes a large value around the crossing points with
the magnetic Brillouin zone (MBZ)-boundary,
which we call {\it hot spots} as often referred to in literatures
 \cite{Rice,Pines-Hall}.
On the other hand, $\gamma_\k$ takes the minimum value 
at the points where the distance from the MBZ-boundary is the largest,
which are called {\it cold spots}.
(see Fig. \ref{fig:FS}.)
Transport phenomena for lower frequencies are governed by
the electronic properties around the cold spot.

As shown in fig.\ref{fig:FS} (a),
the location of the cold spot for 
hole-doped systems is around A.
Whereas the cold spot for electron-doped systems 
locates around B:
This fact was first predicted by ref. \cite{Kontani-Hall}
theoretically based on the FLEX approximation,
and it is verified by the ARPES measurement later
 \cite{Armitage,Armitage2}.
The difference of the position of the cold spot
explains the remarkable difference in
transport phenomena between hole-doped systems
and electron-doped ones
 \cite{Kontani-Hall}.
For example, 
both $R_{\rm H}$ and $S$ are positive in YBCO and LSCO
whereas they are negative in NCCO.

In the present FLEX
calculation for LSCO ($n=0.90$),
the maximum (minimum) value of $\gamma_\k$
on the FS is 0.38 (0.12) at $T=0.02$.
The ratio of anisotropy is 3.2,
which is consistent with ARPES measurements
in slightly under-doped compounds.
Such a small anisotropy cannot 
account for the enhancement of $R_{\rm H}$ in 
under-doped systems 
within the RTA
\cite{Pines-Hall}.
In the presence of strong AF fluctuations, however,
the total current ${\bf J}_\k$ becomes quite
different from the quasiparticle velocity ${\bf v}_\k$
due to the back-flow, which is totally dropped 
in the RTA
 \cite{Kontani-Hall}.
In fact, various anomalous transport phenomena in HTSC's
are brought by the nontrivial momentum-dependence of
${\bf J}_\k$ around the cold spot.

% AF-fluctuation model
Finally, we discuss the 
dynamical spin susceptibility given by the 
FLEX approximation.
Its phenomenological form is expressed as
\begin{eqnarray}
\chi_\k^{\rm s}(\w+i\delta)
 \approx \frac{\chi_Q}{1+\xi^2(\q-\Q)^2 -i\w/\w_{\rm sf}} ,
 \label{eqn:AFmodel}
\end{eqnarray}
where $\xi$ is the AF correlation length, and
$\Q$ is the nesting vector.
$\Q=(\pi,\pi)$ in both YBCO and NCCO
 \cite{Kontani-Hall}.
$\chi_Q\propto 1/\w_{\rm sf} \propto \xi^2$, and
$\xi^2 \propto T^{-1}$ in the FLEX approximation,
which is equivalent to the SCR theory
 \cite{Moriya}. 
Here we call eq.(\ref{eqn:AFmodel}) the AF-fluctuation model.

%\end{multicols}
%\begin{widetext}
%%%%%%%%%%%%%%%%%%%%%%%
\subsection{Conserving approximation for $\s_{\mu\nu}(\w)$}
%%%%%%%%%%%%%%%%%%%%%%%

Here, we explain how to calculate the
optical conductivity based on the Kubo formula.
To satisfy the conservation laws,
we have to take the current vertex correction (CVC)
into account in accordance with the Ward identity.

According to the Kubo formula,
the optical conductivity $\s_{xx}(\w)\equiv\s(\w)$
and the optical Hall conductivity $\s_{xy}(\w)$
are given by
\begin{eqnarray}
\s_{\mu\nu}(\w)=\frac1{i\w}\left[
 K_{\mu\nu}^{\rm R}(\w)- K_{\mu\nu}^{\rm R}(0) \right] ,
 \label{eqn:Kubo}
\end{eqnarray}
where $K_{\mu\nu}^{\rm R}(\w)$ is the retarded
current-current correlation function, which
are given by the analytic continuations of
the following thermal Green functions
 \cite{Eliashberg,Kohno}:
\begin{widetext}
\begin{eqnarray}
K_{xx}(i\w_l)
 &=& -2e^2 T\sum_{n,\k} v_{\k x}^0 g_\k(\e_n;\w_l)
  \Lambda_{\k x}(\e_n; \w_l) ,
 \label{eqn:Eliashberg1} \\
K_{xy}(i\w_l)
 &=& i\cdot 2e^3 T\sum_{n,\k} \Lambda_{\k x}(\e_n; \w_l)
 g_\k(\e_n;\w_l) 
 \nonumber \\
& &\times\frac12 \left\{ [ G_\k(\e_n+\w_l)v_{\k x}(\e_n+\w_l) - 
   G_\k(\e_n)v_{\k x}(\e_n) ]\frac{\d}{\d k_y}
 - \langle x \leftrightarrow y \rangle \right\}
  \Lambda_{\k y}(\e_n; \w_l) ,
 \label{eqn:Kohno1}
\end{eqnarray}
\end{widetext}
where $-e (e>0)$ represents the charge of an electron.
Here we put $\hbar=c=1$.
$g_\k(\e_n;\w_l)\equiv G_\k(\e_n+\w_l) G_\k(\e_n)$,
$v_{\k \mu}^0\equiv \d \e_\k / \d k_\mu$ is the
velocity of the non-interacting electron with $\k$, 
and $\Lambda_{\k \mu}(\e_n; \w_l)$ is the dressed current
which is given by
\begin{eqnarray}
{\bf \Lambda}_{\k}(\e_n; \w_l)
 &=& {\bf v}_{\k}^0 + T\sum_{m,\p}
 \Gamma_{\k,\p}(\e_n,\e_m; \w_l) g_\p(\e_m;\w_l) {\bf v}_{\p}^0
  \nonumber \\
 &=& {\bf v}_{\k}^0 + T\sum_{m,\p}
 \Gamma_{\k,\p}^{\rm I}(\e_n,\e_m; \w_l) g_\p(\e_m;\w_l) 
 \nonumber \\
& &\times {\bf \Lambda}_{\p}(\e_m; \w_l) ,
 \label{eqn:BS} 
\end{eqnarray}
where $\Gamma_{\k,\p}(\e_n,\e_m; \w_l)$ is the full
four-point vertex function, and $\Gamma_{\k,\p}^{\rm I}$ 
is the irreducible four-point vertex with respect to
the particle-hole channel.

Equation (\ref{eqn:Eliashberg1})
was derived by Eliashberg in 1962
 \cite{Eliashberg}.
Equation (\ref{eqn:Kohno1})
was first derived 
within the Born approximation 
 \cite{Fukuyama}.
later it was proved to be exact for any Fermi liquid system
up to the order of $O(\gamma^{-2})$, which is
the most divergent term with respect to $\gamma^{-1}$
 \cite{Kohno}.
In the same way,
the general formulae for the
magnetoresistance \cite{Kontani-MR-form},
the thermoelectric power and the Nernst coefficient
\cite{Kontani-S-form}
have been derived from the linear-response theory
(Kubo formula) on the basis of the microscopic Fermi 
liquid theory.

In the numerical study of transport coefficients,
one have to take account of the irreducible four point vertex
$\Gamma^{\rm I}$ 
to satisfy the conservation laws,
which is given by $\delta \Sigma^{\rm FLEX}/\delta G$
 \cite{Baym-Kadanoff}.
In the FLEX approximation, $\Gamma^{\rm I}$ is expressed as
\begin{widetext}
\begin{eqnarray}
\Gamma_{\k,\k'}^{\rm I}(\e_n,\e_{n'}; \w_l) 
 &=& V_{\k-\k'}^{\rm FLEX}(\e_n-\e_{n'})
 \nonumber \\
 & &- T\sum_{\p m} \left( G(k+k'+p)+G(k-p) \right)
 G(-p+k')W(p;p+\w_l) ,
 \label{eqn:G-I-FLEX} \\
W(p;p+\w_l)
 &=& \frac32 U^2(1+U\chi^{\rm s}(p))(1+U\chi^{\rm s}(p+\w_l))
 + \frac12 U^2(1-U\chi^{\rm c}(p))(1+U\chi^{\rm c}(p+\w_l)) - U^2 ,
\end{eqnarray}
\end{widetext}
where $k,k',p=(\k,\e_n),(\k',\e_{n'}),(\p,\w_l)$.
In the light hand side of eq.(\ref{eqn:G-I-FLEX}),
the first and the second terms are called 
the Maki-Thompson (MT) and the Aslamazov-Larkin (AL)
terms, respectively, in literature.

Using the Green function given by the FLEX approximation,
we numerically solve the Bethe-Salpeter equation for 
${\bf \Lambda}$, eq.(\ref{eqn:BS}), by iteration.
The kernel function
$\Gamma^{\rm I}$ is given in eq.(\ref{eqn:G-I-FLEX}).
Then, we obtain $K_{\mu\nu}(i\w_l)$
by inserting the obtained ${\bf \Lambda}$
into eqs.(\ref{eqn:Eliashberg1}) and (\ref{eqn:Kohno1}).
The retarded function
$K_{\mu\nu}^{\rm R}(\w)$ is derived from the
analytic continuation of $K_{\mu\nu}(i\w_l)$
with $\w_l \ge 0$, using the numerical Pade approximation.

Pade approximation is less reliable
when the function under consideration
is strongly $\w$-dependent,
and when the temperature is high because
the Matsubara frequency is sparse.
To increase the accuracy of the Pade approximation,
both $\Sigma(\e_n)$ and ${\bf \Lambda}(\e_n;\e_l)$ have to be obtained
with high accuracy; their relative errors should
be $10^{-8}\sim10^{-10}$.
In performing the Pade approximation,
we utilize the fact that the $i\w$-linear term
of $K_{\mu\nu}(\w)$ is equal to the DC value of $\s_{\mu\nu}$,
which can be obtained within the FLEX approximation
with high accuracy, as performed in refs.
\cite{Kontani-review,Kontani-MR,Kontani-Hall,Kontani-S,Kontani-N}.
By imposing this constraint on the Pade approximation, 
we succeed in deriving the
$\s_{\mu\nu}(\w)$ with enough accuracy in the present study.

In the present numerical study, 
we take the infinite series of the 
MT-terms in ${\bf \Lambda}$, whereas
we drop all the AL-terms in eq.(\ref{eqn:G-I-FLEX}).
This simplification is justified for DC-conductivities
when the AF fluctuations for ${\bf Q}\approx(\pi,\pi)$
are dominant, as proved in ref.\cite{Kontani-Hall}.
In the same way,
AL-terms would be also negligible 
for IR optical conductivities.
In fact, we have checked that the $f$-sum rule
both for $\s(\w)$ and $\s_{xy}(\w)$
are well satisfied even if all the AL-terms are dropped,
as will be shown in Fig.\ref{fig:sum}.
This results ensure the reliability of the present numerical study.
(Although we have also tried to include the AL-terms,
then the accuracy of the numerical Pade approximation 
became worse, unfortunately.)

In a Fermi liquid,
the relaxation time $\tau_\k$
in the relaxation time approximation (RTA)
is $1/2\gamma_\k \equiv 1/2{\rm Im}\Sigma_\k(-i\delta)$.
Transport coefficients can be expanded
in terms of $\gamma^{-1}$, which diverges 
as the temperature approaches zero.
Equation ({\ref{eqn:Eliashberg1}) is formally an exact expression.
On the other hand, eq.({\ref{eqn:Kohno1}) 
gives an exact expression for $\s_{xy}(\w)$ 
up to the order of $O(\gamma^{-2})$,
which is the most divergent term with respect to $\gamma^{-1}$.
Less singular terms, which are given as C and D
in ref.\cite{Kohno} (p.636) or the last two terms
in Fig. 3 of ref.\cite{Kontani-S-form}, 
are dropped in $K_{xy}$ given by eq.({\ref{eqn:Kohno1}).
In terms of the FLEX approximation,
eq.({\ref{eqn:Kohno1}) is exact beyond $O(\gamma^{-2})$
because both C and D vanish within this approximation
 \cite{Kontani-RH-kappa}.

%%%%%%%%%%%%%%%%%%%%%%%%%%%%%%%%%%%%%%%%%
\section{$\w$-linear term of $\s_{\mu\nu}(\w)$: 
The role of the CVC}
%%%%%%%%%%%%%%%%%%%%%%%%%%%%%%%%%%%%%%%%%

Based on the Kubo formula,
we derive the general formula for the 
$\w$-linear term of $\s_{\mu\nu}(\w)$,
which we denote $\Delta\s_{\mu\nu}$ hereafter.
The relation $\s_{\mu\nu}(\w)=\{ \s_{\mu\nu}(-\w) \}^\ast$
tells that $\Delta\s_{\mu\nu}$ is pure imaginary.
The most divergent terms of $\Delta\s \equiv \Delta\s_{xx}$
and $\Delta\s_{xy}$ are order of $O(\gamma^{-2})$ 
and $O(\gamma^{-3})$, respectively.
The derived expressions in this section are exact
up to the most divergent terms.
Based on the derived expression,
we discuss the role of the current vertex correction (CVC)
for $\Delta\s_{xy}$, and find that it is strongly 
enhanced when the AF fluctuations ate strong.
Readers who are not interested in the microscopic
derivation can skip to the next section,
where we will show the numerical results for $\s_{\mu\nu}(\w)$
given by the CVC-FLEX approximation.

%%%%%%%%%%%%%%%%%%%%%%%%%%
\subsection{Exact expression for the $\w$-linear term
of $\s_{\mu\nu}(\w)$}
%%%%%%%%%%%%%%%%%%%%%%%%%%

Here, we perform the analytic continuations
of eq.(\ref{eqn:Eliashberg1}) and (\ref{eqn:Kohno1})
according to refs.\cite{Eliashberg} and \cite{Kohno},
and derive the general formula for 
$\Delta\s_{\mu\nu}(\w)$.
In preparation for analyzing the CVC in later sections,
we derive the expressions without CVC,
which corresponds to the relaxation time approximation (RTA)
with $\k$- and $\e$-dependent relaxation time $\tau_\k(\e)$.
After the analytic continuation of eq.(\ref{eqn:Eliashberg1}),
\begin{eqnarray}
\s^0(\w)&=& e^2\sum_\k\int\frac{d\e}{\pi}
% \left(-\frac{\d f}{\d\e} \right)
 \frac1{2\w} \left[{\rm th}\frac{\e_{+}}{2T}
 -{\rm th}\frac{\e_{-}}{2T} \right]
 \nonumber \\
& &\times G_\k^{\rm R}(\e_{+})G_\k^{\rm A}(\e_{-}) \{v_{\k x}^0\}^2 ,
 \label{eqn:sxx0}
\end{eqnarray}
where $\e_{\pm}\equiv \e \pm \w/2$,
and we have dropped the terms with 
$G^{\rm R}(\e_{+})G^{\rm R}(\e_{-})$ and 
$G^{\rm A}(\e_{+})G^{\rm A}(\e_{-})$ 
because they are less divergent with respect to
$\gamma^{-1}$, and they vanish at $T=0$
 \cite{Eliashberg}.
Here we expand eq.(\ref{eqn:sxx0}) with respect to $\w$:
\begin{eqnarray}
& &{\rm th}\frac{\e_{+}}{2T}-{\rm th}\frac{\e_{-}}{2T}
 = 2\w\left(-\frac{\d f}{\d\e} \right)+ O(\w^2) ,
 \\
& &G_\k^{\rm R}(\e_{+})G_\k^{\rm A}(\e_{-})
 = |G_\k(\e)|^2 + iz_\k^{-1}(\e)\w 
 \\
& &\ \ \ \ \ \ \ \ \ \ \ \ \ \ \ \ \ \ \ \ \ \ 
 \times |G_\k(\e)|^2 |{\rm Im}G_\k(\e)|+ O(\w^2) ,
 \nonumber 
\end{eqnarray}
where $z_\k(\e)=(1-\frac{\d}{\d\e}{\rm Re}\Sigma_\k(\e))^{-1}$
is the renormalization factor.
At sufficiently lower temperature,
the Green function for $\w\sim 0$ and $\e_\k\sim \mu$
is well approximated as
\begin{eqnarray}
 G_\k^{\rm R}(\w)=\frac{z_\k}{\w-\e_\k^\ast +i\gamma_\k^\ast} ,
 \label{eqn:QPrep}
\end{eqnarray}
where $\e_\k^\ast=z_\k(\e_\k+ {\rm Re}\Sigma_\k(0)-\mu)$
and $\gamma_\k^\ast= z_\k\gamma_\k$.
Equation 
is called the quasiparticle representation of Green function,
whose validity is assured by the microscopic Fermi liquid theory.
When eq.(\ref{eqn:QPrep}) is valid,
\begin{eqnarray}
& & |{\rm Im}G_\k(\w)|
 = \pi z_\k\delta(\w-\e_\k^\ast) ,
 \nonumber \\
& &|G_\k(\w)|^2
 = \frac{\pi}{\gamma_\k}z_\k\delta(\w-\e_\k^\ast) ,
 \nonumber \\
& &|G_\k(\w)|^2|{\rm Im}G_\k(\e)|
 = \frac{\pi}{2\gamma_\k^2}z_\k\delta(\w-\e_\k^\ast) ,
 \label{eqn:Gdelta}\\
& &|G_\k(\w)|^4
  = \frac{\pi}{2\gamma_\k^3}z_\k\delta(\w-\e_\k^\ast) ,
 \nonumber \\
& &|G_\k(\w)|^2 {\rm Re}G_\k^2(\e) 
  = -\frac{\pi}{4\gamma_\k^3}z_\k\delta(\w-\e_\k^\ast) ,
 \nonumber 
\end{eqnarray}
for $\w,\e_\k^\ast\sim0$.
As a result,
the expression for $\s^0(\w)$ 
expanded with respect to $\w$ is 
\begin{eqnarray}
\s^0(\w)
 &=& e^2\sum_\k \left(-\frac{\d f}{\d\e} \right)
 \frac{\{v_{\k x}^0\}^2}{\gamma_\k}
 \label{eqn:sxx0-2} \\
& & + i\w \cdot e^2\sum_\k \left(-\frac{\d f}{\d\e} \right)
 \frac{z_\k^{-1}\{v_{\k x}^0\}^2}{2\gamma_\k^2} + O(\w^2) .
 \nonumber
\end{eqnarray}
In a free-dispersion model $\e_\k=k^2/2m$,
eq.(\ref{eqn:sxx0-2}) becomes
\begin{eqnarray}
\s^0(\w)
&=& \frac{e^2n}{m\cdot 2\gamma}
 +iz^{-1}\w \frac{e^2n}{m\cdot (2\gamma)^2}
 \nonumber \\
&\approx& \frac{e^2n}{m(2\gamma-iz^{-1}\w)} ,
\end{eqnarray}
which is equal to the Drude form for $\s(\w)$
given by the RTA if we replace $2\gamma$ with $\tau^{-1}$.

In the same way,
$\s_{xy}(\w)$ without CVC is given by the 
analytic continuation of eq.(\ref{eqn:Kohno1}):
\begin{eqnarray}
\s_{xy}^0(\w)
&=& -e^3\sum_\k\int\frac{d\e}{\pi}
% \left(-\frac{\d f}{\d\e} \right)
 \frac1{2\w} [{\rm th}\frac{\e_{+}}{2T}-{\rm th}\frac{\e_{-}}{2T}]
 G_\k^{\rm R}(\e_{+})G_\k^{\rm A}(\e_{-})
 \nonumber \\
& &\times  
 \left[G_\k^{\rm R}(\e_{+})-G_\k^{\rm A}(\e_{-})\right]
 \frac i2 A_{\k} ,
 \label{eqn:sxy0} \\
A_{\k}&=& v_{\k x}^0 \left(v_{\k x}^0\frac{\d}{\d k_y}
 - v_{\k y}^0\frac{\d}{\d k_x} \right) v_{\k y}^0 ,
 \label{eqn:Ak0}
\end{eqnarray}
where we have dropped CVC for the quasiparticle velocity
given by the momentum derivative of the self-energy.
Up to the order of $O(\w)$, we see that
%
%\begin{widetext}
\begin{eqnarray}
& &G_\k^{\rm R}G_\k^{\rm A} 
 \left[G_\k^{\rm R}-G_\k^{\rm A}\right]\frac i2
=|G_\k(\e)|^2|{\rm Im} G_\k(\e)|
 \nonumber \\
& & \ \ \ \ \ \ \ \ \ \ \ \ \ \ \ 
 + iz_\k^{-1}\w \left\{ \frac12 |G_\k(\e)|^4
 - |G_\k(\e)|^2 {\rm Re} G_\k^2(\e) \right\}
 \nonumber \\
& &\ \ \ \ \ \ \ \ \ \ \ \ \ 
 \approx \frac{\pi}{2\gamma_\k^2}
 z_\k^{-1}\delta(\e-\e_\k^\ast)
 + iz_\k^{-1}\w \frac{\pi}{2\gamma_\k^3}\delta(\e-\e_\k^\ast) .
 \nonumber \\
\end{eqnarray}
%\end{widetext}
%
As a result,
the expression for $\s_{xy}^0(\w)$ within the order of
$O(\w)$ is
\begin{eqnarray}
\s_{xy}^0(\w)
 &=& -e^3\sum_\k \left(-\frac{\d f}{\d\e} \right)
 \frac{A_\k}{2\gamma_\k^2}
\nonumber \\
& & + i\w \cdot (-e^3)\sum_\k \left(-\frac{\d f}{\d\e} \right)
 \frac{z_\k^{-1}A_\k}{2\gamma_\k^3} .
 \label{eqn:sxy0-2}
\end{eqnarray}
In a free-dispersion model, eq.(\ref{eqn:sxy0-2}) becomes
\begin{eqnarray}
\s_{xy}^0(\w)
&=& \frac{-e^3n}{m^2\cdot(2\gamma)^2}
 +2iz^{-1}\w \frac{-e^3n}{m^2\cdot (2\gamma)^3}
 \nonumber \\
&\approx& \frac{-e^3n}{m^2(2\gamma-iz^{-1}\w)^2} ,
\end{eqnarray}
which is also equal to the Drude formula given by the RTA.

In the next stage, we derive the general expression
for $\Delta\s_{\mu\nu}$ by taking all the CVC's into account.
After the analytic continuations of eqs.(\ref{eqn:Eliashberg1})
and (\ref{eqn:Kohno1})
 \cite{Eliashberg,Kohno},
\begin{widetext}
\begin{eqnarray}
\s(\w)
&=& e^2\sum_\k\int\frac{d\e}{\pi}\left(-\frac{\d f}{\d\e} \right)
 {\tilde {v}}_{\k x}(\e;\w) g_\k^{(2)}(\e;\w) J_{\k x}(\e;\w) ,
 \label{eqn:sxx-w} \\
\s_{xy}(\w)
&=& -e^3 \sum_\k\int\frac{d\e}{\pi}\left(-\frac{\d f}{\d\e} \right)
 J_{\k x}(\e;\w) g_\k^{(2)}(\e;\w)
 \nonumber \\
& &\times \frac{i}{2} \left[ \left(
 [G_\k^{\rm R}v_{\k x}^{\rm R}]_{\e_{+}} - 
 [G_\k^{\rm A} v_{\k x}^{\rm A}]_{\e_{-}} \right)\frac{\d}{\d k_y}
 - \langle x \leftrightarrow y \rangle \right] J_{\k y}(\e;\w) ,
 \label{eqn:sxy-w}
\end{eqnarray}
\end{widetext}
where 
$g_\k^{(1)}(\e;\w)= G_\k^{\rm R}(\e_{+})G_\k^{\rm R}(\e_{-})$,
$g_\k^{(2)}(\e;\w)= G_\k^{\rm R}(\e_{+})G_\k^{\rm A}(\e_{-})$
and
$g_\k^{(3)}(\e;\w)= G_\k^{\rm A}(\e_{+})G_\k^{\rm A}(\e_{-})$.
$v_{\k \mu}^{\rm R}(\e)= v_{\k \mu}^0 
 + \frac{\d}{\d k_\mu} \Sigma_\k^{\rm R}(\e)$.
$G^{\rm R}$ and $G^{\rm A}$ are the retarded and advanced
Green functions, respectively.
${\bf J}$, ${\bf v}$ and ${\bf \tilde{v}}$
are given by
\begin{widetext}
\begin{eqnarray}
{\bf J}_{\k}(\e;\w)
&=& {\bf v}_{\k}(\e;\w) + \sum_{\q} \int\frac{d\e'}{4\pi i}
 {\cal T}_{22}(\k\e,\q\e';\w) g_{\q}^{(2)}(\e';\w)
 {\bf v}_{\k}(\e';\w)
 \nonumber \\
&=& {\bf v}_{\k}(\e;\w) + \sum_{\q} \int\frac{d\e'}{4\pi i}
 {\cal T}_{22}^{(0)}(\k\e,\q\e';\w) g_{\q}^{(2)}(\e';\w)
 {\bf J}_{\k}(\e';\w) ,
 \label{eqn:Jw} \\
{\bf v}_{\k}(\e;\w)
&=& {\bf v}_{\k}^0 + \sum_{\q,j=1,3} \int\frac{d\e'}{4\pi i}
 {\cal T}_{2j}^{(0)}(\k\e,\q\e';\w) g_{\q}^{(j)}(\e';\w)
 {\bf v}_{\q}^0 ,
 \label{eqn:v}\\
{\bf \tilde{v}}_{\k}(\e;\w)
&=& {\bf v}_{\k}^0 + \sum_{\q,j=1,3}(j-2)\int\frac{d\e'}{4\pi i}
 {\bf v}_{\q}^0 \cdot {\rm th}\left(\frac{\e'+(j-2)\w/2}{2T}\right)
 g_{\q}^{(j)}(\e';\w)\Gamma_{j2}(\q\e',\k\e;\w) ,
 \label{eqn:vtilde}
\end{eqnarray}
\end{widetext}
where the definition of ${\cal T}_{22}^{(0)}(\k\e,\q\e';\w)$
is given in ref.\cite{Eliashberg}.
${\cal T}_{22}^{(0)}$ is a subgroup of
${\cal T}_{22}$ which is irreducible with respect to 
$g_{\q}^{(2)}$, whereas it is reducible 
with respect to $g_{\q}^{(1,3)}$.
The following Bethe-Salpeter equation holds;
${\cal T}_{22} = {\cal T}_{22}^{(0)}
 + \sum\int {\cal T}_{22}^{(0)} g^{(2)}{\cal T}_{22}$.

As explained in ref.\cite{Eliashberg},
${v}_{\k x}(\e;\w=0)={\tilde {v}}_{\k x}(\e;\w=0)
= {\rm Re}v_\k^{\rm R}(\e) \equiv v_\k(\e)$ 
is well satisfied in a Fermi liquid.
As a result
 \cite{Eliashberg,Kohno},
\begin{widetext}
\begin{eqnarray}
\s(0)
&=& e^2\sum_\k\int\frac{d\e}{\pi}\left(-\frac{\d f}{\d\e} \right)
  |G_\k(\e)|^2 {v}_{\k x}(\e) J_{\k x}(\e) ,
 \label{eqn:sigw0} \\
\s_{xy}(0)
&=& -e^3\sum_\k\int\frac{d\e}{\pi}\left(-\frac{\d f}{\d\e} \right)
 |G_\k(\e)|^2|{\rm Im}G_\k(\e)|
 |{\bf v}_{\k}(\e)| {J}_{\k x}(\e) \frac{\d}{\d k_{\parallel}} J_{\k y}(\e)
 \nonumber \\
&=& -e^3\sum_\k\int\frac{d\e}{2\pi}\left(-\frac{\d f}{\d\e} \right)
 |G_\k(\e)|^2|{\rm Im}G_\k(\e)|
 |{\bf v}_{\k}(\e)| \left( {\bf J}_{\k}(\e) \times 
 \frac{\d {\bf J}_{\k}(\e)}{\d k_{\parallel}} \right)_z ,
 \label{eqn:sxyw0} 
\end{eqnarray}
\end{widetext}
where $k_{\parallel} \equiv
({\hat e}_z\times {\bf v})/|{\bf v}|$, which is
parallel to the Fermi surface.
In deriving the first line in eq.(\ref{eqn:sxyw0}),
we have used the relation
$v_{\k x}\frac{\d}{\d k_y}
 - v_{\k y}\frac{\d}{\d k_x}
= ({\hat e}_z\times {\bf v} )\nabla
= |{\bf v}|\frac{\d}{\d k_{\parallel}}$.
The Onsager's relation
$\s_{xy}=-\s_{yx}$ is used in deriving the
second line in eq.(\ref{eqn:sxyw0}).

Here we expand ${\bf J}_{\k}(\e;\w)$ with respect to $\w$
as ${\bf J}_{\k}(\e=0;\w)= {\bf J}_{\k}(0)
 +i\w {\bf J}_{\k}^{(1)}(0)$.
From eq.(\ref{eqn:Jw}),
one can check that the most divergent term of
${\bf J}_{\k}^{(1)}$ is proportional to $\gamma^{-1}$, which
comes from the $\w$-derivative of $g^{(2)}$
or that of the thermal factor in ${\cal T}_{22}^{(0)}$;
$\frac{\d}{\d \w}{\cal T}_{22}^{(0)}(\e,\e';\w)
\approx 2(-\d f/\d\e'){\rm Re}\Gamma(0,0)+ O(\w^2)$.
For simplicity, we denote hereafter
$J(\e;\w=0) \equiv J(\e)$,
$\Gamma(\k\e,\p\e';\w=0) \equiv \Gamma(\k\e,\p\e')$, and so on.
On the other hand,
the $\w$-linear term of eq.(\ref{eqn:v}) or (\ref{eqn:vtilde})
is not singular with respect to $\gamma^{-1}$,
so we put $\w=0$ in eqs.(\ref{eqn:v}) and (\ref{eqn:vtilde})
hereafter.

Using the relations in eq.(\ref{eqn:Gdelta}),
${\bf J}_{\k}^{(1)} \propto \gamma^{-1}$ is given by
\begin{eqnarray}
{\bf J}_{\k}^{(1)}
&=& \sum_{\q} \frac{1}{4 i}
 {\cal T}_{22}(\k0,\q\e_{\q}^\ast) 
 \frac{z_\q}{\gamma_{\q}} \cdot
 \frac{{\bf J}_{\q}}{2\gamma_{\q}^\ast}
 \nonumber \\
& & -\sum_{\q}\frac{1}{4 i}
 \left( 4 i \delta_{\k,\q}
 + {\cal T}_{22}(\k0,\q\e_{\q}^\ast)
 \frac{ z_{\q}}{\gamma_{\q}} \right)
 \nonumber \\
& &\times \sum_{\q'}{\rm Re}\Gamma(\q,\q')
 \left(-\frac{\d f}{\d\e}\right)_{\e_{\q'}^\ast} z_{\q'}
 \frac{{\bf J}_{\q'}}{2\gamma_{\q'}}
 \nonumber \\
&\equiv& {\bf L}_{\k}+{\bf M}_{\k} ,
 \label{eqn:J1}
\end{eqnarray}
where the first and the second terms come from
the $\w$-derivatives of $g^{(2)}$ and ${\cal T}_{22}^{(0)}$,
respectively.
Then, the $\w$-linear term of the conductivity,
which we denote as $\Delta\s_{\mu\nu}$, is given by
\begin{eqnarray}
\Delta\s
&=& i\w\cdot e^2\sum_\k \left(-\frac{\d f}{\d\e} \right)
 \left[ \frac{ z_\k^{-1} v_{\k x} J_{\k x}}{2\gamma_\k^2}
 + \frac{ v_{\k x} J_{\k x}^{(1)}}{\gamma_\k} \right] ,
 \nonumber \\
 \\
\Delta\s_{xy}
&=& i\w\cdot (-e^3)\sum_\k \left(-\frac{\d f}{\d\e} \right)
 \left[ \frac{z_\k^{-1}A_\k}{2\gamma_\k^3}
 + \frac{B_\k}{2\gamma_\k^2} \right] ,
 \label{eqn:Dsxy}
 \\
%A_\k &=& J_{\k x} \left(v_{\k x}\frac{\d}{\d k_y}
% - v_{\k y}\frac{\d}{\d k_x} \right) J_{\k y} ,
A_\k &=& J_{\k x}
 |{\bf v}_\k|\frac{\d}{\d k_{\parallel}} J_{\k y} ,
 \\ 
B_\k &=& J_{\k x}^{(1)} 
 |{\bf v}_\k|\frac{\d}{\d k_{\parallel}} J_{\k y}
 + J_{\k x} |{\bf v}_\k|\frac{\d}{\d k_{\parallel}}
 J_{\k y}^{(1)} ,
%B_\k &=& J_{\k x}^{(1)} \left(v_{\k x}\frac{\d}{\d k_y}
% - v_{\k y}\frac{\d}{\d k_x} \right) J_{\k y}
% + J_{\k x} \left(v_{\k x}\frac{\d}{\d k_y}
% - v_{\k y}\frac{\d}{\d k_x} \right) J_{\k y}^{(1)} ,
%\langle J_{\k x}^{(1)} \rightarrow J_{\k x},
% J_{\k y} \rightarrow J_{\k y}^{(1)} \rangle
\end{eqnarray}
where
$|{\bf v}_\k|\frac{\d}{\d k_{\parallel}}
= v_{\k x}\frac{\d}{\d k_y}
 - v_{\k y}\frac{\d}{\d k_x}$ as explained before.
They are exact with respect to the most divergent terms
with respect to $\gamma^{-1}$.
Note that $\Delta\s_{\mu\nu}$ is pure imaginary.

%\begin{multicols}{2}
%\end{widetext}

Here, we further analyze ${\bf J}_\k^{(1)}$ given 
in eq. (\ref{eqn:J1}). 
First, 
${\bf L}_{\k}$ is given by the following Bethe-Salpeter equation:
\begin{eqnarray}
{\bf P}_{\k}
&=& \frac{{\bf J}_{\k}}{2\gamma_\k^\ast}
+ \sum_{\q} \frac{1}{4 i}
 {\cal T}_{22}(\k0,\q\e_{\q}^\ast) 
 \frac{z_{\q}}{\gamma_{\q}} \cdot
 \frac{{\bf J}_{\q}}{2\gamma_{\q}^\ast} 
 \nonumber \\
&=& \frac{{\bf J}_{\k}}{2\gamma_\k^\ast}
+ \sum_{\q} \frac{1}{4 i}
 {\cal T}_{22}^{\rm I}(\k0,\q\e_{\q}^\ast) 
 \frac{z_{\q}}{\gamma_{\q}} \cdot {\bf P}_{\q} ,
 \label{eqn:P} \\
{\bf L}_{\k}
&=& {\bf P}_{\k}- \frac{{\bf J}_{\k}}{2\gamma_\k^\ast} ,
 \label{eqn:L}
\end{eqnarray}
where 
${\cal T}_{22}^{\rm I}(\k0,\q\e_{\q}^\ast)
 = 2i {\rm Im}V_{\k-\q}^{\rm R}(\e_{\q}^\ast)
 ( {\rm cth}(\e_{\q}^\ast/2T)-{\rm th}(\e_{\q}^\ast/2T))$
in the FLEX approximation.
In a similar way,
${\bf M}_{\k}$ is rewritten as
\begin{eqnarray}
{\bf M}_{\k}
&=& {\bf N}_{\k}+ \sum_{\q} \frac{1}{4 i}
 {\cal T}_{22}^{\rm I}(\k0,\q\e_{\q}^\ast) 
 \frac{z_{\q}}{\gamma_{\q}} \cdot {\bf M}_{\q} ,
 \label{eqn:M} \\
{\bf N}_{\k}
&=& -\sum_{\q}{\rm Re}\Gamma(\k,\q)
 \left(-\frac{\d f}{\d\e}\right)_{\e_{\q}^\ast} z_{\q}
 \frac{{\bf J}_{\q}}{2\gamma_{\q}} .
 \label{eqn:N}
\end{eqnarray}

At sufficiently lower temperatures,
the expression for $\Delta\s_{\mu\nu}(\w)$ in eq.(\ref{eqn:Dsxy})
is rewritten by using ${\bf {\tilde J}}_{\k}^{(1)}$ 
(instead of ${\bf {J}}_{\k}^{(1)}$) as,
\begin{eqnarray}
\Delta\s
&=& i\w\cdot e^2\sum_\k \left(-\frac{\d f}{\d\e} \right)
 \frac{v_{\k x} {\tilde J}_\k^{(1)}}{\gamma_\k} ,
 \\
\Delta\s_{xy}
&=& i\w\cdot (-e^3)\sum_\k \left(-\frac{\d f}{\d\e} \right)
 \frac{z_\k^{-1}{\tilde B}_\k}{2\gamma_\k^2} ,
 \\
{\tilde B}_\k 
 &=& {\tilde J}_{\k x}^{(1)} 
 |{\bf v}_\k|\frac{\d}{\d k_{\parallel}} J_{\k y}
 + J_{\k x} |{\bf v}_\k|\frac{\d}{\d k_{\parallel}}
 {\tilde J}_{\k y}^{(1)} ,
% &=& {\tilde J}_{\k x}^{(1)} \left(v_{\k x}\frac{\d}{\d k_y}
% - v_{\k y}\frac{\d}{\d k_x} \right) J_{\k y}
% + J_{\k x} \left(v_{\k x}\frac{\d}{\d k_y}
% - v_{\k y}\frac{\d}{\d k_x} \right) {\tilde J}_{\k y}^{(1)} ,
 \\
{\bf \tilde J}_{\k}^{(1)} &=& {\bf P}_{\k}+{\bf M}_{\k}
 = {\bf J}_{\k}^{(1)}+ \frac{{\bf J}_{\k}}{2\gamma_\k^\ast} .
 \label{eqn:J1tilde}
\end{eqnarray}
In the next subsection,
we will discuss the temperature dependence of
$\Delta\s_{xy}$ when the AF fluctuations are strong,
by analyzing the $\k$-dependence of ${\bf \tilde J}_{\k}^{(1)}$.
We will approximately solve the Bethe-Salpeter equations
(\ref{eqn:P}) and (\ref{eqn:M}) based on the
AF-fluctuation model.

In a Fermi liquid, $\gamma_\k$ and $z_\k$
are expressed as
\begin{eqnarray}
\gamma_\k
&=& \frac1{4i} \sum_{\q}\int d\e {\cal T}_{22}^{\rm I}(\k0,\q\e)
 \rho_{\q}(\e)
 \nonumber \\
&\approx& \frac1{4i} \sum_{\q} 
 {\cal T}_{22}^{\rm I}(\k0,\q\e_{\q}^\ast) z_{\q} ,
 \label{eqn:gamma}\\
z_\k^{-1}-1 
&=& -\left. \frac{\d}{\d\e}{\rm Re}\Sigma_\k(\e) \right|_{\e=0}
 \nonumber \\
&=& \sum_\q {\rm Re}\Gamma(\k,\q)
 \left(-\frac{\d f}{\d\e}\right)_{\e_{\q}^\ast}z_{\q}
 \nonumber \\
& & - \frac{\d}{\d\mu}{\rm Re}\Sigma_\k(0) .
 \label{eqn:z}
\end{eqnarray}
Note that
the uniform charge susceptibility
in a Fermi liquid is given by
$\displaystyle \chi_{\rm c}=
 [1-{\d}{\rm Re}\Sigma_\k(0)/{\d\mu}]\chi_{\rm c}^0$,
where $\chi_{\rm c}^0$ is the susceptibility for $U=0$.
Because $\chi_{\rm c} \ll \chi_{\rm c}^0$
is expected in strongly correlated systems
(like in heavy Fermion systems), the relation
${\d}{\rm Re}\Sigma_\k(0)/{\d\mu} \simle 1$ 
should be satisfied.
This relation is also expected to be realized 
in HTSC according to the FLEX approximation.

%%%%%%%%%%%%%%%%%%%%%%%%%%
\subsection{Role of the CVC in the presence of 
AF fluctuations}
%%%%%%%%%%%%%%%%%%%%%%%%%%

Using the general expression for $\Delta\s_{\mu\nu}$
derived in the previous subsection,
we discuss its temperature dependence
when the AF fluctuations are strong.
For that purpose,
we approximately analyze the CVC included
in the expression for $\s_{\mu\nu}(\w)$
based on the AF fluctuation model
given in eq.(\ref{eqn:AFmodel}).

First, we explain
%discuss the role of the CVC on $J_\k$ for $\w=0$, which is 
the total current for the DC-conductivity.
The Bethe-Salpeter equation eq.(\ref{eqn:Jw})
is rewritten at lower temperatures as
 \cite{Kontani-Hall}
\begin{eqnarray}
{\bf J}_{\k}= {\bf v}_{\k}
+ \sum_{\p}{\cal T}_{22}^{\rm I}(\k0,\p\e_\p^\ast)z_{\p}\cdot 
 \frac 1{\gamma_\p} \cdot {\bf J}_{\p}
 \label{eqn:JJJ}
\end{eqnarray}
for $\e=\w=0$.
In the FLEX approximation,
${\cal T}_{22}(\k0,\p\e)\approx
 3U^2\cdot i{\rm Im}\chi_{\k-\p}^{\rm s}(\e+i\delta)
 [ {\rm cth}(\e/2T)-{\rm th}(\e/2T) ]$.
Due to the thermal factor,
${\cal T}_{22}(\k0,\p\e)$ takes large value
only when $|\e|\simle T$. 
If we apply the AF-fluctuation model, eq.(\ref{eqn:AFmodel}),
the main contributions of the $\p$-summation
in eq.(\ref{eqn:JJJ}) come from the region
$|\p-\k'|\simle 1/\xi$,
where $\k'$ is the momentum on the FS defined as
$(k_x',k_y')=-{\rm sgn}( k_x k_y)\cdot(k_y,k_x)$,
as shown in Fig. \ref{fig:FS} (b).
We see the relation $\k-\k'\approx\Q$ is satisfied
on the FS.
Here we assume $|\Q-(\k-\k')|\simle \xi^{-1}$ 
even at the cold spot, which is in fact satisfied
in the present FLEX approximation for hole-doped systems
 \cite{Kontani-Hall}.
In this case,
$\gamma_{\k_{\rm c}}\propto T$ is satisfied.

Taking account of the expression for $\gamma_\k$
given in eq.(\ref{eqn:gamma}),
we obtain a simplified Bethe-Salpeter  equation
 \cite{Kontani-Hall},
\begin{eqnarray}
{\bf J}_\k= {\bf v}_\k + \a_\k\cdot{\bf J}_{\k'} .
 \label{eqn:BS_simple}
\end{eqnarray}
where $\a_\k= |\sum_{\p}{\cal T}_{22}^{\rm I}(\k0,\p\e_\p^\ast)z_{\p}
 {\bf J}_{\p}/\gamma_\p |/|{\bf J}_{\k}|$.
According to the AF-fluctuation model, 
$\a_\k \approx \langle \ \cos(\theta_J(\q)\!-\!\theta_J(\k')) 
\ \rangle_{|\q-\k'|<1/\xi} \approx (1-c/\xi^2)<1$ where
$c\sim O(1)$ is a constant.
$\a_\k$ takes the maximum value around hot spots.
The solution of eq. (\ref{eqn:BS_simple}) is 
 \cite{Kontani-Hall}
\begin{eqnarray}
{\bf J}_{\k}= \frac 1{1-\a_\k^2} 
 \left( {\bf v}_{\k} + \a_\k \cdot {\bf v}_{\k'} \right) ,
  \label{eqn:Jy}
\end{eqnarray}
whose schematic behavior is shown in fig. \ref{fig:FS} (b).
Note that 
$(v_{\k'x},v_{\k'y})=-{\rm sgn}( k_x k_y)
 \cdot(v_{\k y},v_{\k x})$ and
$(J_{\k'x},J_{\k'y})=-{\rm sgn}( k_x k_y)
 \cdot(J_{\k y},J_{\k x})$.

We stress that 
the same vertex functions ${\cal T}_{22}(\k0,\p\e)$
appear in eqs.(\ref{eqn:Jw}) and (\ref{eqn:gamma}),
which is the consequence of the Ward identity
and is satisfied in any conserving approximation.
This fact assures that the relation
$1-\a \propto \xi^2$ holds even beyond the FLEX approximation.

In the same way, 
we study ${\bf {\tilde J}}_\k^{(1)}={\bf P}_\k+{\bf M}_\k$.
The Bethe-Salpeter equation for ${\bf P}_\k$,
eq. (\ref{eqn:P}), is simplified as
\begin{eqnarray}
{\bf P}_{\k}
\approx  \frac{{\bf J}_{\k}}{2\gamma_\k^\ast}
 + \a_{\k} {\bf P}_{\k'} ,
\end{eqnarray}
where $\a_{\k}$ is the same as that in eq.(\ref{eqn:BS_simple}).
The solution is given by
\begin{eqnarray}
{\bf P}_{\k}
\approx  \frac{1}{2\gamma_\k^\ast}
 \frac{{\bf J}_\k+ \a_\k {\bf J}_{\k'}}{1-\a_\k^2} .
 \label{eqn:PP}
\end{eqnarray}

Next, we analyze ${\bf N}_\k$ in eq.(\ref{eqn:N}).
Considering the relation 
${\rm Re}\Gamma_{\k,\q} 
 \sim z^{-1}{\rm Re}V_{\k-\q}^{\rm FLEX}(0)$, 
we obtain
\begin{eqnarray}
{\bf N}_{\k}
&\approx& -\frac{{\bf J}_{\k'}}{2\gamma_{\k}}
 \cdot {\bar \a}_\k \sum_{\q} {\rm Re}\Gamma(\k,\q)
 \left(-\frac{\d f}{\d\e}\right)_{\e_{\q}^\ast} z_{\q} ,
 \label{eqn:N-approx}
\end{eqnarray}
where ${\bar \a}_\k \simle {\a}_\k$ is expected
in general because the momentum dependence of 
${\rm Im}V_{\k-\p}^{\rm FLEX}(\e)/\e$ is
much prominent than that of 
${\rm Re}V_{\k-\p}^{\rm FLEX}(0)$, which are included 
in eqs.(\ref{eqn:JJJ}) and (\ref{eqn:N-approx}), respectively.
Actually, ${\rm Im}\chi_\q^{\rm s}(\w)/\w
\propto \{\chi_\q^{\rm s}(0)\}^{2}
\propto \xi^4 (1+\xi^2({\bf Q}-\q)^2)^{-2}$
according to eq. (\ref{eqn:AFmodel}).
In fact, in the present FLEX
calculation for LSCO ($n=0.90$),
the maximum (minimum) value of Im$\Sigma_\k(-i\delta)$
on the FS is 0.38 (0.12) at $T=0.02$; 
the ratio of anisotropy is 3.2, 
reflecting the sharp $\q$-dependence of  
${\rm Im}\chi_\q^{\rm s}(\w)/\w$.
On the other hand, the maximum (minimum) value of 
$z_\k^{-1}-1$
on the FS is 5.0 (3.8); the ratio of anisotropy
is only 1.4.

Considering eq.(\ref{eqn:z}),
we rewrite eq.(\ref{eqn:N-approx}) as
\begin{eqnarray}
{\bf N}_{\k}
&\approx& -\frac{{\bf J}_{\k'}}{2\gamma_{\k}}
 \cdot {\bar \a}_\k \left( z_\k^{-1}-\chi_{\rm c}/\chi_{\rm c}^0 \right)
 \equiv -{\tilde \a}_\k \frac{{\bf J}_{\k'}}{2\gamma_{\k}^\ast} .
\end{eqnarray}
${\tilde \a}_\k \simle {\bar \a}_\k \simle {\a}_\k$
should be satisfied because 
$\chi_{\rm c}/\chi_{\rm c}^{0}$ is positive.
We note again that
$\chi_{\rm c}/\chi_{\rm c}^0 \ll 1$ in strongly 
correlated systems.
Then, the approximate solution of eq.(\ref{eqn:M}) is 
\begin{eqnarray}
{\bf M}_{\k}
\approx  -\frac{{\tilde \a}_\k}{2\gamma_\k^\ast}
 \frac{\a_\k {\bf J}_\k+{\bf J}_{\k'}}{1-\a_\k^2} .
 \label{eqn:MM}
\end{eqnarray}

In conclusion, an approximate expression for 
$\s_{xy}(\w)$ up to the order of $O(\w)$ is given by
\begin{eqnarray}
\s_{xy}(\w)
&=& -e^3\sum_\k \left(-\frac{\d f}{\d\e} \right)
 |{\bf v}_\k|\frac{1}{4\gamma_\k^2}
 \nonumber \\
& & \times \left( {\bf \tilde J}_\k(0;\w)\times 
\frac{\d {\bf\tilde J}_\k(0;\w)}{\d k_\parallel} \right)_z ,
 \label{eqn:sxy-ap} \\
{\bf \tilde J}_\k(0;\w)
&=& {\bf J}_\k(0;\w)+ i\w\frac{{\bf J}_\k}{2\gamma_\k^\ast}
 = {\bf J}_\k + i\w {\bf\tilde J}_\k^{(1)} .
 \label{eqn:JwJw} 
\end{eqnarray}
where ${\bf\tilde J}_\k^{(1)}$ is given by eqs.
(\ref{eqn:J1tilde}), (\ref{eqn:PP}) and (\ref{eqn:MM}). 
After a simple but lengthy calculation,
${\bf \tilde J} \times \d {\bf\tilde J}/\d k_\parallel$
in eq.(\ref{eqn:sxy-ap}) is rewritten as 
\begin{eqnarray}
& &{\bf \tilde J}_\k(0;\w)\times 
\frac{\d {\bf\tilde J}_\k(0;\w)}{\d k_\parallel} 
= {\bf J}_\k\times 
 \frac{\d {\bf J}_\k}{\d k_\parallel} 
 \nonumber \\
& &\ \ \ \ \ \ \ \ \ \ \ \ \ 
 + \frac{i\w}{\gamma_\k^\ast}
 \frac{1-{\tilde \a}_\k{\a}_\k}{1-\a_\k^2}
 {\bf J}_\k\times 
 \frac{\d {\bf J}_\k}{\d k_\parallel} 
 \nonumber \\
& &\ \ \ \ \ \ \ \ \ \ \ \ \ 
 + i\w \frac{\d}{\d k_\parallel}
 \left( \frac1{2\gamma_\k^\ast} 
 \frac{\a_\k-{\tilde \a}_\k}{1-\a_\k^2} \right)
 {\bf J}_\k \times {\bf J}_{\k'} 
 \nonumber \\
& &\ \ \ \ \ \ \ \ \ \ \ \ \ 
 \ \ \ + O(\w^2) ,
 \label{eqn:JxJ}
\end{eqnarray}
where the second and the third terms
contribute to $\Delta\s_{xy}$.
Note that 
\begin{eqnarray}
& &\left( {\bf J}_\k\times \frac{\d {\bf J}_\k}{\d k_\parallel} 
 \right)_z = \left| {\bf J}_\k\ \right|^2
 \left( \frac{\d\theta_\k^J}{\d k_\parallel} \right) ,
 \label{eqn:JdJ} \\
& &\left( {\bf J}_\k \times {\bf J}_{\k'} \right)_z
= {\rm sgn}(k_x k_y) \frac{v_{\k y}^2 - v_{\k x}^2}{1-\a_\k^2} ,
\end{eqnarray}
where $\theta^J=\tan^{-1}(J_y/J_x)$.
At the cold spot in hole-doped systems, 
eq.(\ref{eqn:JdJ}) is proportional to $\xi^2 \propto T^{-1}$
because $|{\bf J}_{\k}| \simle |{\bf v}_{\k}|$
and $(\d\theta^J/\d k_\parallel) \propto 
\xi^2 \cdot (\d\theta^v/\d k_\parallel)$ 
around the cold-spot in hole-doped systems
(point A in Fig.\ref{fig:FS}),
which we denote as $\k_{\rm c}$ hereafter.
Note that $(\d\theta^v/\d k_\parallel)$
represents the curvature of the FS
 \cite{Kontani-Hall}.

Let us consider the hole-doped system, 
where the cold spot locates on the XY line
in Fig.\ref{fig:FS} (a).
Because ${\bf J}_\k \times {\bf J}_{\k'}=0$
at the cold spot,
the second term of the right-hand-side of
eq.(\ref{eqn:JxJ}) gives the main contribution
to $\Delta\s_{xy}$.
This fact immediately tells that
\begin{eqnarray}
& &\s_{xy}(\w) = \frac{a}{2\gamma}
 + iz^{-1}\w \frac{b}{(2\gamma)^2},
  \label{eqn:expand-sxy} \\
& &\ \ \ \
 a \propto \xi^2, \ \ \ b \propto \xi^m ,
 \nonumber 
\end{eqnarray}
whereas $a=b=1$ if all the CVC's are dropped, 
i.e., in the RTA.
$\k_{\rm c}$ represents the cold spot.
Considering eq.(\ref{eqn:JxJ}) and
the relation $1-\a \propto \xi^{-2}$,
we obtain that
$m\approx 4$ when ${\tilde \a}_\k \ll {\a}_\k$, 
and $m\approx 2$ when ${\tilde \a}_\k \approx {\a}_\k$.
As discussed above,
${\tilde \a}_\k \simle {\a}_\k$
is expected by the present analysis for the CVC.
In the next section,
we will show that the relation $m\approx 3$ 
holds for hole-doped systems in the numerical study.

%In the case of electron-doped systems,
%on the other hand,
%the third term of the right-hand-side of
%eq.(\ref{eqn:JxJ}) will contribute
%to $\Delta\s_{xy}$, too.
%By this reason, the relation (\ref{eqn:sgn-Dsxy})
%will not always hold.

In a similar way, we also discuss the role of 
the CVC in $\s(\w)$ for hole-doped systems.
Because ${\bf J}_{\k}=-{\bf J}_{\k'}$
at the cold spot $\k_{\rm c},$ we obtain
\begin{eqnarray}
{\bf {\tilde J}}_{\k_{\rm c}}^{(1)}
&\approx&
 \frac {1+{\tilde \a}_{\k_{\rm c}}}{1+\a_{\k_{\rm c}}}
 \frac{{\bf J}_{\k_{\rm c}}}{2\gamma_{\k_{\rm c}}^\ast},
% \nonumber \\
%&\sim& \frac {{\bf v}_{\k_{\rm c}}}{2\gamma_{\k_{\rm c}}^\ast} ,
\end{eqnarray}
which is close to 
${\bf J}_{\k_{\rm c}}/2\gamma_{\k_{\rm c}}^\ast$
because $\a_{\k_{\rm c}},{\tilde \a}_{\k_{\rm c}} \simle 1$.
We note that $|{\bf J}_{\k_{\rm c}}| \simle |{\bf v}_{\k_{\rm c}}|$
because ${\bf J}_{\k_{\rm c}} \approx 
 {\bf v}_{\k_{\rm c}}/(1+\a_{\k_{\rm c}})$.
This result suggests that the CVC changes the values of 
$\s$ and $\Delta\s$ only slightly.
As a result,
\begin{eqnarray}
& &\s(\w) \propto v_{\k_{\rm c}} J_{\k_{\rm c}}
 \left( \frac{1}{2\gamma_{\k_{\rm c}}}
 + iz^{-1}\w \frac{1}{(2\gamma_{\k_{\rm c}})^2} \right),
  \label{eqn:expand-sig} 
\end{eqnarray}
In conclusion, $\s(\w)$ is insensitive 
against the CVC, as the case of 
DC-conductivity within the FLEX approximation
 \cite{Kontani-Hall}.
We will show in the next section that 
$\s(\w) \approx \s^{\rm RTA}(\w)$ 
holds for hole-doped systems in the numerical study.

According to eqs. (\ref{eqn:expand-sxy}) and 
(\ref{eqn:expand-sig}), 
the Hall coefficient and the Hall angle are given by
\begin{eqnarray}
R_{\rm H}(\w) &\propto&
 a+ i\w z^{-1}
 \left( b-a \right) \frac1{\gamma_{\k_{\rm c}}}
 + O(\w^2) ,
 \label{eqn:expand-RH} \\
\theta_{\rm H}^{-1}(\w) &\propto&
 \frac{\gamma_{\k_{\rm c}}}{a}-i\w z^{-1}
 \left( \frac{2b}{a^2}-\frac1{a} \right) 
 + O(\w^2) ,
 \label{eqn:expand-IHA}
\end{eqnarray}
where $a=b=1$ in the absence of the CVC,
that is, in the RTA.

As a result, 
we can conclude that the origin of 
the anomalous behaviors of $R_{\rm H}(\w)$
and $\theta_{\rm H}(\w)$, 
that is, prominent deviations form the 
extended-Drude (ED) formula,
is the strong temperature dependences of
$a$ and $b$ which originate from $\xi$.
We will discuss this mechanism in more detail
in later sections.

%%%%%%%%%%%%%%%%%%%%%%%%%%%%%%%%%%%%%%%%%
\section{Numerical Results}
%%%%%%%%%%%%%%%%%%%%%%%%%%%%%%%%%%%%%%%%%

In this section, we show the optical conductivities
obtained by the FLEX approximation with full MT-type CVC's.
This kind of calculation has been performed for the first time.
Hereafter, we call this scheme ``the CVC-FLEX approximation''.
We will see that $\s_{xy}(\w)$ shows striking deviation from 
the ED-form, which is highly consistent with experimental results.
%Unit
Here, the unit of energy is the nearest-neighbor hopping
integral $t_0$, which corresponds to $\sim4000$K 
according to LDA band calculation.
Thus, $T=0.01 \sim40$K and $\w=0.1 \sim 300$cm$^{-1}$
in the present study.

%%%%%%%%%%%%%%%%%%%%%%%%%%%%%
\subsection{DC transport coefficients}
%%%%%%%%%%%%%%%%%%%%%%%%%%%%%

%%%%%%%%%%%%%%%%%%%%%%%%%%%%%%%%%%%%%%%%%%%%%%%%%%%%%%
\begin{figure}
%\vspace{10mm}
\begin{center}
\epsfig{file=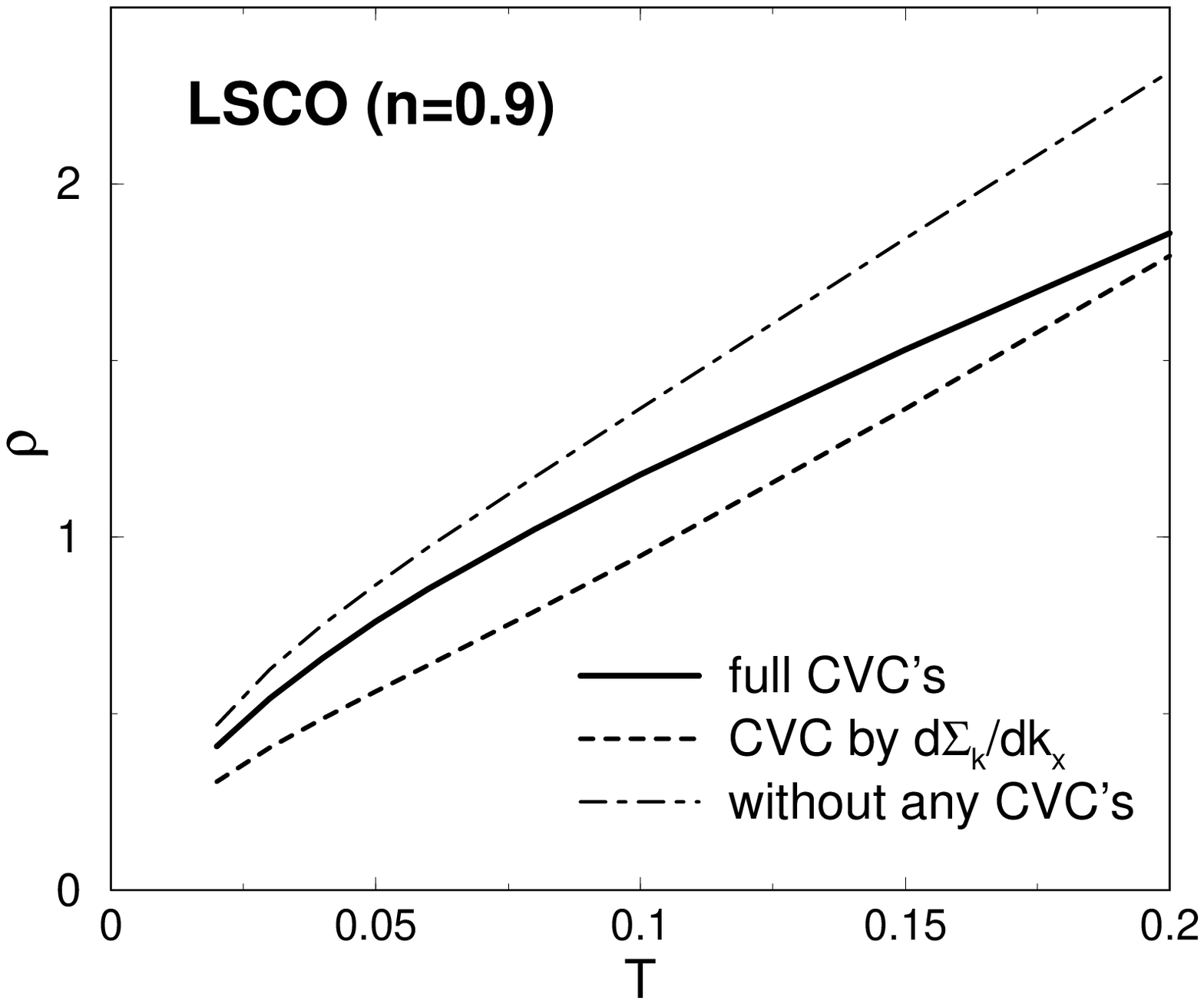,width=7cm}
\epsfig{file=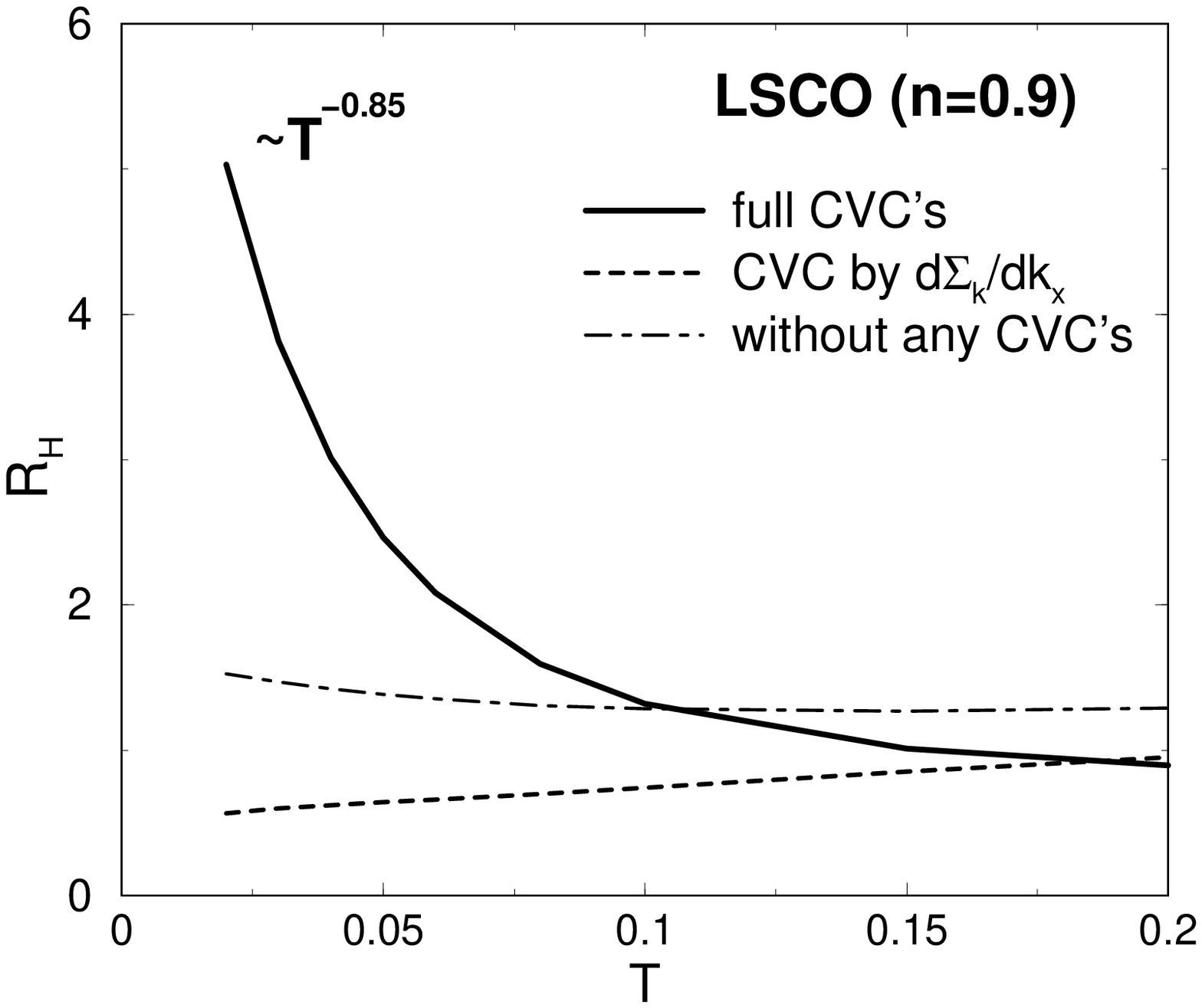,width=7cm}
\end{center}
%\vspace{10mm}
\caption{
Resistivity and the Hall coefficient for $\w=0$
obtained by the FLEX approximation:
(i) without any CVC's, 
(ii) with CVC given by $d\Sigma_\k(0)/dk_\mu$,
(iii) with all the CVC's (CVC-FLEX).
At lower temperatures,
$R_{\rm H}\propto T^{-0.9}$ for (iii)
and $R_{\rm H}^0\propto T^{-0.2}$ for (i),
respectively.
}
  \label{fig:DC}
\end{figure}
%%%%%%%%%%%%%%%%%%%%%%%%%%%%%%%%%%%%%%%%%%%%%%%%%%%%%

Before discussing the optical conductivities, 
we shortly explain the DC transport phenomena
given by the CVC-FLEX approximation
 \cite{Kontani-Hall}.
Obtained $\rho$ and $R_{\rm H}$ are shown in Fig.\ref{fig:DC}.
Results by the CVC-FLEX approximation,
which are calculated using eqs.(\ref{eqn:sigw0}) 
and (\ref{eqn:sxyw0}),
are denoted as ``full CVC's'' in figures.
Curie-Weiss like behavior of $R_{\rm H}$ 
(more precisely $R_{\rm H}\propto T^{-0.85}$)
is reproduced due to the CVC.
In the Fermi liquid theory,
the CVC is divided into (i) the back-flow
which is expressed by ${\cal T}_{22}$, and 
(ii) the renormalization of ${\bf v}_\k$ given by $d\Sigma_\k/dk_x$.
To clarify the effect of the back-flow,
we calculate the conductivities by 
replacing all the $J_{\k \mu}(\e)$'s
with $v_{\k \mu}(\e)$'s in 
eqs.(\ref{eqn:sigw0}) and (\ref{eqn:sxyw0}).
The obtained results are denoted as
``CVC by $d\Sigma_\k/dk_x$'' in fig.\ref{fig:DC}.
%which means the fact that corrections for the velocity
%by the self-energy are taken.
They correspond to the ``without VC'' in ref.
\cite{Kontani-Hall}.
We see that the resistivity increases
to some extent due to the back-flow (${\cal T}_{22}$).

Furthermore, 
we calculate the conductivities by
replacing all the $J_{\k \mu}(\e)$'s and $v_{\k \mu}(\e)$'s
with $v_{\k \mu}^0$ in eqs.(\ref{eqn:sigw0}) and (\ref{eqn:sxyw0}).
The results 
are shown as ``without any CVC's'' in fig.\ref{fig:DC}.
Then, the resistivity takes the smallest value
because self-energy correction for the velocity
enhances the conductivity; 
$|{\bf v}_\k(0;0)| > |{\bf v}_\k^0|$ at the cold spot.
Hereafter, ``RTA'' in figures 
represents the results by ``without any CVC's''.
As will be shown in figs. \ref{fig:L-sigma} and \ref{fig:Y-sigma},
the DC-conductivity by RTA is smaller than 
that by the conserving approximation,
because the effect of the velocity correction
dominates the back-flow effect.

%%%%%%%%%%%%%%%%%%%%%%%%%%%%%
\subsection{$\s(\w)$ and $\s_{xy}(\w)$}
%%%%%%%%%%%%%%%%%%%%%%%%%%%%%

Here, we perform the numerical calculation for
the complex optical conductivities,
$\s(\w)$ and $\s_{xy}(\w)$,
using the FLEX approximation.
The CVC is taken into account in the conserving way.
We calculate $\s_{\mu\nu}(\w)$
by eq.(\ref{eqn:Kubo}), where
$K_{\mu\nu}(\w)$ is derived from eqs.
(\ref{eqn:Eliashberg1}) and (\ref{eqn:Kohno1})
using the Pade approximation.
As explained above, we utilize the values of $\s(\w=0)$ and 
$\s_{xy}(\w=0)$, which are derived from eqs. (\ref{eqn:sigw0})
and (\ref{eqn:sxyw0}) as shown in ref.\cite{Kontani-Hall},
in the course of the Pade approximation.
This procedure is highly demanded to achieve enough accuracy.
$64\times64$ $\k$-meshes and 512 Matsubara frequencies
are used in the present FLEX approximation.

%%%%%%%%%%%%%%%%%%%%%%%%%%%%%%%%%%%%%%%%%%%%%%%%%%%%%
\begin{figure}
%\vspace{10mm}
\begin{center}
\epsfig{file=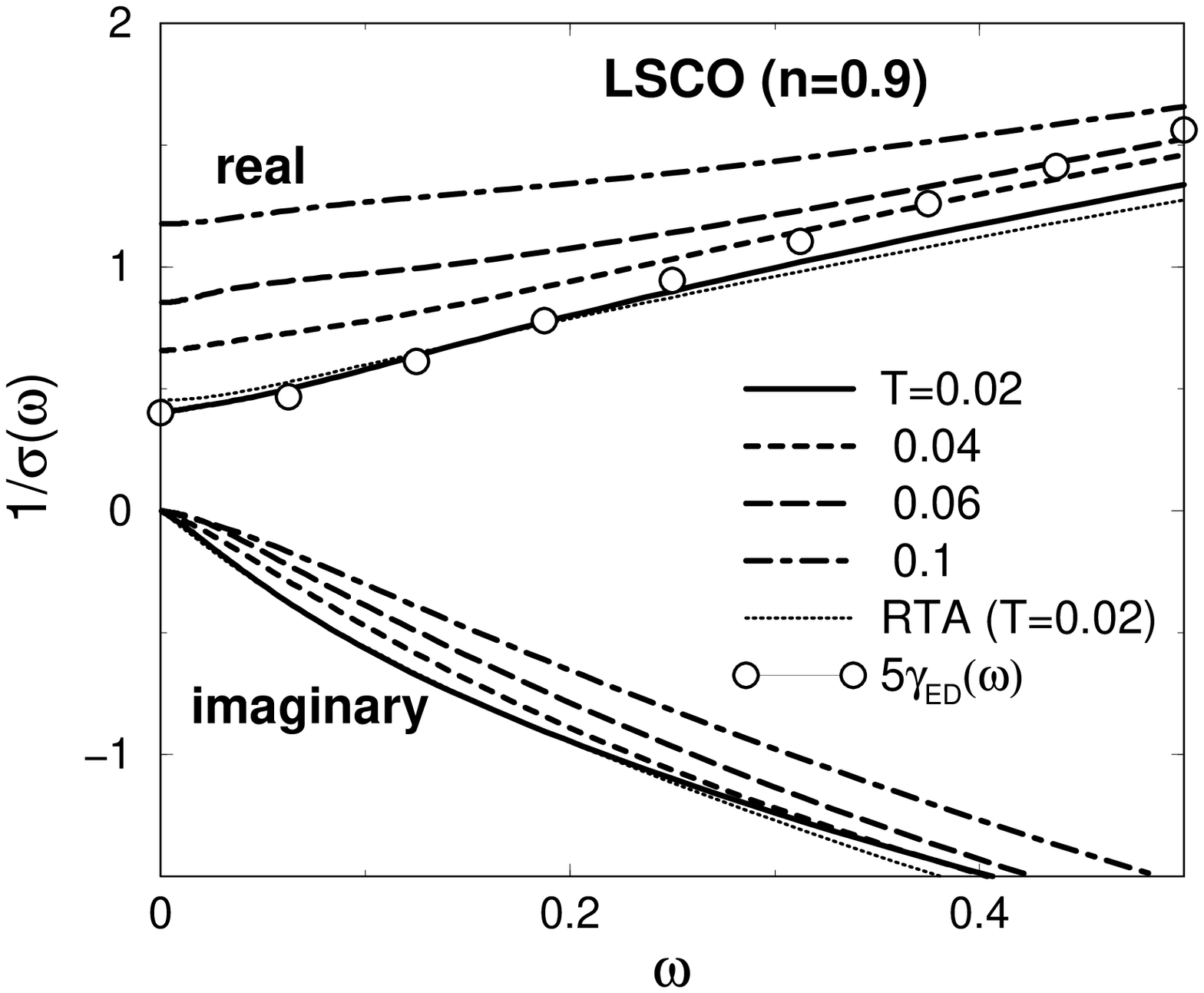,width=7cm}
\epsfig{file=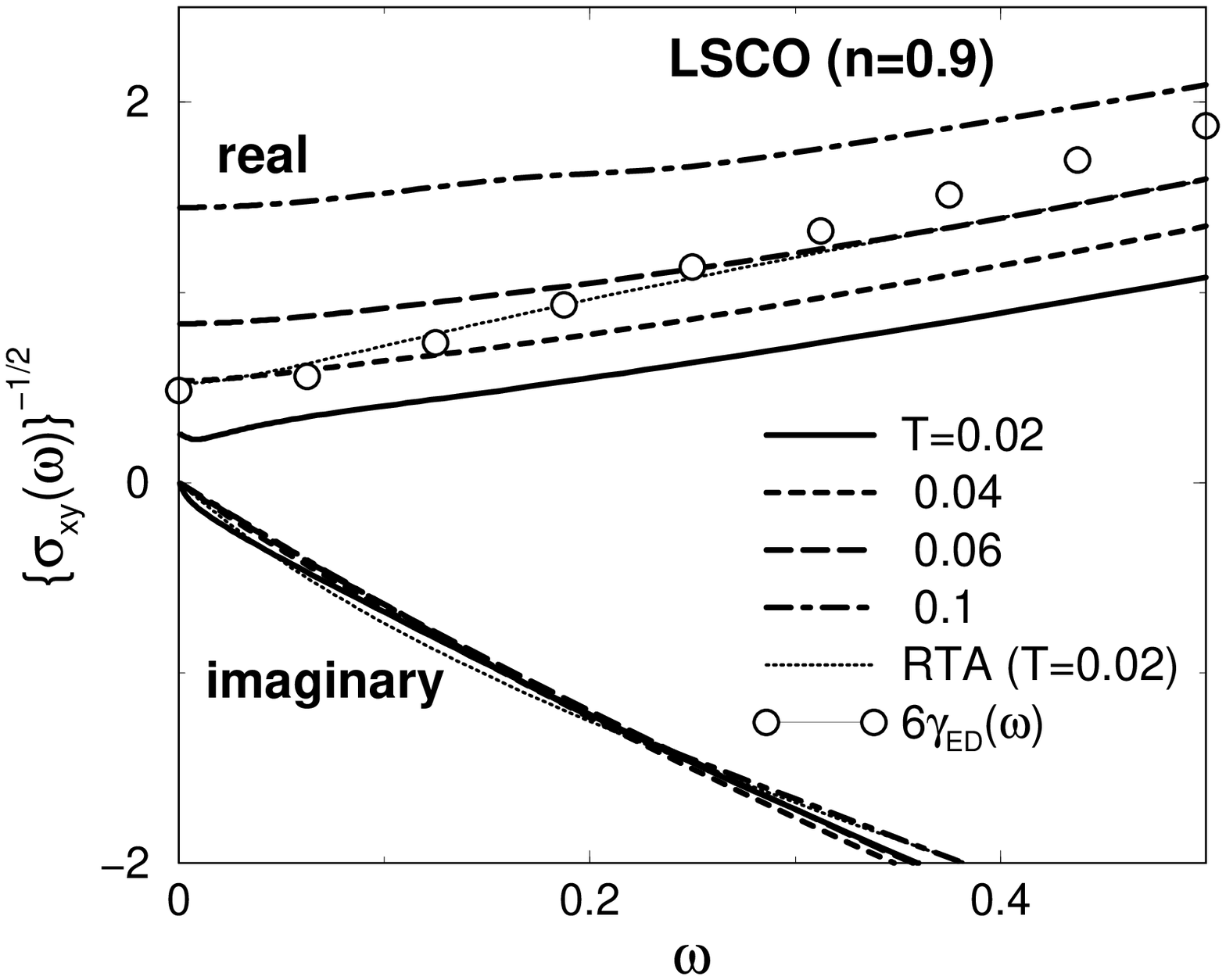,width=7cm}
\end{center}
%\vspace{10mm}
\caption{
Obtained $\w$-dependence of 
(a) $\{ \s(\w) \}^{-1}$ and (b) $\{ \s_{xy}(\w) \}^{-0.5}$
by CVC-FLEX approximation.
Suffix '0' represents the results by RTA.
Real parts of 
$\{ \s(\w) \}^{-1}$, $\{ \s^0(\w) \}^{-1}$ and 
$\{ \s_{xy}^0(\w) \}^{-0.5}$ are approximately
proportional to $\g_{\rm ED}(\w)$ for $\w < 0.2$,
as expected by the extended-Drude expression.
However, $\w$-dependence of
Re$\{ \s_{xy}(\w) \}^{-0.5}$ is much weaker 
than that of $\g_{\rm ED}(\w)$ 
due to the CVC.
}
  \label{fig:INVSIG}
\end{figure}
%%%%%%%%%%%%%%%%%%%%%%%%%%%%%%%%%%%%%%%%%%%%%%%%%%%%%

Here we derive an extended-Drude (ED) forms for 
$\s_{\mu\nu}^0(\w)$ from the Kubo formula within the RTA, 
where the suffix 0 means the result by the RTA hereafter.
At zero temperature, $\s^0(\w)$ 
for smaller $\w$ $(\w\simle\gamma)$ is given by,
\begin{eqnarray}
\s^0(\w)&=& \frac{e^2}{\w}\sum_\k\int_{-\w}^{0}\frac{d\e}{\pi}
 \frac{z}{\w+\e-\e_\k^\ast+i\gamma_\k^\ast(\w+\e)}
\nonumber \\
& &\times \frac{z}{\e-\e_\k^\ast-i\gamma_\k^\ast(\e)}v_{\k x}^2
 \nonumber \\
&\sim& N(0)\frac{e^2}{\w}\int_{-\w}^{0}d\e
 \frac{2 v_{\k_{\rm c}}^2}{\gamma_{\k_{\rm c}}(\e)
 +\gamma_{\k_{\rm c}}(\w+\e) 
 - iz_{\k_{\rm c}}^{-1}\w} ,
 \nonumber \\
 \label{eqn:sig-est}
\end{eqnarray}
where $\k_{\rm c}$ represents the cold spot. 
$\w$-dependence of $z(\w)$ has been neglected.
In deriving eq.(\ref{eqn:sig-est}),
we take only the contribution comes from the cold spot
into account.
In the same way,
\begin{eqnarray}
\s_{xy}^0(\w)
&\sim& -N(0)\frac{e^3}{\w}
 \nonumber \\
& &\times \int_{-\w}^{0}d\e
 \frac{A_{\k_{\rm c}}}{\left(\gamma_{\k_{\rm c}}(\e)
 +\gamma_{\k_{\rm c}}(\w+\e) 
 - iz_{\k_{\rm c}}^{-1}\w \right)^2} ,
 \nonumber \\
 \label{eqn:sxy-est}
\end{eqnarray}
where $A_{\k}$ is given in eq.(\ref{eqn:Ak0}).
In a crude expectation,
$\gamma_{\k_{\rm c}}(\e)+\gamma_{\k_{\rm c}}(\w+\e)$
in eqs.(\ref{eqn:sig-est}) and (\ref{eqn:sxy-est})
would take the minimum value around $\e\sim -\w/2$.
As a result, we obtained the following 
ED expressions for smaller $\w$:
\begin{eqnarray}
\s^{\rm ED}(\w) &=&
 \frac{\Omega}{2\gamma_{\rm ED}(\w)+ iz^{-1}\w} ,
 \label{eqn:ED1} \\
\s_{xy}^{\rm ED}(\w)&=& 
 \frac{\Omega_{xy}}{(2\gamma_{\rm ED}(\w)+ iz^{-1}\w)^{2}} ,
 \label{eqn:ED2} 
\end{eqnarray}
where $\gamma_{\rm ED}(\w)(>0)$ is approximately given by
\begin{eqnarray}
\gamma_{\rm ED}(\w)&\equiv& 
 \frac12 \left[ \gamma_{\k_{\rm c}}(\w/2)
 +\gamma_{\k_{\rm c}}(-\w/2) \right] .
 \label{eqn:AV}
\end{eqnarray}
Below, we will show that the above ED formulae for $\s(\w)$ 
still holds even if CVC is taken into account, 
whereas it completely fails for $\s_{xy}(\w)$
owing to the CVC, which is the origin of
anomalous behaviors of $R_{\rm H}(\w)$ and 
$\theta_{\rm H}(\w)$.

%%%%%%%%%%%%%%%%%%%%%%%%%%%%%%%%%%%%%%%%%%%%%%%%%%%%%
%\begin{figure}
%%\vspace{10mm}
%\begin{center}
%\epsfig{file=L-invsxy-T.eps,width=7cm}
%\end{center}
%%\vspace{10mm}
%\caption{
%Obtained $T$-dependence of $\{ \s_{xy}(\w) \}^{-0.5}$.
%Its real part is almost $T$-linear as observed experimentally.
%(H.D. Drew, private communication.)
%}
%  \label{fig:INVSIG2}
%\end{figure}
%%%%%%%%%%%%%%%%%%%%%%%%%%%%%%%%%%%%%%%%%%%%%%%%%%%%%%

Figure \ref{fig:INVSIG} (a) shows the obtained 
$\s^{-1}(\w)$ for LSCO.  
We see that
${\rm Re}\{\s^{-1}(\w)\}$ possesses strong 
$\w$-dependence so the simple Drude formula is violated.
It is approximately
proportional to $\gamma_{\rm ED}(\w)$
for $\w\simle 0.2$, which suggests that
the ED-form in eq.(\ref{eqn:ED1})
is well satisfied, even if CVC is taken into account.
We also see that ${\rm Im}\{\s(\w)\}^{-1}$
shows moderate $\w$- and temperature-dependences,
which is proportional to $\w z^{-1}(\w)$
according to the ED-model.
As shown in fig. \ref{fig:INVSIG},
its gradient decreases as $\w$ and/or $T$ increases,
which is naturally explained
as the $\w$- and $T$-dependences of
$z^{-1}(\w)$ ($\approx m^\ast/m$).

Figure \ref{fig:INVSIG} (b) shows 
$\{\s_{xy}(\w)\}^{-0.5}$ for LSCO.
We recognize that 
Re$\{\s_{xy}^0(\w)\}^{-0.5}\propto \gamma_{\rm ED}(\w)$
within the RTA, whereas
Re$\{\s_{xy}(\w)\}^{-0.5}$ given by the CVC-FLEX 
approximation possesses much moderate $\w$-dependence.
These results means that
the ED-form in eq.(\ref{eqn:ED2}) 
is well satisfied for $\s_{xy}^0(\w)$,
while it is violated for $\s_{xy}(\w)$
due to the CVC.
We will study the role of the CVC in $\s_{xy}(\w)$
in more detail hereafter.
We also see that 
Im$\{\s_{xy}(\w)\}^{-0.5}$ is almost unchanged
against the temperature,
whereas its gradient slightly decreases as $\w$ increases.

%%%%%%%%%%%%%%%%%%%%%%%%%%%%%%%%%%%%%%%%%%%%%%%%%%%%%%
\begin{figure}
%\vspace{10mm}
\begin{center}
\epsfig{file=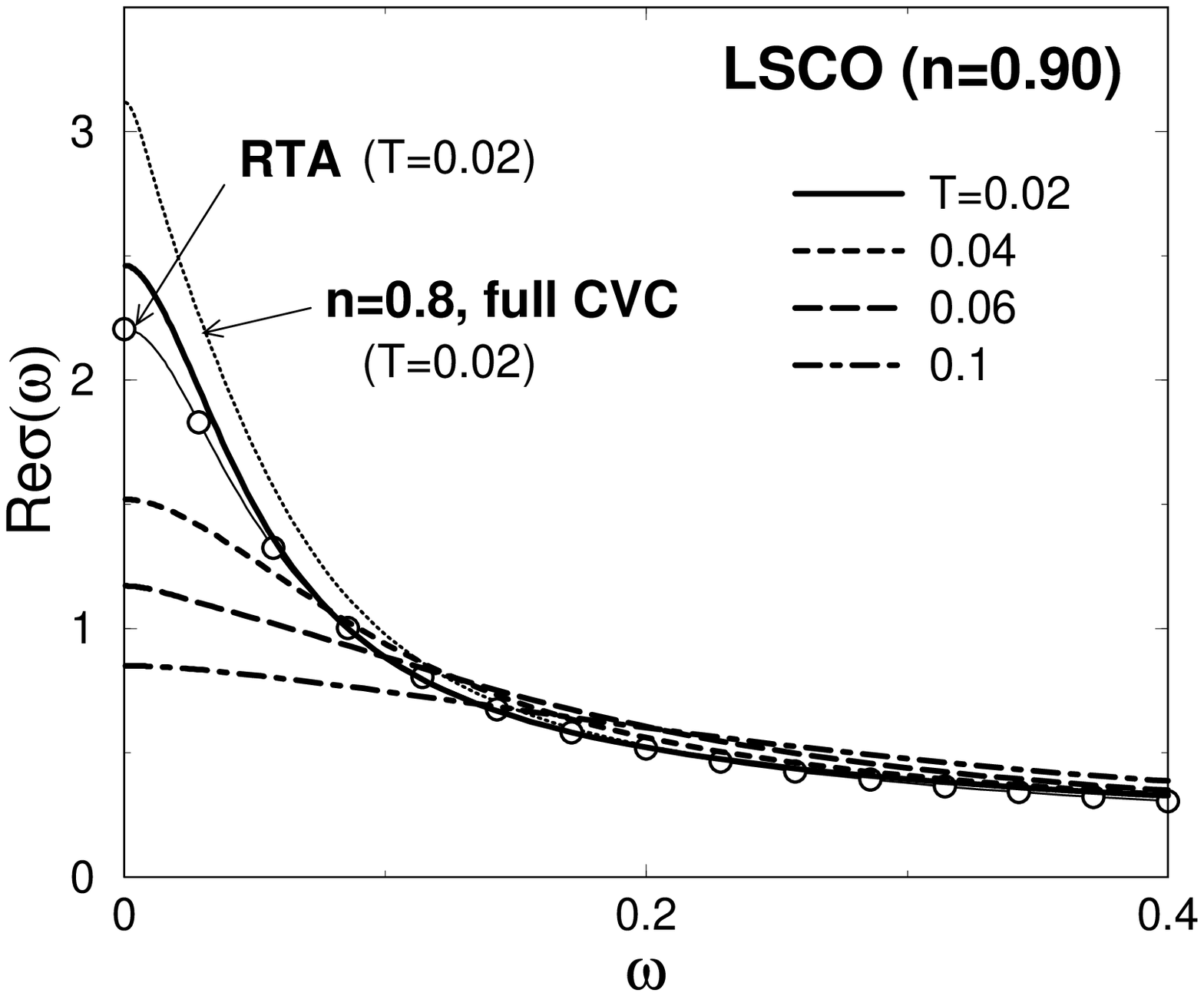,width=7cm}
\epsfig{file=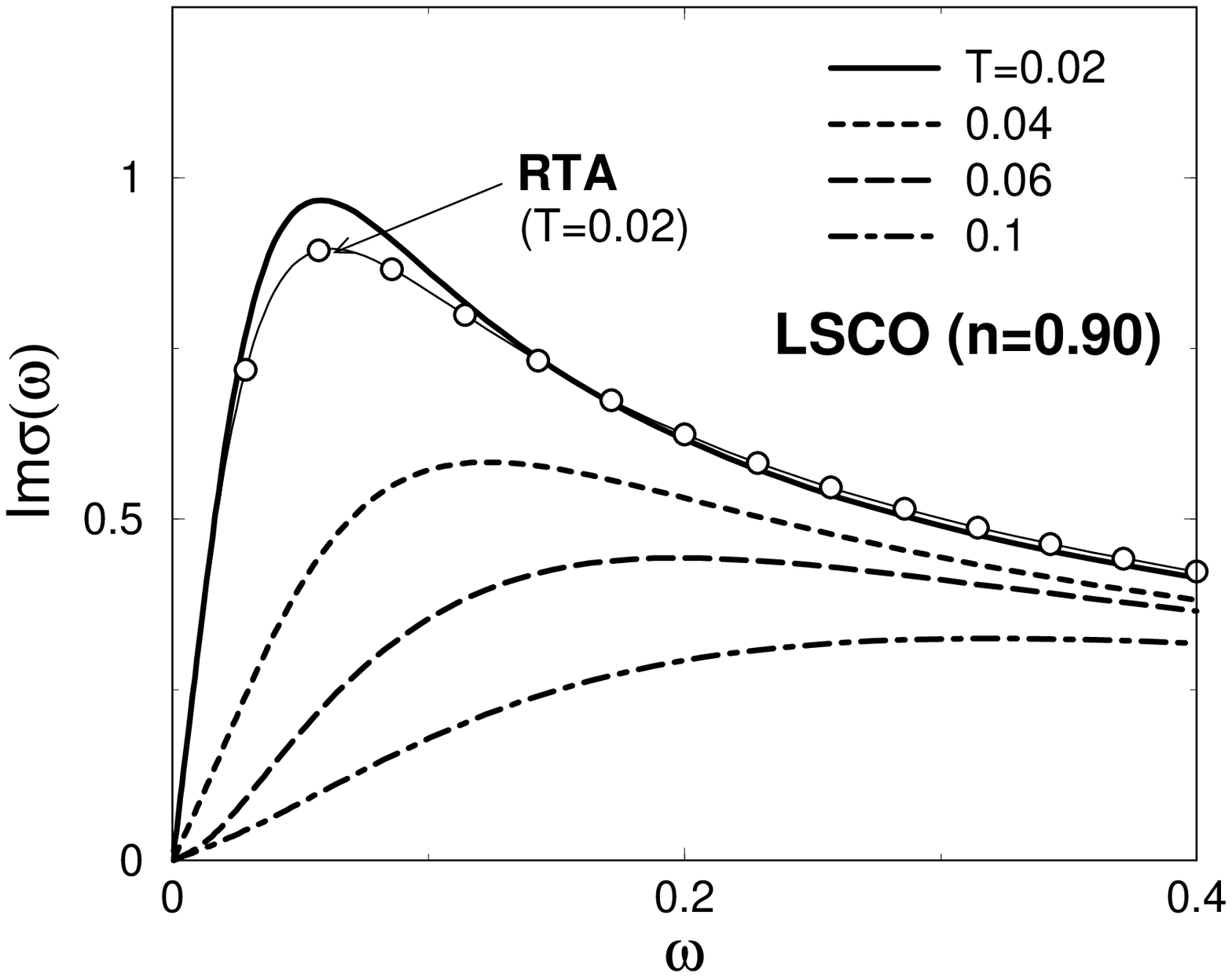,width=7cm}
\end{center}
%\vspace{10mm}
\caption{
Obtained $\sigma(\w)$ for LSCO
by the CVC-FLEX approximation.
The Drude weight is slightly reduced by the CVC.
}
  \label{fig:L-sigma}
\end{figure}
%%%%%%%%%%%%%%%%%%%%%%%%%%%%%%%%%%%%%%%%%%%%%%%%%%%%%

%%%%%%%%%%%%%%%%%%%%%%%%%%%%%%%%%%%%%%%%%%%%%%%%%%%%%%
\begin{figure}
%\vspace{10mm}
\begin{center}
\epsfig{file=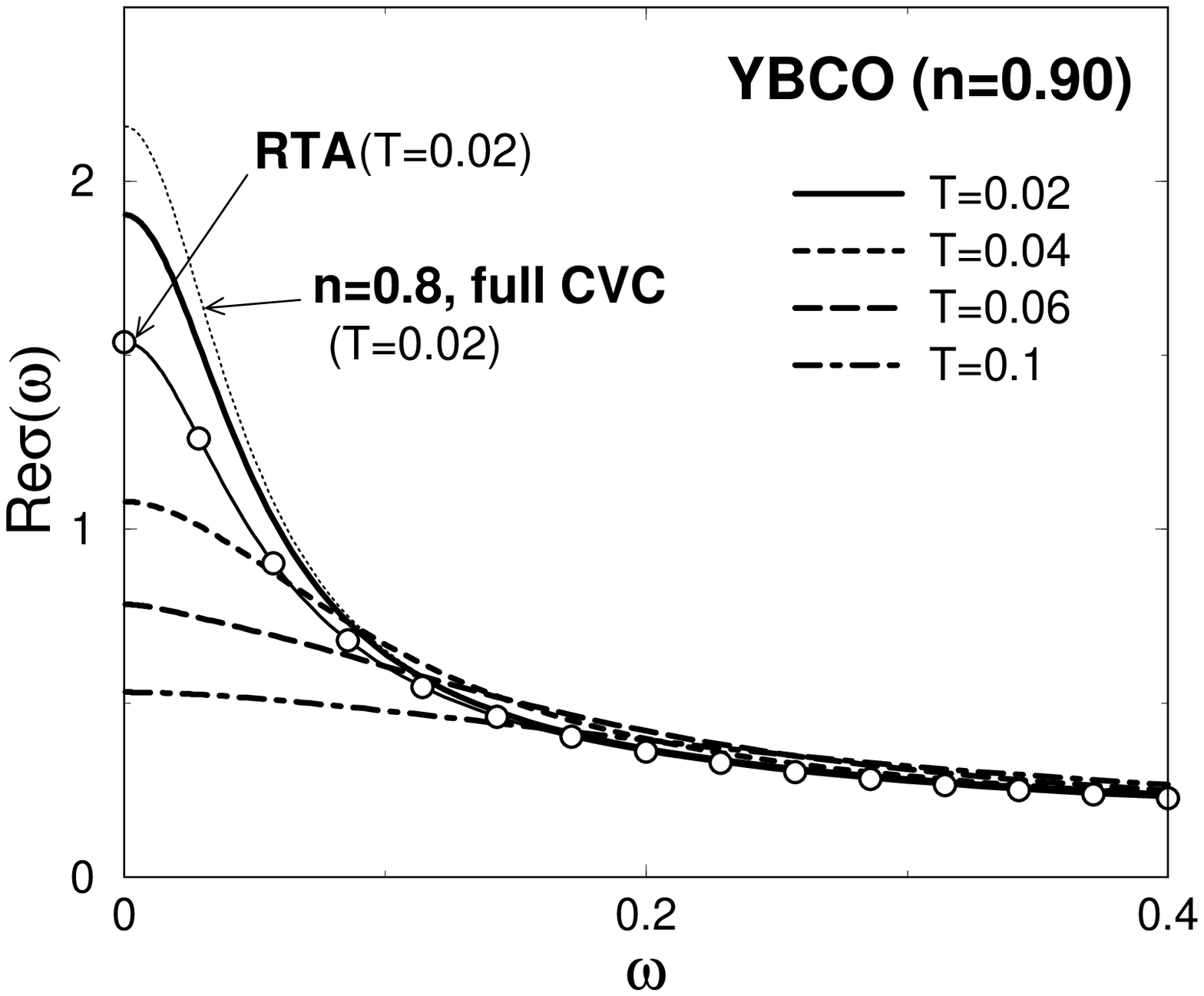,width=7cm}
\epsfig{file=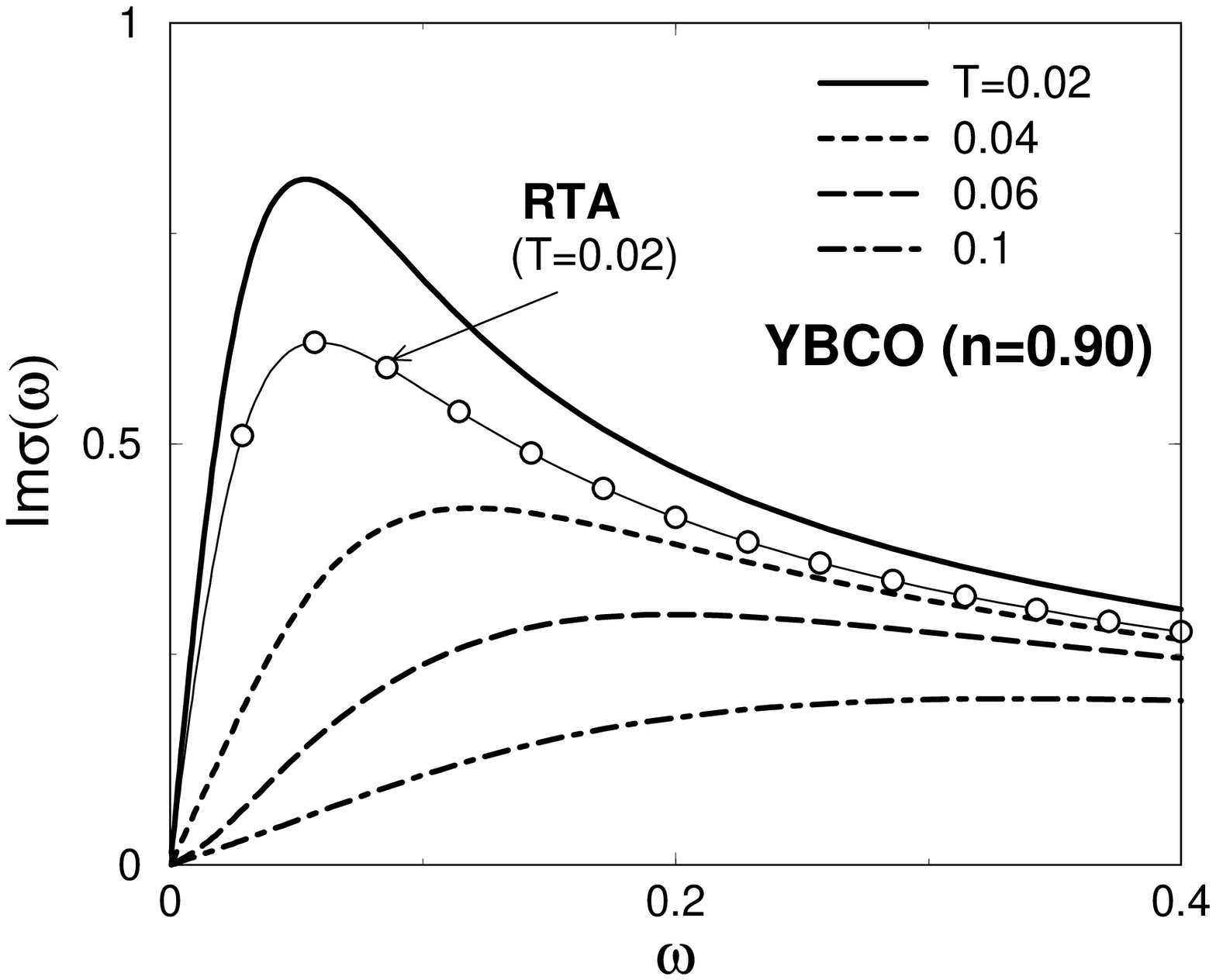,width=7cm}
\end{center}
%\vspace{10mm}
\caption{
Obtained $\sigma(\w)$ for YBCO by the CVC-FLEX approximation.
}
  \label{fig:Y-sigma}
\end{figure}
%%%%%%%%%%%%%%%%%%%%%%%%%%%%%%%%%%%%%%%%%%%%%%%%%%%%%

%Here, we discuss the optical conductivity.
Figures \ref{fig:L-sigma} and \ref{fig:Y-sigma}
show the $\w$-dependence of $\s(\w)$ for 
LSCO ($n=0.90$) and YBCO ($n=0.90$), respectively.
Both of them are qualitatively similar.
Parameters for each compound are explained
below eq.(\ref{eqn:Hub}). 
In both cases, $\s(\w)$ given by the CVC-FLEX approximation
is slightly larger than that by RTA.
In more detail, 
$\s(\w)$ decreases due to the back-flow,
whereas it increases due to $\d\Sigma_\k/\d k_\mu$
in ${\bf v}_\k(\e)$;
the latter slightly dominates in the present model parameters.
Re$\s(\w)$ apparently decreases much slower than Lorentzian
for larger $\w$, because $\gamma_{\rm ED}(\w)$ increases
with $\w$.
The Drude weight is a little sharper in LSCO because 
a smaller value of $U$ is used.
We note that the Drude weight increases 
as the system moves away from the half-filling ($n=1$).
For LSCO at $T=0.02$, Im$\s(\w)$ takes the maximum value
at $\w_{xx}\sim 0.06$, which is about three times larger
than $2\gamma_{\k_{\rm c}}(0)$ because of the
$\w$-dependence of $\gamma_{\k}(\w)$.

%%%%%%%%%%%%%%%%%%%%%%%%%%%%%%%%%%%%%%%%%%%%%%%%%%%%%%
\begin{figure}
%\vspace{10mm}
\begin{center}
\epsfig{file=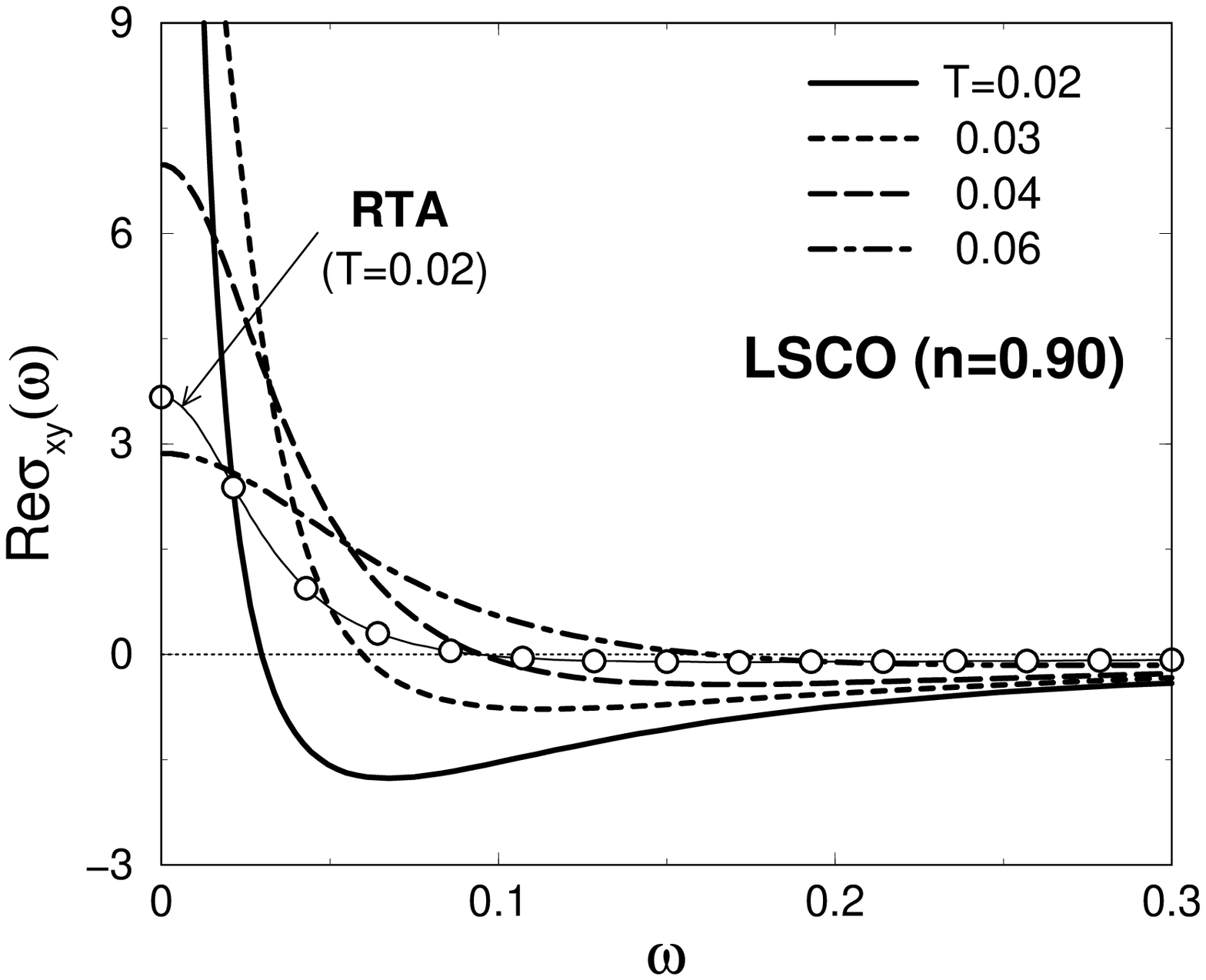,width=7.5cm}
\epsfig{file=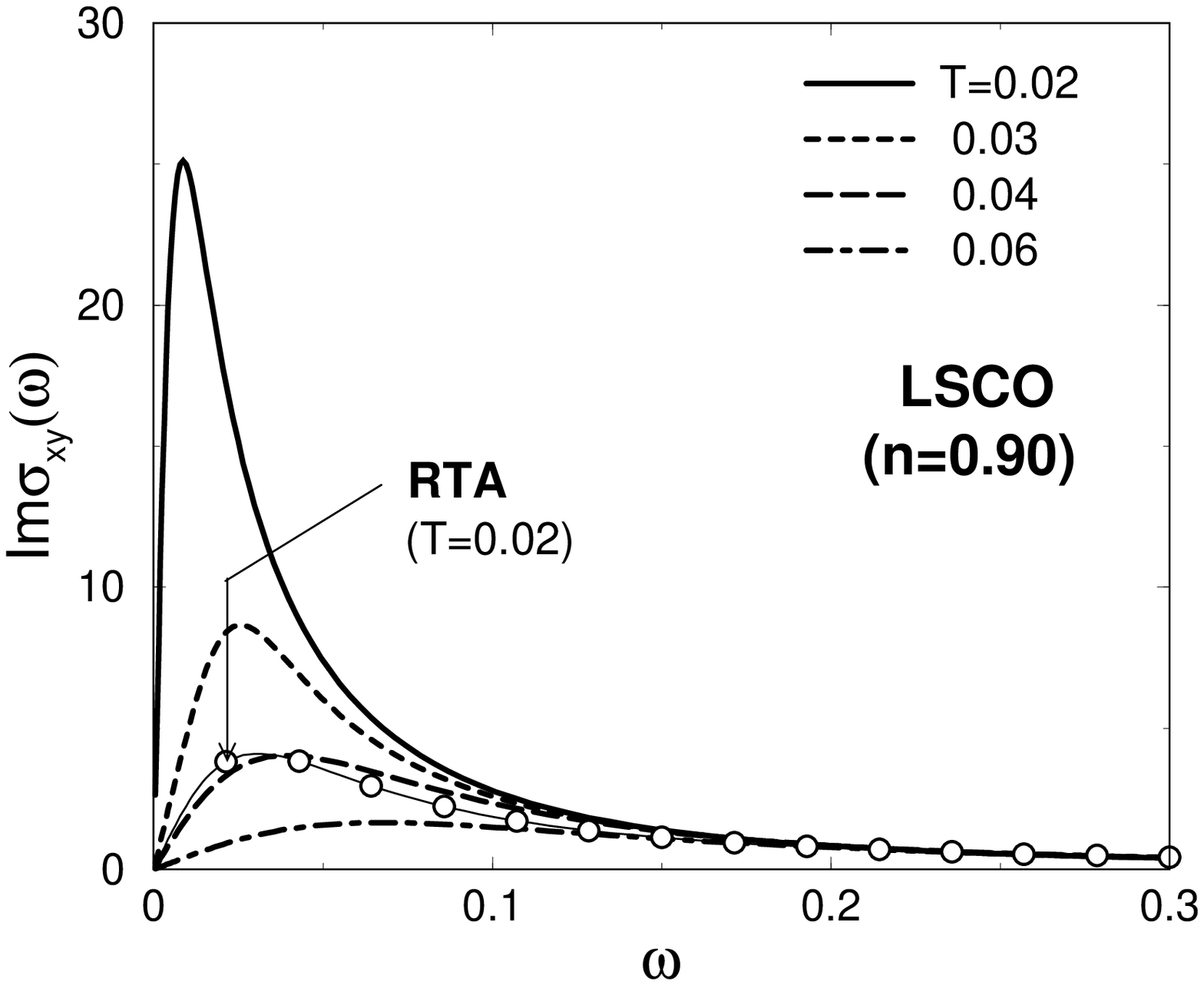,width=7.5cm}
\end{center}
%\vspace{10mm}
\caption{
Obtained $\sigma_{xy}(\w)$ for LSCO
by the CVC-FLEX approximation.
The hight of the Drude weight is remarkably 
magnified by the CVC.
At $T=0.02$,
Re$\s_{xy}(\w)$ is negative for $\w>0.3$
due to the CVC, which is consistent with experiments.
}
  \label{fig:L-sxy}
\end{figure}
%%%%%%%%%%%%%%%%%%%%%%%%%%%%%%%%%%%%%%%%%%%%%%%%%%%%%

%%%%%%%%%%%%%%%%%%%%%%%%%%%%%%%%%%%%%%%%%%%%%%%%%%%%%%
\begin{figure}
%\vspace{10mm}
\begin{center}
\epsfig{file=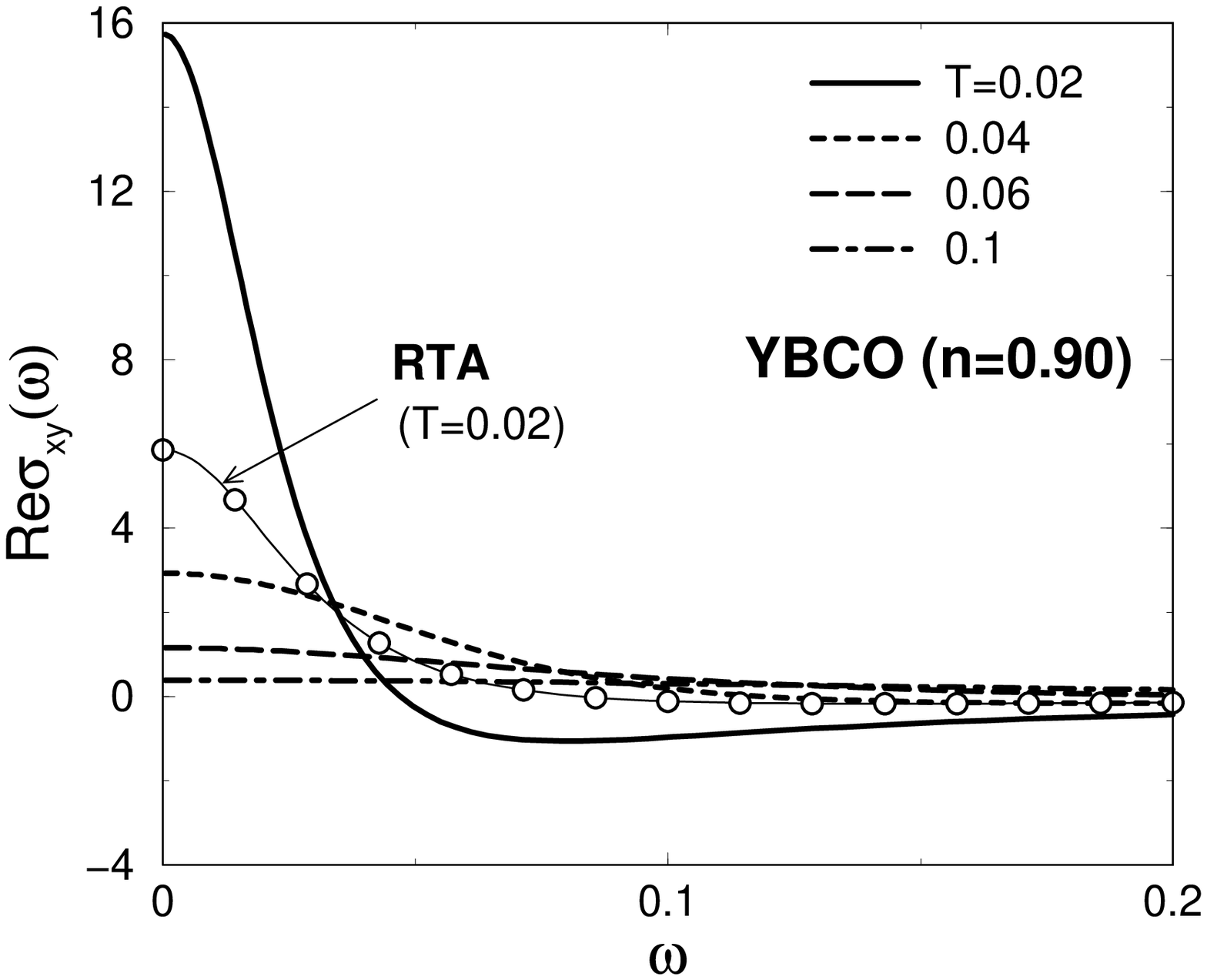,width=7.5cm}
\epsfig{file=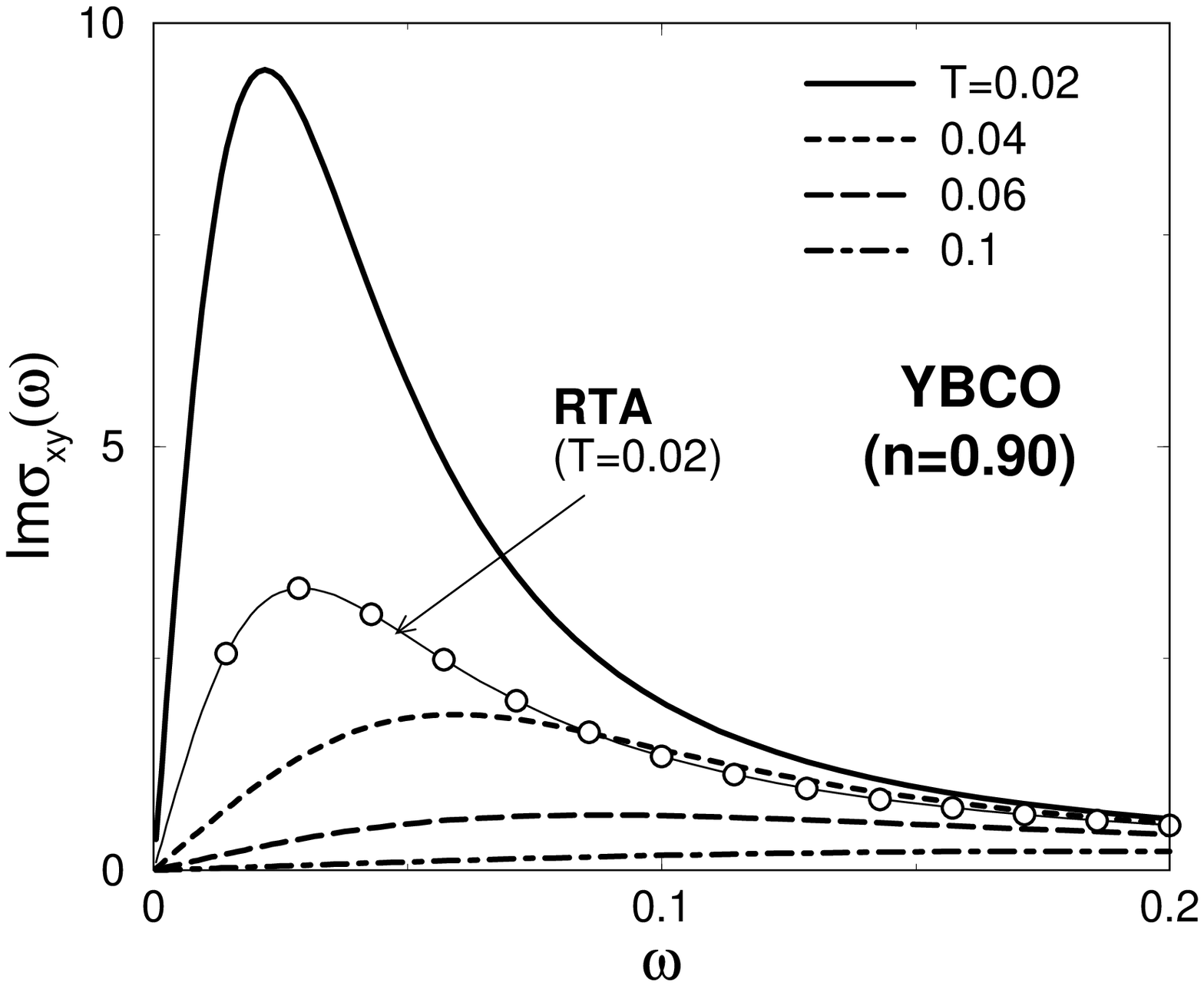,width=7.5cm}
\end{center}
%\vspace{10mm}
\caption{
Obtained $\sigma_{xy}(\w)$ for YBCO
by the CVC-FLEX approximation.
}
  \label{fig:Y-sxy}
\end{figure}
%%%%%%%%%%%%%%%%%%%%%%%%%%%%%%%%%%%%%%%%%%%%%%%%%%%%%

In contrast to $\s(\w)$,
$\s_{xy}(\w)$ with full CVC is quite different
from $\s_{xy}^0(\w)$ given by the RTA:
Figures \ref{fig:L-sxy} and \ref{fig:Y-sxy}
show the $\w$-dependence of $\s_{xy}(\w)$ for 
LSCO and YBCO, respectively.
The $\w$-dependence of Re$\s_{xy}(\w)$ becomes
prominent due to the CVC.
For LSCO (YBCO), Re$\s_{xy}(\w)$ at $T=0.02$ 
takes a large negative value for 
$\w>0.03\sim 120{\rm cm}^{-1}$
($\w>0.045\sim 180{\rm cm}^{-1}$),
which is consistent with experimental observations
 \cite{Drew04,Drew02,Drew00,Drew00-c,Drew96}.
Although Re$\s_{xy}^0(\w)$ also changes its sigh
for $\w>0.1$, its absolute value is very small.
It is naturally understood from the ED-form
because $\gamma(\w)$ increases with $\w$.
This large dip in Re$\s_{xy}(\w)$ is naturally
understood in terms of the $f$-sum rule, 
eq.(\ref{eqn:sum-sxy}), because Re$\s_{xy}(0)>0$ 
takes an enhanced value due to the CVC.
We also stress that 
${\rm Im}\s_{xy}(\w)/\w|_{\w\rightarrow0}$ 
is strongly enhanced due to the CVC, 
which is consistent with
the analysis in the previous section.
Later, we will discuss its temperature dependence
in more detail.
The overall behavior of $\s_{xy}(\w)$ for YBCO
is qualitatively similar to that for LSCO.
For LSCO at $T=0.02$, Im$\s_{xy}(\w)$ takes the maximum value
at $\w_{xy}\sim 0.01$, which is about six times larger
than $\w_{xx}$ for Im$\s(\w)$.

%%%%%%%%%%%%%%%%%%%%%%%%%%%%%%%%%%%%%%%%%%%%%%%%%%%%%%
\begin{figure}
%\vspace{10mm}
\begin{center}
\epsfig{file=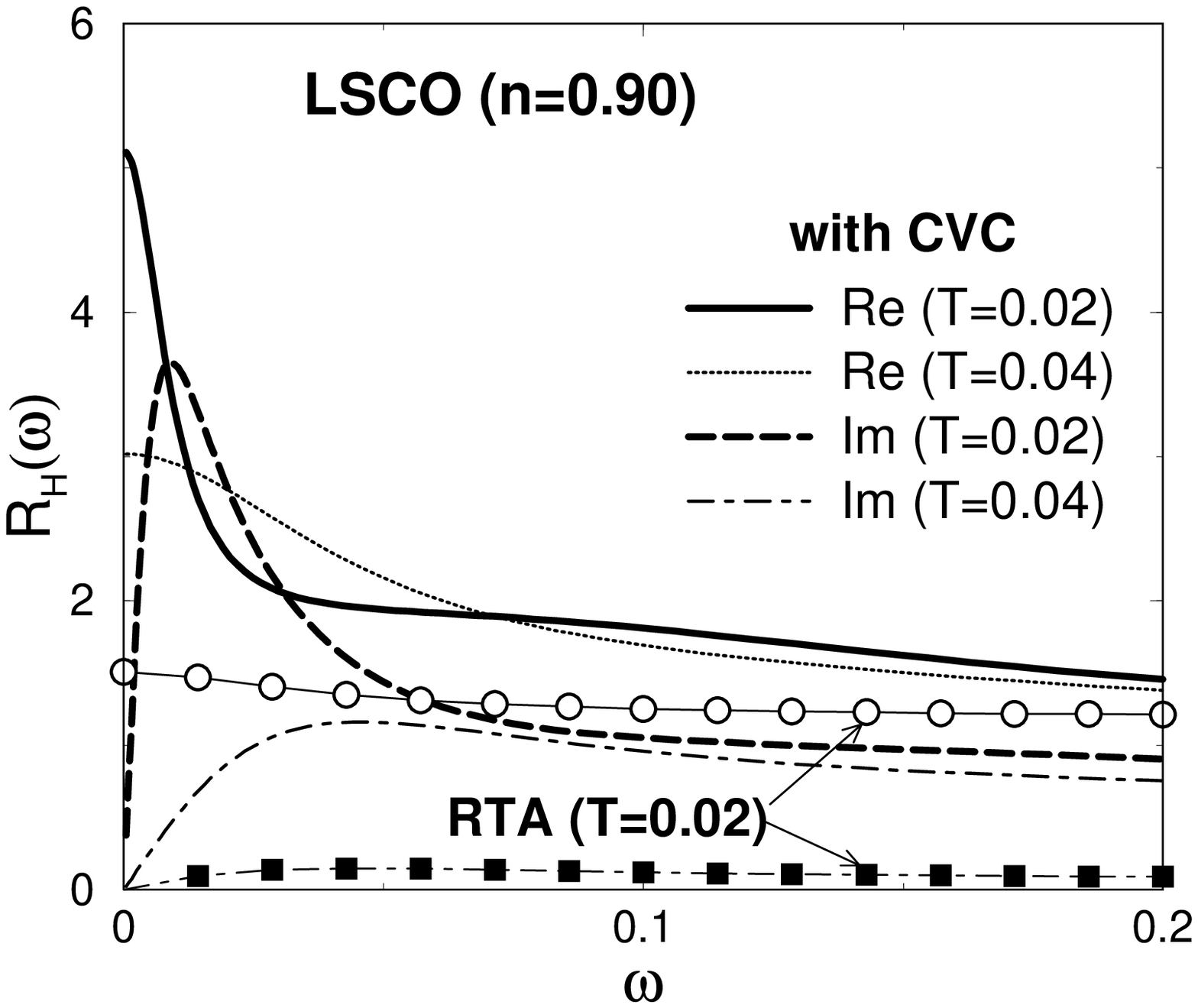,width=7.5cm}
\epsfig{file=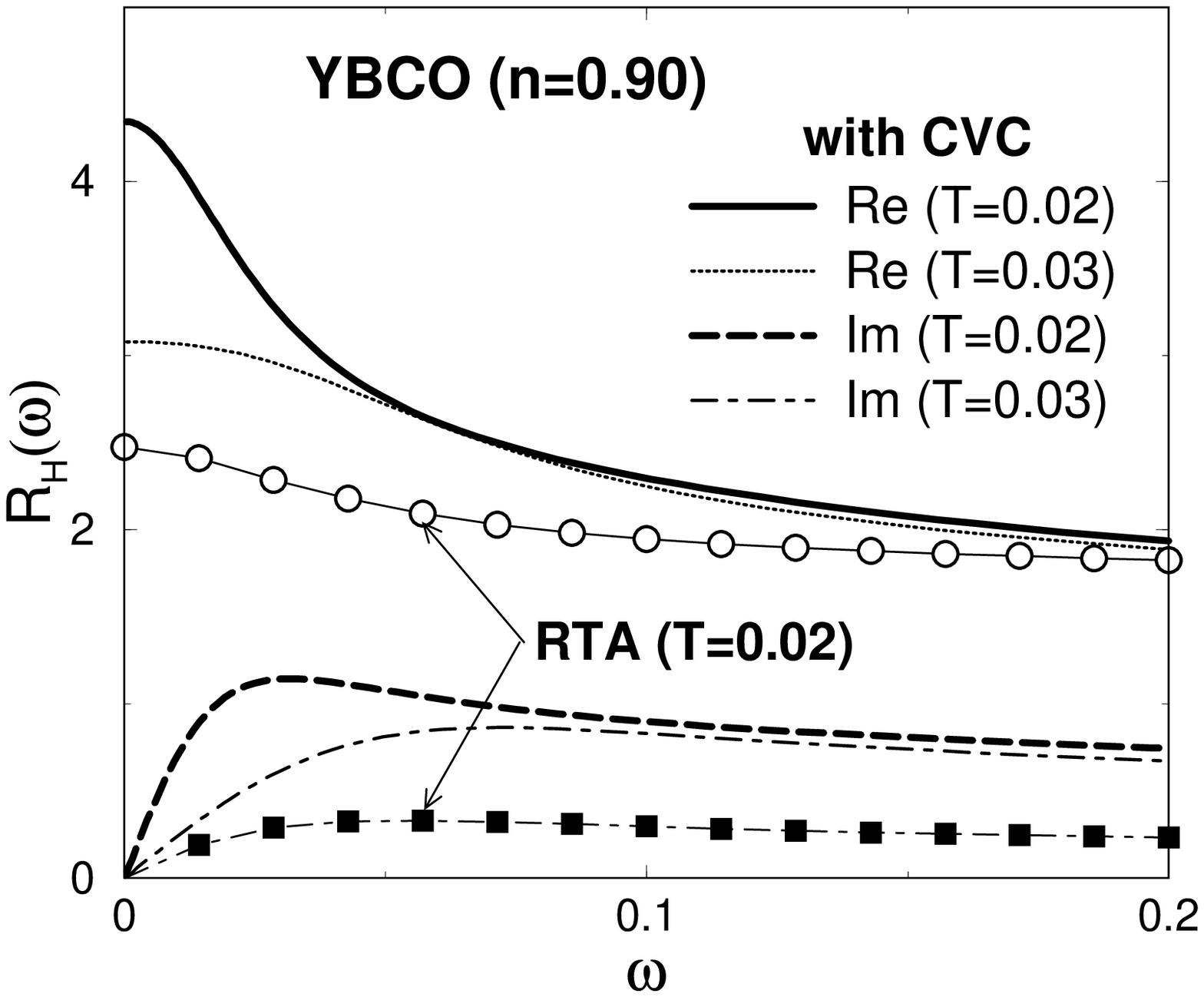,width=7.5cm}
\end{center}
%\vspace{10mm}
\caption{
Obtained $R_{\rm H}(\w) \equiv \sigma_{xy}(\w)/(\sigma_{xy}(\w))^2$ 
by the CVC-FLEX approximation.
Im$R_{\rm H}(\w)$ takes large values
because the ED-form for $\s_{xy}(\w)$ is violated due to the CVC.
We note that the magnitude of $R_{\rm H}^{\rm RTA}$
in this figure is too large since the renormalization to the QP velocity by $d\Sigma_\k/d k_x$ has been dropped.
The correct value is $R_{\rm H}^{\rm RTA} \;\stackrel{<}{\sim}\; 1$
according to ref.\cite{Kontani-Hall}.
}
  \label{fig:RH}
\end{figure}
%%%%%%%%%%%%%%%%%%%%%%%%%%%%%%%%%%%%%%%%%%%%%%%%%%%%%%

The deviation of $\s_{xy}(\w)$ from the ED-form
gives a prominent $\w$-dependence of the optical Hall coefficient 
$R_{\rm H}(\w)=\s_{xy}(\w)/\s^2(\w)$,
which is shown in fig.\ref{fig:RH}.
For LSCO, the $\w$-dependence of 
$R_{\rm H}^0(\w)$ given by the RTA, 
$\s_{xy}^0(\w)/(\s^0(\w))^2$,
is very weak, and its imaginary part is tiny.
This fact gives the conclusive evidence that
both $\s_{xy}^0(\w)$ and $\s^0(\w)$ follow
the ED-form.
On the other hand,
$R_{\rm H}(\w)$ given by the conserving approximation
shows prominent frequency as well as 
temperature dependences.
For LSCO at $T=0.02$, Im$R_{\rm H}(\w)$ takes the maximum value
at $\w_{\rm RH}\sim 0.01$, which is similar to $\w_{xy}$ for 
Im$\s_{xy}(\w)$ and is six times larger than 
$\w_{xx}$ for Im$\s(\w)$.
The relation $\w_{xx}\gg\w_{xy},\w_{\rm RH}$
obtained in the present study,
which is consistent with
experimental observation \cite{Drew04,Drew96},
cannot be reproduced by the RTA:
It can be explained only when
the back-flow is taken into account.
Qualitatively similar results are obtained
for YBCO, although its anomalies are more moderate.
The observed $R_{\rm H}(\w)$ in YBa$_2$Cu$_3$O$_7$ 
at 95K in ref. \cite{Drew96} 
looks similar to the present
result for LSCO at $T=0.04$ in fig.\ref{fig:RH}.
The model parameters for YBCO used here
may not really appropriate for a quantitative study.

%%%%%%%%%%%%%%%
% beki
%%%%%%%%%%%%%%%
In order to elucidate the reason why
$\s_{xy}(\w)$ deviates from the ED-form due to the CVC,
we analyze $\s_{xy}(\w)$ in the 
low frequency limit. 
In the RTA where CVC is absent, relations
$\s_{xy}^0 \propto \gamma_{\k_{\rm c}}^{-2}$ and
$\s_{xy}^0(\w)/i\w \propto z^{-1} \gamma_{\k_{\rm c}}^{-3}$ 
are expected.
As shown in fig.\ref{fig:beki} (a),
following relations are held by the RTA 
in the present numerical study: 
\begin{eqnarray}
\s_{xy}^0/(\s^0)^2 &\propto& {\rm const.} ,
 \label{eqn:beki4} \\ 
{\rm Im}\s_{xy}^0(\w_0)/(\s^0)^3 
 &\propto& z^{-1} \propto T^{-0.4} ,
 \label{eqn:beki5}
\end{eqnarray}
where $\w_0$ is a small constant ($\w_0\sim10^{-4}$).
These temperature dependences is drastically changed
due to the CVC, as discussed in the previous section.
Actually, when the CVC's are fully taken into account,
we obtain
\begin{eqnarray}
\s_{xy}/\s^2 &\propto& a \propto T^{-0.9} ,
 \label{eqn:beki6} \\
{\rm Im}\s_{xy}(\w_0)/\s^3 &\propto& z^{-1}b \propto T^{-1.7} ,
 \label{eqn:beki7}
\end{eqnarray}
where coefficients $a$ and $b$ had been introduced 
in eq. (\ref{eqn:expand-sxy}).
Thus, both of which are enhanced by the CVC's
as the temperature decreases.
We find that
${\rm Im}\s_{xy}(\w_0)/\s^3 \propto 
(\s_{xy}/\s^2)^2$ is approximately realized
in the present CVC-FLEX approximation.

%%%%%%%%%%%%%%%%%%%%%%%%%%%%%%%%%%%%%%%%%%%%%%%%%%%%%
\begin{figure}
%\vspace{10mm}
\begin{center}
\epsfig{file=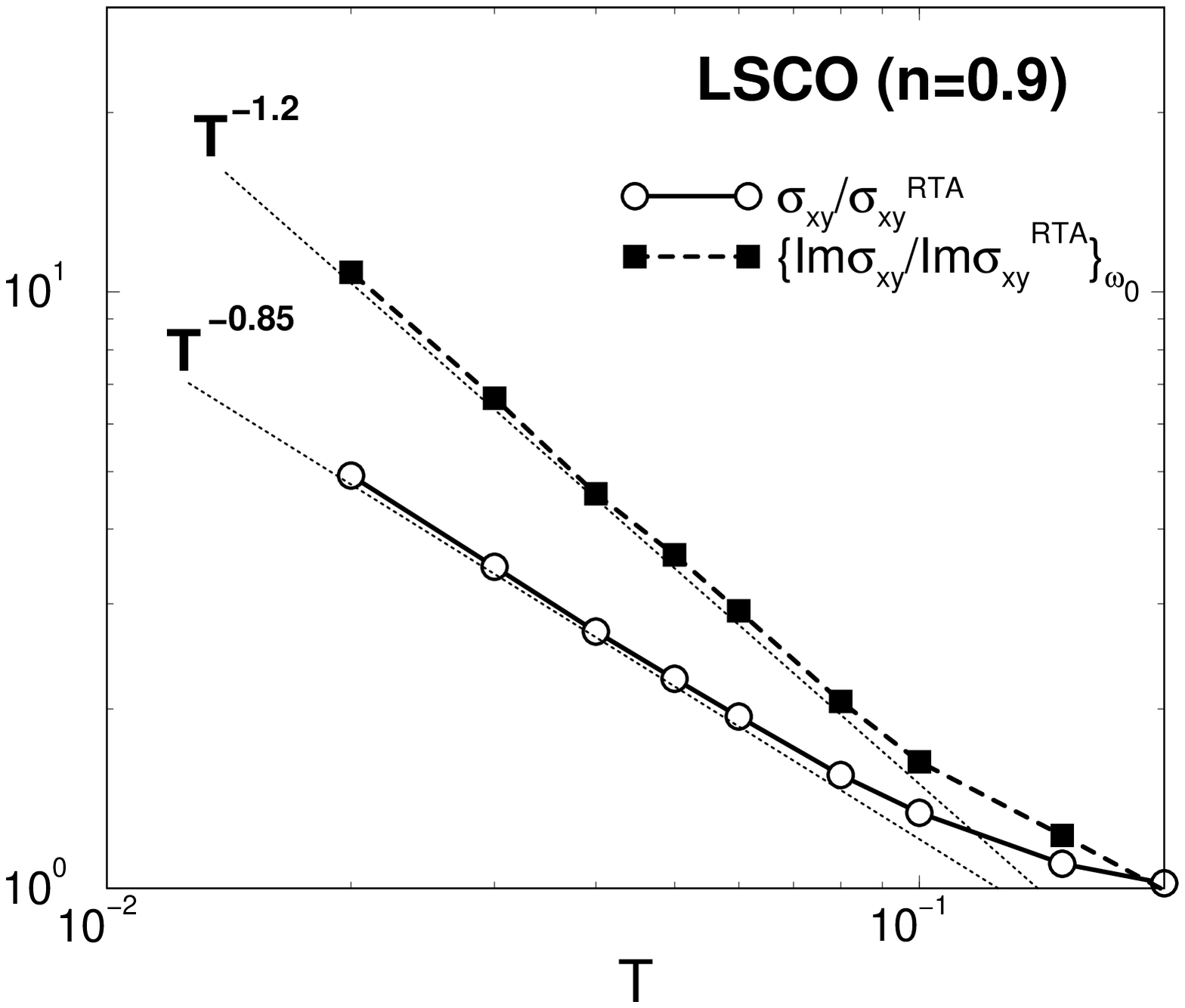,width=7cm}
\epsfig{file=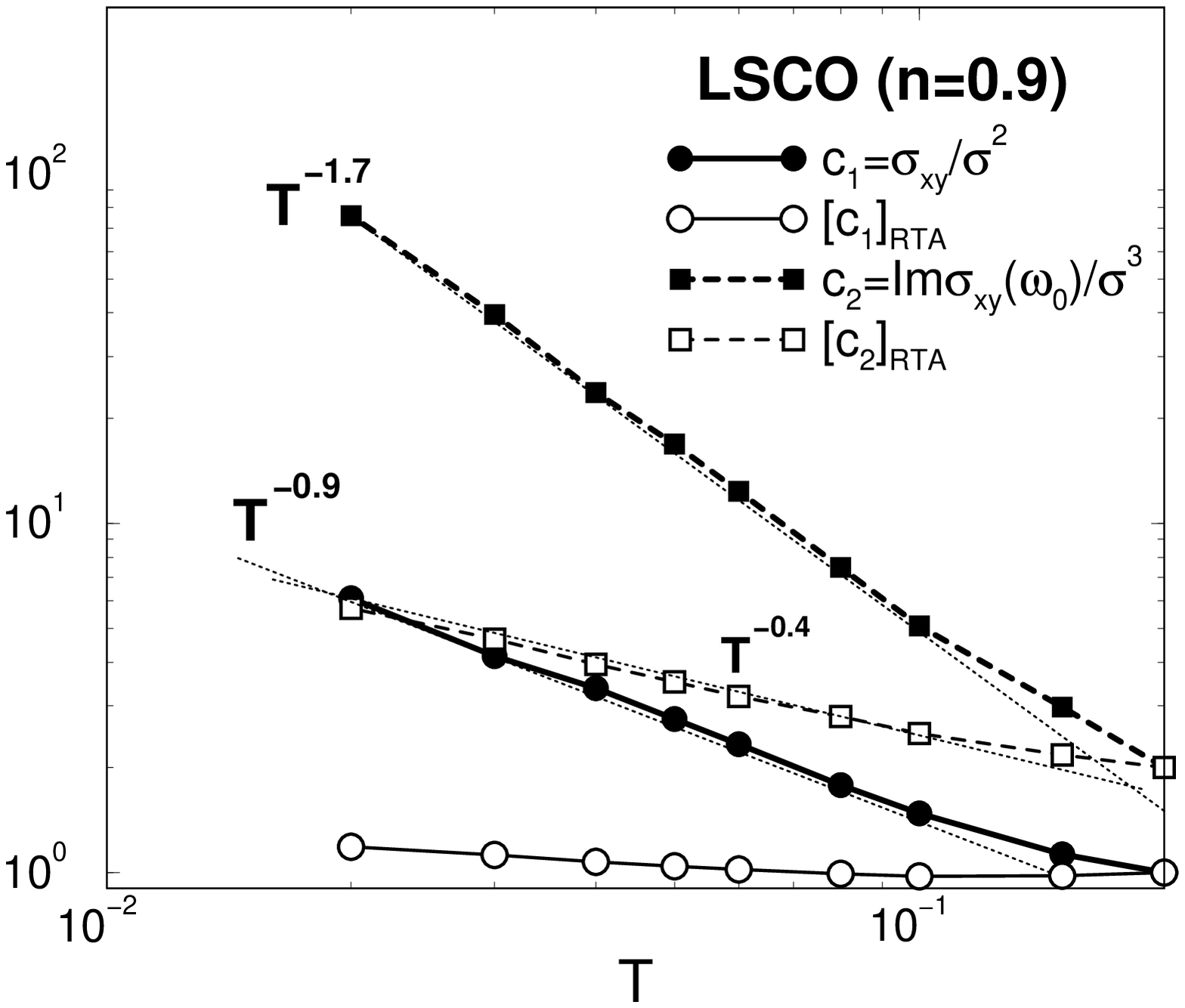,width=7cm}
\end{center}
%\vspace{10mm}
\caption{
(a) $\s_{xy}/\s_{xy}^0 \propto T^{-0.85}$ at $\w=0$ and
Im$\s_{xy}/{\rm Im}\s_{xy}^0\propto T^{-1.2}$ 
for small $\w_0$.
This result suggests that
they are proportional to $\xi^2$ and $\xi^3$
($\xi$ being the AF correlation length)
for $T=0.03\sim 0.1$.
(b) $\s_{xy}/\s^2 \propto T^{-0.9}$ at $\w=0$ and
Im$\s_{xy}/\s_{xy}^3 \propto T^{-1.7}$ for small $\w_0$.
}
  \label{fig:beki}
\end{figure}
%%%%%%%%%%%%%%%%%%%%%%%%%%%%%%%%%%%%%%%%%%%%%%%%%%%%%

In the previous section, relations
$a\propto \xi^2$ and $b=\xi^m$ ($m=2\sim4$)
are derived from the analysis of the CVC.
By eqs.(\ref{eqn:beki4})-(\ref{eqn:beki7}),
relations $\xi^2\approx T^{0.9}$ and
$m\approx3$ are concluded.
To confirm these results more completely,
we perform another plot shown in fig. \ref{fig:beki} (b),
\begin{eqnarray}
 \s_{xy}/\s_{xy}^0 &\propto& \xi^{2} \propto T^{-0.85} ,
 \label{eqn:beki8} \\
 {\rm Im}\s_{xy}(\w_0)/{\rm Im}\s_{xy}^0(\w_0)
 &\propto& \xi^{m} \propto T^{-1.2} ,
 \label{eqn:beki9}
\end{eqnarray}
As a result, the relation $m\approx 3$ is also derived
from eqs.(\ref{eqn:beki8}) and (\ref{eqn:beki9}).
Note that the exponents of $T$ in eqs.(\ref{eqn:beki6})
and (\ref{eqn:beki8}) are slightly different 
because $(\s^0)^2/\s_{xy}^0$ shows a subtle
temperature dependence.

According to eq. (\ref{eqn:expand-RH}),
$\lim_{\w\rightarrow0}R_{\rm H}(\w)/i\w= 
2z^{-1}( b-a ) (2\gamma_{\k_{\rm c}})^{-1}$.
As we discussed above,
$b-a$ is positive and is enhanced
as $T$ decreases.
As a result, in nearly AF Fermi liquid,
$i\w$-linear term of $R_{\rm H}(\w)$
is strongly enhanced by the CVC, which is consistent
with experiments
 \cite{Drew04,Drew02,Drew00,Drew00-c,Drew96}.

%%%%%%%%%%%%%%%%%%%%%%%%%%%%%%%%%%%%%%%%%%%
\subsection{$f$-sum rule}
%%%%%%%%%%%%%%%%%%%%%%%%%%%%%%%%%%%%%%%%%%%
The $f$-sum rule for $\s_{\mu\nu}(\w)$
gives a rigorous relation between 
the conductivity and the electron density
 \cite{Coleman,Kotliar,Yanase}.
It is violated in the RTA
because the conservation laws are not satisfied.
On the other hand,
$f$-sum rule is automatically satisfied 
in the conservation approximation,
if all the CVC's given by the Ward identity are taken.
Thus, $f$-sum rule is a useful check for the
reliability of the numerical study.

The $f$-sum rules for $\s(\w)$ and $\s_{xy}(\w)$
in an anisotropic system are given by
\begin{eqnarray}
\int_0^\infty d\w {\rm Re}\s(\w)
&=& \pi e^2 \sum_{k} \frac {\d^2 \e_\k}{\d k_x^2}n_\k,
% = \frac{\pi e^2}{2} \left\langle \frac{\d^2 \e_\k}{\d k_x^2} 
% \right\rangle
 \label{eqn:sum-sig}
 \\
\int_0^\infty d\w {\rm Re}\s_{xy}(\w)
&=& 0 .
 \label{eqn:sum-sxy}
\end{eqnarray}
Equation (\ref{eqn:sum-sig}), which
can be derived directly from the Kubo formula
 \cite{Kubo}, represents the
contribution by the diamagnetic current.
Equation (\ref{eqn:sum-sxy}) 
is easily recognized form the fact that
$\s_{xy}(\w)\sim |\w|^{-2}$ 
as $|\w|\rightarrow\infty$, and
it is analytic in the upper-half plane
of the complex $\w$-space.
In the case of  $t'=t''=0$,
the right hand side of eq.(\ref{eqn:sum-sig})
is equal to $-(\pi e^2)\langle \e_\k \rangle$,
where $\langle \e_\k \rangle=\sum_\k \e_\k n_\k$ 
gives the kinetic energy.

%%%%%%%%%%%%%%%%%%%%%%%%%%%%%%%%%%%%%%%%%%%%%%%%%%%%%
\begin{figure}
%\vspace{10mm}
\begin{center}
\epsfig{file=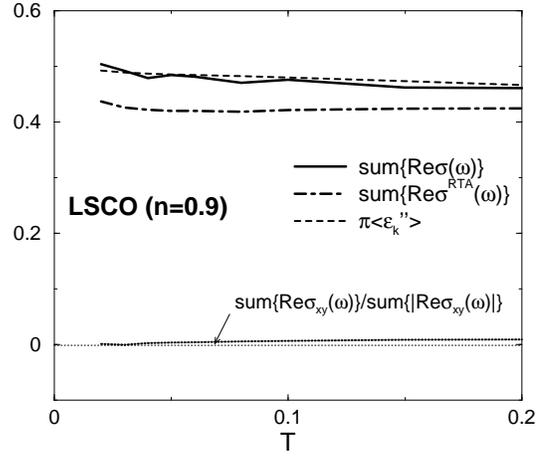,width=7cm}
\end{center}
%\vspace{10mm}
\caption{
$f$-sum rules both for $\s(\w)$ and $\s_{xy}(\w)$
are well satisfied if the CVC is correctly taken into account.
This fact assures the reliability of the
present numerical study.
}
  \label{fig:sum}
\end{figure}
%%%%%%%%%%%%%%%%%%%%%%%%%%%%%%%%%%%%%%%%%%%%%%%%%%%%%

The numerical check for the $f$-sum rule
is shown in Fig. \ref{fig:sum}.
Sum$\{{\rm Re}\s(\w)\}$ and $\pi\langle \e_\k'' \rangle$ 
represent the left- and right-hand-side 
of eq.(\ref{eqn:sum-sig}), respectively.
Sum$\{{\rm Re}\s(\w)\}$ is obtained by performing 
the numerical $\w$-integration form 0 to 100.
We see that the $f$-sum rule (\ref{eqn:sum-sig})
holds well, within the relative error $\sim 2\%$.
This results assure the high reliability
of the present numerical study when $\w$ is not so large.
In general, the Pade approximation for larger $\w$ is less reliable
because the distance form the imaginary axis is large.
On the other hand, Sum$\{{\rm Re}\s^{\rm RTA}(\w)\}$
within the RTA (without any CVC) is smaller
than the correct value,  whose relative error 
is more than 12$\%$:
This discrepancy is due to the violation
of the conservation laws in the RTA.

We also plot
$\displaystyle \int_0^X d\w {\rm Re}\s_{xy}(\w)/
\int_0^X d\w |{\rm Re}\s_{xy}(\w)|$ 
in fig.\ref{fig:sum}, where we put $X=100$.
It should vanish identically when $X=\infty$
according to the $f$-sum rule (\ref{eqn:sum-sxy})
in the conserving approximation.
It becomes less than 0.02 as shown in fig.\ref{fig:sum},
which also suggests the high reliability
of the present numerical study.
This result means that the unessential
poles of $\s_{xy}(\w)$ in the upper-half-plane
of the complex $\w$-plane, which arises from
$g^{(1)}=G^{\rm R}G^{\rm R}$ in the presence of interaction,
are correctly cancelled by the vertex corrections.

The realization of $f$-sum rules
confirmed in the present numerical study 
is better than expected, despite that all the AL-type
vertex corrections are dropped.
This result strongly suggests that
the AL terms are insignificant for the
quantitative study of $\s(\w)$ and $\s_{xy}(\w)$,
as they are for $\s(0)$ and $\s_{xy}(0)$
 \cite{Kontani-Hall}.

%%%%%%%%%%%%%%%%%%%%%%%%%%%%%
\subsection{Inverse Hall Angle}
%%%%%%%%%%%%%%%%%%%%%%%%%%%%

IR optical Hall angle $\theta_{\rm H}(\w)=\s_{xy}(\w)/\s(\w)$
($\w\simle 1000{\rm cm}^{-1}=1440$K)
has been intensively measured by Drew et al
 \cite{Drew04,Drew00}.
They concluded that
(I) Im$\theta_{\rm H}(\w)/\w$ is almost independent of $\w$
and $T$, and (II) Re$\theta_{\rm H}(\w)$ is also
independent of $\w$, while its $T$-dependence is large.
In contrast to (I),
$|{\rm Im}\s(\w)/\w|$ monotonously decreases as $\w$ increases,
as shown in fig. \ref{fig:INVSIG} (a).
As a result, the Hall angle in HTSC follows a simple Drude expression
IR range ($\w\simle1000{\rm cm}^{-1}$):
\begin{eqnarray}
\theta_{\rm H}^{\rm SD}(\w)
 &=& \frac{\Omega_{\rm H}^\ast}{2\gamma_{H}^\ast-i\w} ,
 \label{eqn:D-HA}\\
\gamma_{H}^\ast &\propto& T^{-d}; \ \ d=1.5\sim2
 \nonumber \\
\Omega_{\rm H}^\ast &\propto& T^0 ,
 \nonumber 
\end{eqnarray}
where
$\gamma_{H}^\ast$ and $\Omega_{\rm H}^\ast$
are $\w$-independent,
In contrast, 
$\gamma_{\rm ED}(\w)$ deduced from the optical conductivity
is approximately $\gamma_{\rm ED}(\w)\propto \max\{\w,\pi T \}$
 \cite{Pines-opt}:
It is proportional to $\w$ and is temperature-independent
has a large $\w$-dependence when $\w > \pi T$,
as recognized in fig. \ref{fig:INVSIG}.
$\Omega_{\rm H}^\ast$ is a constant independent
of $\w$ and $T$, whereas it increases as the doping decreases.
This unexpected behavior of the Hall angle
puts very severe constraints on theories of HTSC.
%For example, it declines the Luttinger liquid model
%with two kinds of relaxation times
%proposed in ref.
% \cite{Anderson}.

From now on, we show that
such anomalous behaviors of $\theta_{\rm H}(\w)$
in HTSC's are well understood in terms of the 
Fermi liquid with strong AF fluctuations.
The frequency dependence of back-flow is 
crucial to reproduce the correct results.
Here, we mainly show numerical results only for LSCO,
although similar results are also obtained for YBCO.
In the present numerical study, we derive
$\theta_{\rm H}(\w)$'s from
the analytic continuations of 
$K_{xy}(i\w_l)/(K_{xx}(i\w_l)-K_{xx}(0))$
with $\w_l>0$, which is a analytic function on 
the upper-half complex $\w$-plane
 \cite{Coleman}.
The value of $\theta_{\rm H}(\w)$ obtained by this procedure
is more accurate than dividing $\s_{xy}(\w)$ by $\s(\w)$
after the analytic continuations of 
$K_{xy}(i\w_l)$ and $K_{xx}(i\w_l)$ individually.

%%%%%%%%%%%%%%%%%%%%%%%%%%%%%%%%%%%%%%%%%%%%%%%%%%%%%
\begin{figure}
%\vspace{10mm}
\begin{center}
\epsfig{file=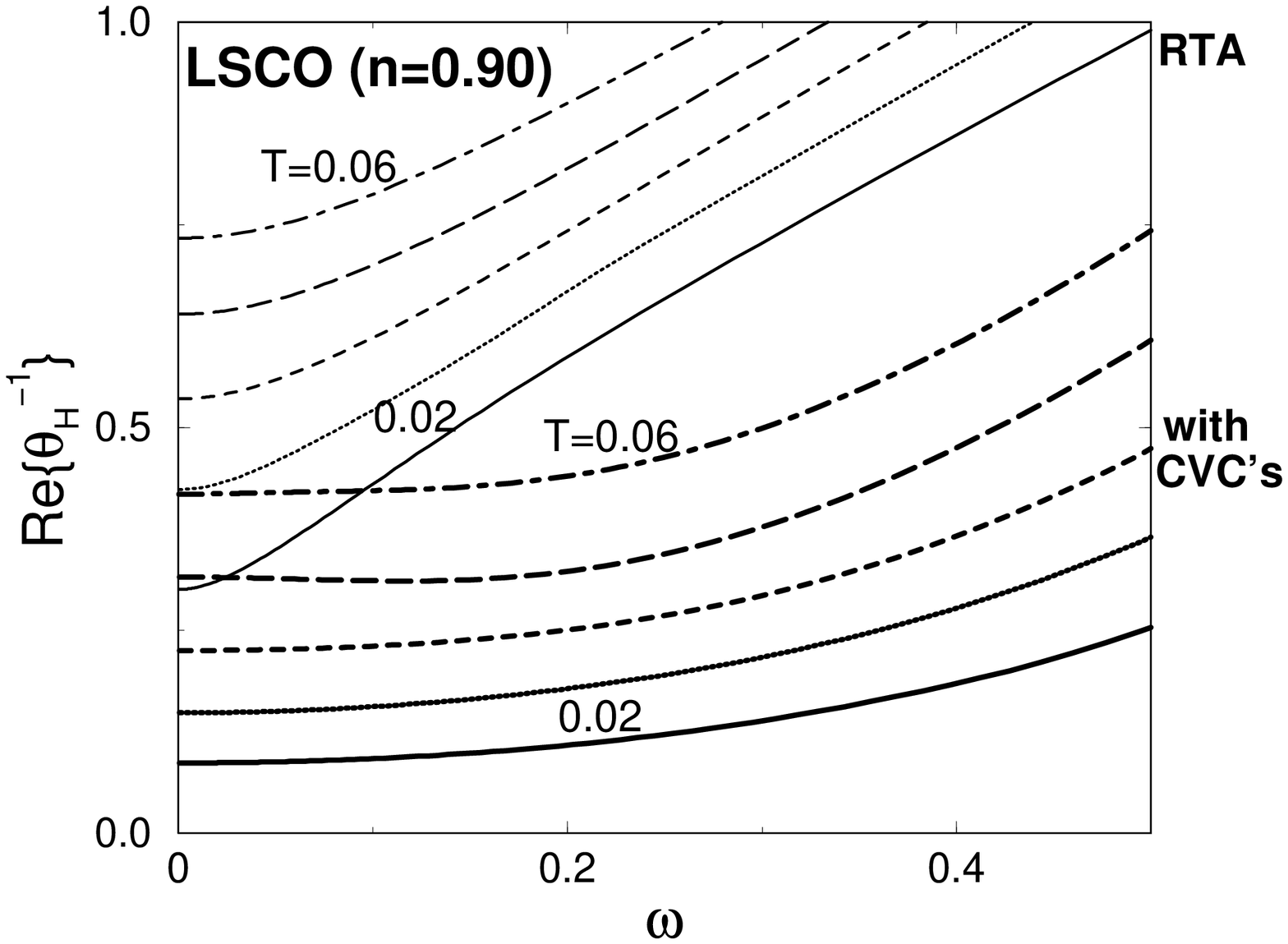,width=7.5cm}
\epsfig{file=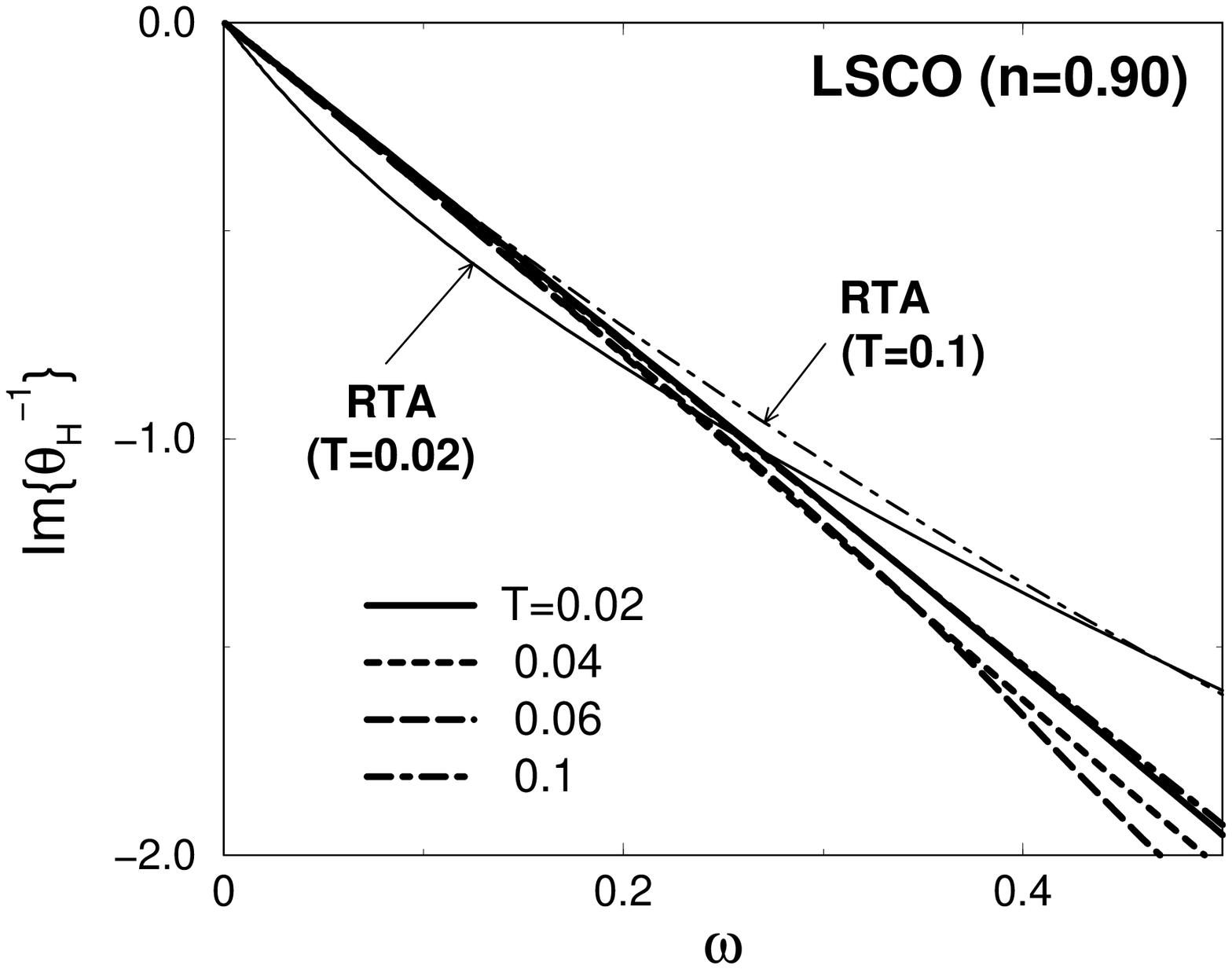,width=7.5cm}
\end{center}
%\vspace{10mm}
\caption{
Obtained $\w$-dependence of the 
inverse Hall angle for several temperatures.
We see that Re$\theta_{\rm H}^{-1}$
is approximately $\w$-independent for $\w<0.2$
while it's temperature dependence is large.
This result, obtained by taking the CVC into account,
is consistent with experiments.
}
  \label{fig:L-IHA}
\end{figure}
%%%%%%%%%%%%%%%%%%%%%%%%%%%%%%%%%%%%%%%%%%%%%%%%%%%%%

%%%%%%%%%%%%%%%%%%%%%%%%%%%%%%%%%%%%%%%%%%%%%%%%%%%%%
\begin{figure}
%\vspace{10mm}
\begin{center}
\epsfig{file=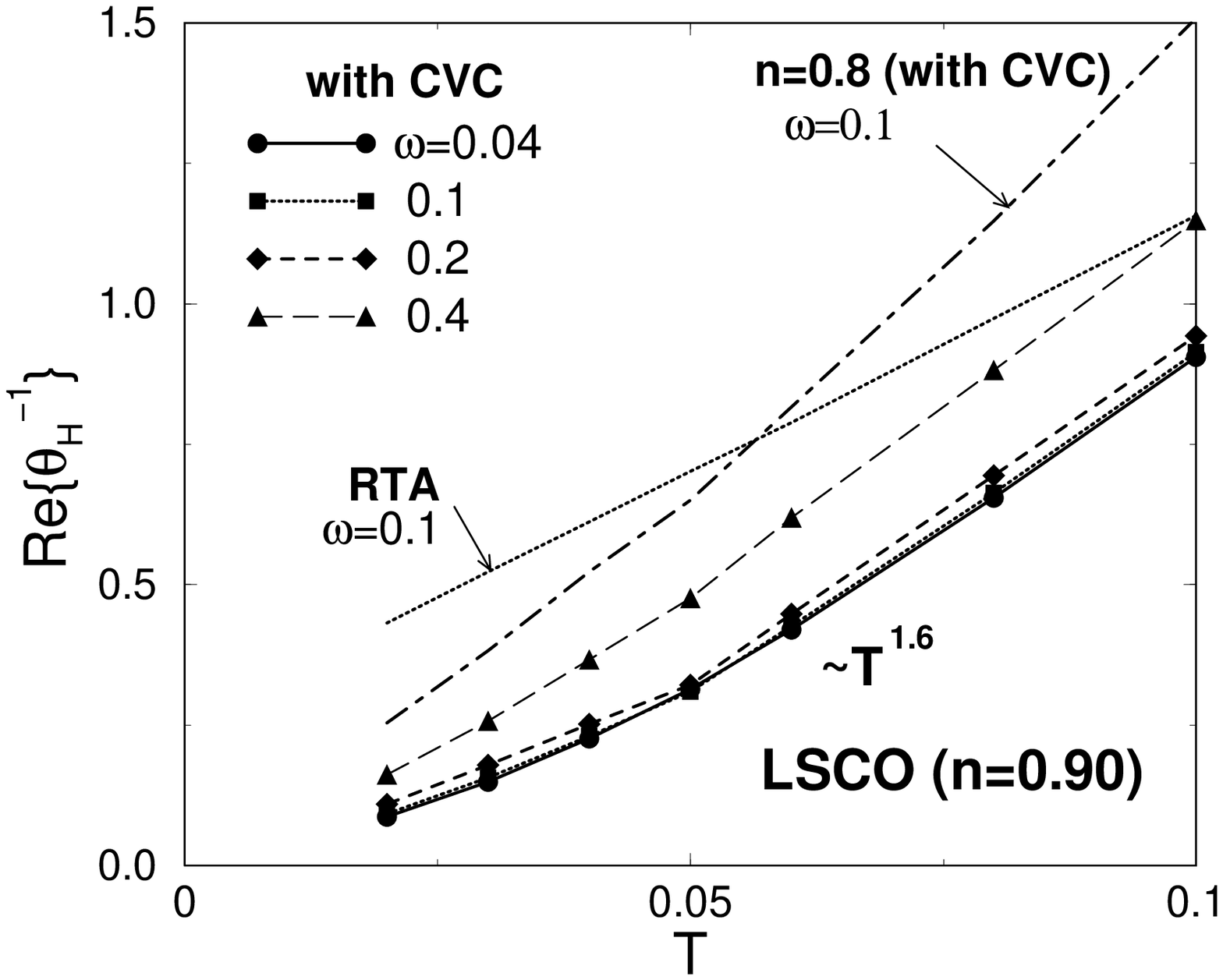,width=7.5cm}
\epsfig{file=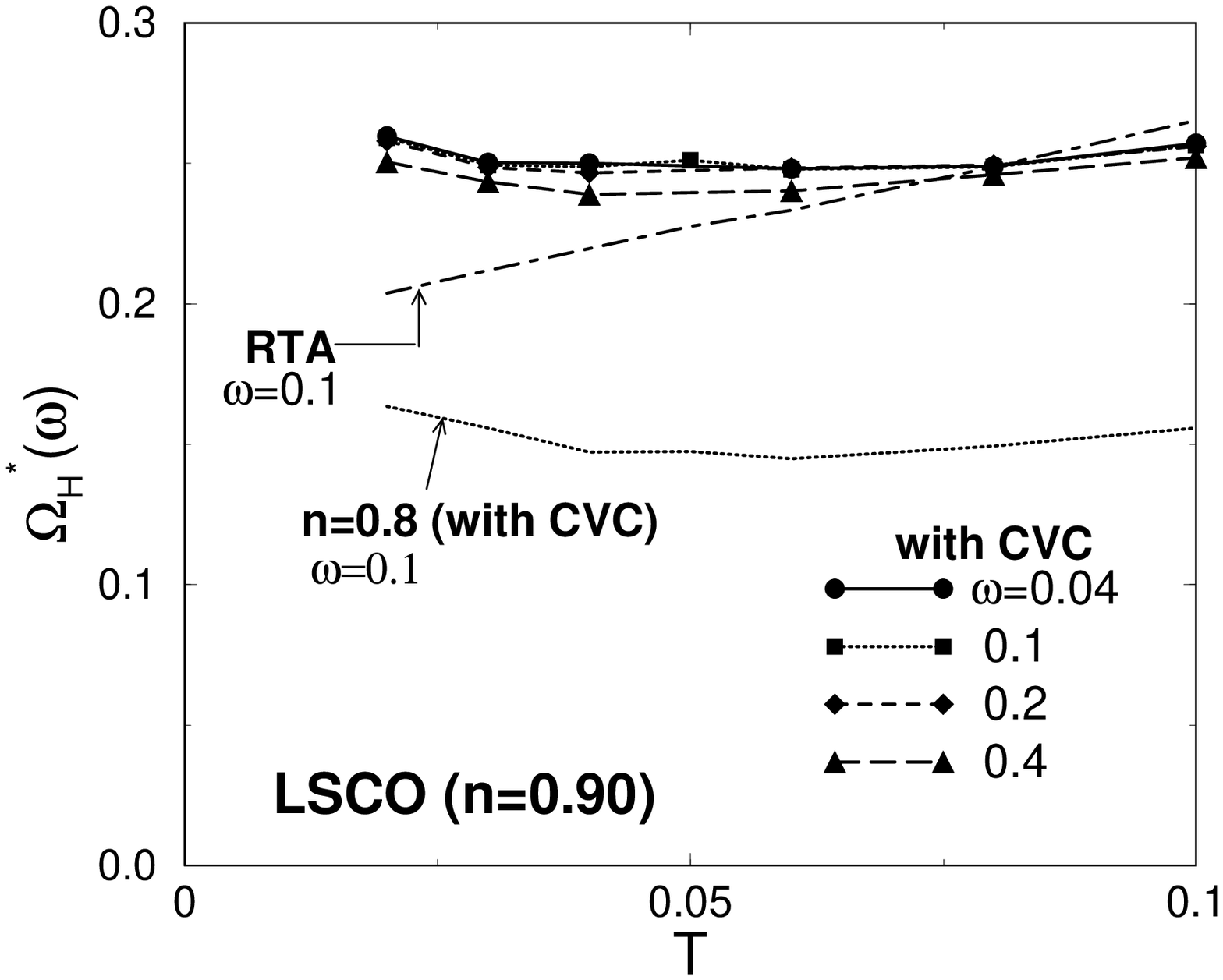,width=7.5cm}
\end{center}
%\vspace{10mm}
\caption{
Obtained temperature dependence of 
$\Omega_{\rm H}^\ast/\gamma_{\rm H}^\ast \equiv \theta_{\rm H}^{-1}(0)$
and $\Omega_{\rm H}^\ast \equiv -{\rm Im}\theta_{\rm H}^{-1}(\w)/\w$
for several $\w$'s.
$\Omega_{\rm H}^\ast$ is almost independent of $\w$ and $T$.
$\gamma_{\rm H}^\ast$ is insensitive to $\w$,
whereas it strongly depends on $T$.
These behaviors of $\Omega_{\rm H}^\ast$ and $\gamma_{\rm H}^\ast$
are the most prominent experimental results.
}
  \label{fig:L-IHA2}
\end{figure}
%%%%%%%%%%%%%%%%%%%%%%%%%%%%%%%%%%%%%%%%%%%%%%%%%%%%%

Here we discuss the inverse Hall angle
by the CVC-FLEX approximation for LSCO in more detail.
Figures \ref{fig:L-IHA}
shows the $\w$-dependence of 
$\theta_{\rm H}^{-1}(\w)$.
We can see that Re$\theta_{\rm H}^{-1}(\w)$
given by the CVC-FLEX approximation
is almost $\w$-independent for $\w<0.2$,
which is the main experimental finding
as explained above.
On the other hand, 
Re$\{\theta_{\rm H}^0(\w)\}^{-1}$ by RTA
shows sizeable $\w$-dependences,
which is proportional to $\gamma_{\rm AV}(\w)$
given in eq.(\ref{eqn:AV}).
We stress that the effect of the CVC on
Re$\theta_{\rm H}^{-1}(\w)$ is prominent
till frequencies much larger than $\gamma(0)$.

Figure \ref{fig:L-IHA} shows the
imaginary part of the inverse Hall angle.
Im$\theta_{\rm H}^{-1}(\w)$ given by the CVC-FLEX 
approximation shows an almost complete $\w$-linear
behavior, and its gradient stays unchanged against 
the temperature ($T=0.02\sim0.1$).
On the other hand,
Im$\{\theta_{\rm H}^{0}(\w)\}^{-1}$
by the RTA shows a sub-linear behavior
with respect to $\w$ as shown in fig. \ref{fig:L-IHA},
which is inconsistent with experiments.
Its temperature dependence is also inconsistent,
which will be shown in fig. \ref{fig:L-IHA2}.
Such excess $\w$- and $T$-dependences 
of $\{\theta_{\rm H}^{0}(\w)\}^{-1}$ by RTA
come from $z^{-1}(\w)$, which will be discussed below.

Here we further analyze the temperature
dependence of the inverse Hall angle
to make comparison with experiments.
Figure \ref{fig:L-IHA2} shows obtained 
Re$\theta_{\rm H}^{-1}(\w) 
= \Omega_{\rm H}^\ast/2\gamma_{H}^\ast$ and
$(-{\rm Im}\theta_{\rm H}^{-1}(\w)/\w)^{-1}
 = \Omega_{\rm H}^\ast$.
Apparently, both quantities are almost
$\w$-independent for $\w<0.2$,
which confirms the experimental simple Drude
expression for the Hall angle.
Here, the value of $d$ in Re$\theta_{\rm H}^{-1}$ 
for $\w<0.2$ is $d\approx 1.6$
in the present study for LSCO ($n=0.9$),
whereas $d\approx 1$ by the RTA.
Experimentally, $d\sim 1$ 
for $\w\sim 1000$cm$^{-1}$
in the optimally doped YBa$_2$Cu$_3$O$_{6+x}$ ($x=0.93$),
and $d\sim2$ in a slightly under-doped 
compound ($x=0.65$).
Moreover, the value of Re$\theta_{\rm H}^{-1}(\w)$
by CVC-FLEX approximation is much smaller than that by 
the RTA, which is consistent with experimental observations
 \cite{Drew04,Drew00}.
We note that the value of $d$ in the DC-inverse Hall angle
is approximately 2 in under-doped 
BSCCO \cite{exp-IHA-B} and YBCO \cite{exp-IHA-Y}.
It slightly decreases with doping,
and $d\approx 1.75$ at optimum doped systems.
%The value of $d$ obtained by the CVC-FLEX approximation
%for LSCO and YBCO ($n=0.9$) is also smaller than 2
% \cite{Kontani-Hall}.

Figure \ref{fig:L-IHA2} also shows that
$\Omega_{\rm H}^\ast$ by the CVC-FLEX approximation
is almost $\w$- and temperature-independent, 
which is consistent with experiments.
Contradictory to experiments, however, 
it monotonously increases within the RTA.
We also stress that the experimental doping dependence
of $\Omega_{\rm H}^\ast$, which increases as the doping decreases,
is reproduced well in the present study.
According to eq. (\ref{eqn:expand-IHA}),
$\lim_{\w\rightarrow0}
 \Omega_{\rm H}^\ast = z a^{2}(2b-a)^{-1}$.
In the RTA where $a=b=1$,
we obtain $\{\Omega_{\rm H}^0\}^\ast \propto z$,
which is a increase function with $T$.
The inferred $T$-dependence of $\Omega_{\rm H}^\ast$
by RTA is recognized by the numerical study
in figs. \ref{fig:L-IHA} and \ref{fig:L-IHA2}.
If the CVC's are taken into account,
on the other hand,
the temperature dependence of 
$\Omega_{\rm H}^\ast$ will be small
because $\Omega_{\rm H}^\ast \sim z a^2 b^{-1}$
is almost constant
according to eqs.(\ref{eqn:beki6}) and (\ref{eqn:beki7}).
In fact, $\Omega_{\rm H}^\ast$ by the CVC-FLEX approximation
is insensitive to $T$ and $\w$
as shown in fig. \ref{fig:L-IHA2}.
Experimental observations in HTSC's
support the results by the CVC-FLEX approximation satisfactorily.

Here we discuss 
experimental behavior of the Hall angle 
in HTSC's in more detail.
In the IR ($\w=900\sim1100{\rm cm}^{-1}$) measurement
 \cite{Drew04},
the simple Drude form in eq.(\ref{eqn:D-HA}) is 
satisfied very well.
It is also well recognized in YBCO,
however, the extrapolation of 
Im$\theta_{\rm H}^{-1}$ to $\w=0$
gives a positive intercept, which is 
recognized as a consequence of the chain 
contributions to $\s_{xx}$ in YBCO.
Corresponding to this fact,
reference \cite{Drew02} reports that
the far-IR ($\w=20\sim250{\rm cm}^{-1}$)
Hall angle in YBa$_2$Cu$_3$O$_7$ 
deviates from eq.(\ref{eqn:D-HA}).
One possible origin of this deviation
other than the chain contribution
would be the emergence of the pseudo-gap.
In fact, DC transport coefficients show various
anomalous behaviors in the pseudo-gap region.
They are well reproduced theoretically
in terms of the AF+SC fluctuation theory
if one take the CVC into account
 \cite{Kontani-N}.
It is an important future problem to extend the 
scope of the present study to the pseudo-gap region,
using the AF+SC fluctuation theory.

In summary,
experimentally observed simple Drude form of
$\theta_{\rm H}^{-1}(\w)$ in eq.(\ref{eqn:D-HA})
is satisfactorily well reproduced 
by the CVC-FLEX approximation,
using the parameters for LSCO. 
Both $\gamma_{\rm H}^\ast$ and $\Omega_{\rm H}^\ast$
are constant for $\w\simle0.3$, while the former
is strongly temperature dependent.
Similar results are obtained even if one use parameters 
for YBCO, as shown in fig. \ref{fig:Y-IHA}.
We note that $\theta_{\rm H}^{-1} =1.0$ in the
upper panel of Fig.\ref{fig:L-IHA2} corresponds to 5000 [Tesla/radian],
and $\Omega_{\rm H}^\ast=0.2$ in the lower panel corresponds to 
0.2 [1/cm Tesla], approximately.  
These values seem to be well consistent with experiments
 \cite{Drew00}.

%%%%%%%%%%%%%%%%%%%%%%%%%%%%%%%%%%%%%%%%%%%%%%%%%%%%%
\begin{figure}
%\vspace{10mm}
\begin{center}
\epsfig{file=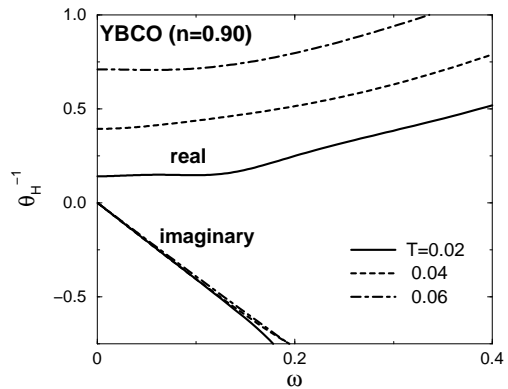,width=6.5cm}
\end{center}
%\vspace{10mm}
\caption{
Inverse Hall angle for YBCO
obtained by the CVC-FLEX approximation.
}
  \label{fig:Y-IHA}
\end{figure}
%%%%%%%%%%%%%%%%%%%%%%%%%%%%%%%%%%%%%%%%%%%%%%%%%%%%%

%%%%%%%%%%%%%%%%%%%%%%%%%%%%%%%%%%%%%
\subsection{Hall Angle}
%%%%%%%%%%%%%%%%%%%%%%%%%%%%%%%%%%%%%
We also discuss $\theta_{\rm H}(\w)$
given by the CVC-FLEX approximation,
and make comparison with experiments. 
Figure \ref{fig:L-HA} shows
Re$\theta_{\rm H}(\w)$ for LSCO.
$\theta_{\rm H}^0(\w)$ by RTA is almost temperature
independent for $\w>0.15$.
In contrast, $\theta_{\rm H}(\w)$
by the CVC-FLEX approximation is $T$-dependent till much 
larger $\w$ due to the $\w$-dependence of the CVC.
Especially, it increases with $T$ for $\w>0.15$,
which is consistent with experimental observation
 \cite{Drew04,Drew00}.
This change of $d {\rm Re}\theta_{\rm H}/ dT$
for larger $\w$ is a natural consequence of
the Lorentzian form of ${\rm Re}\theta_{\rm H}(\w)$,
where $\gamma_{\rm H}$ is $\w$-independent
and is an increase function of temperature.
In contrast,
${\rm Re}\s(\w)$ deviates from the Lorentzian 
due to the $\w$-dependence of $\gamma_{\rm ED}(\w)$,
which is consistent with experiments
 \cite{Drew04,Drew00}.

%%%%%%%%%%%%%%%%%%%%%%%%%%%%%%%%%%%%%%%%%%%%%%%%%%%%%
\begin{figure}
%\vspace{10mm}
\begin{center}
\epsfig{file=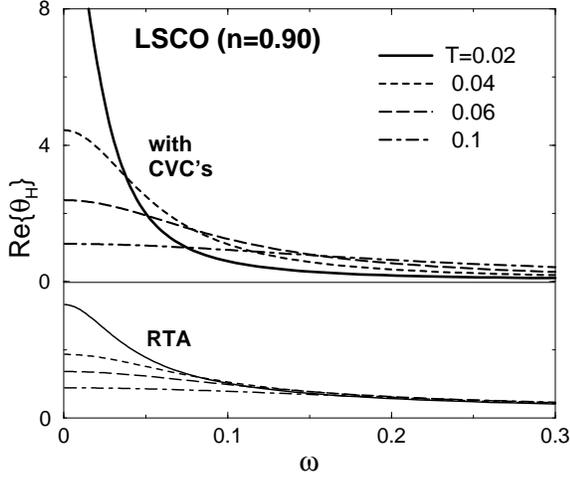,width=7.5cm}
\end{center}
%\vspace{10mm}
\caption{
Obtained $\w$-dependence of the Hall angle
for several temperatures.
}
  \label{fig:L-HA}
\end{figure}
%%%%%%%%%%%%%%%%%%%%%%%%%%%%%%%%%%%%%%%%%%%%%%%%%%%%%

%%%%%%%%%%%%%%%%%%%%%%%%%%%%%%%%%%%%%%%%%%%%%%%%%%%%%
\begin{figure}
%\vspace{10mm}
\begin{center}
\epsfig{file=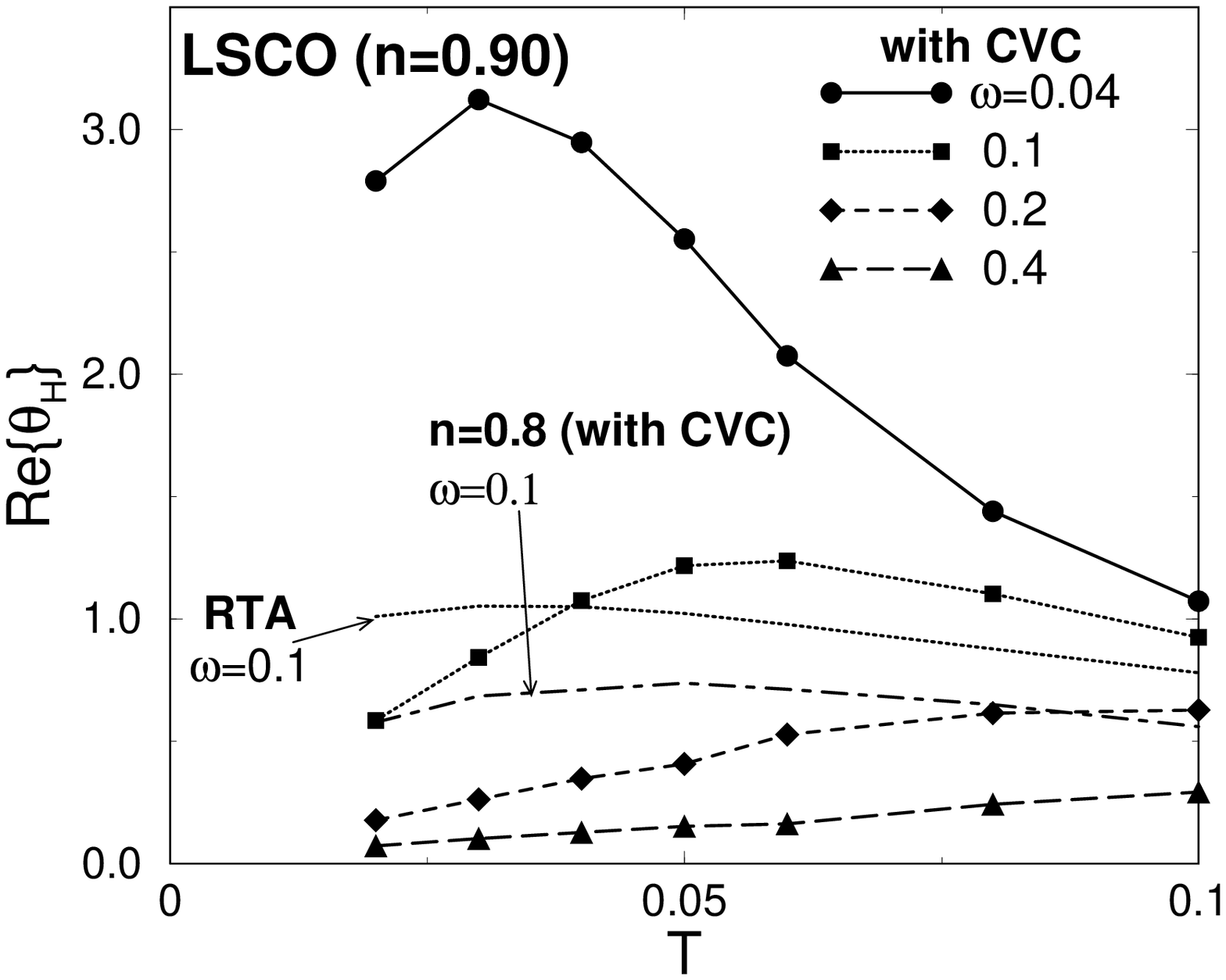,width=7.5cm}
\epsfig{file=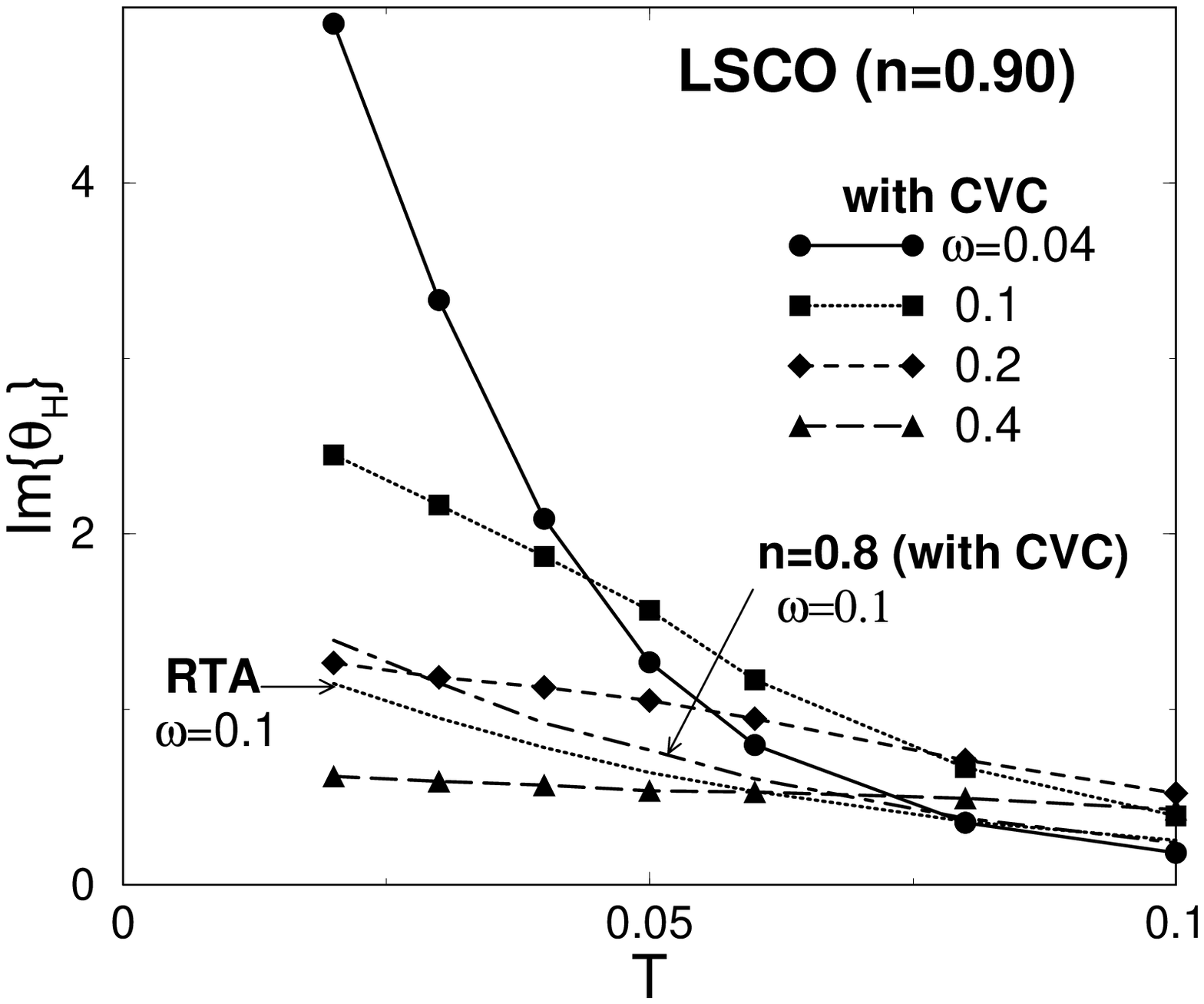,width=7.5cm}
\end{center}
%\vspace{10mm}
\caption{
Obtained temperature dependence of the inverse Hall angle
for several $\w$'s.
Characteristic behaviors of the experimental $\theta_{\rm H}$
are well reproduced.
}
  \label{fig:L-HA2}
\end{figure}
%%%%%%%%%%%%%%%%%%%%%%%%%%%%%%%%%%%%%%%%%%%%%%%%%%%%%

The origin of the Lorentzian form of
${\rm Re}\theta_{\rm H}(\w)$ is
ascribed to the almost perfect cancellation of
$\w$-dependence of $\gamma_{\rm ED}(\w)$
and that of the CVC:
${\rm Re}\theta_{\rm H}(\w)$ for $\w\gg\gamma$
will be enhances by the former effect
because $\gamma_{\rm ED}(\w)$ is a increase function of $\w$,
whereas it will be suppressed by the the latter,
because the back-flow will be less important
for larger $\w$.
It is a nontrivial future problem why
these two effects cancel out almost completely,
which results in the observed Lorentzian form of 
${\rm Re}\theta_{\rm H}(\w)$ for $\w\simle1000{\rm cm}^{-1}$.

Figure \ref{fig:L-HA2}
shows the temperature dependences of
$\theta_{\rm H}(\w)$ for several $\w$'s. 
The obtained results for $\w=0.2$ or $0.4$
look similar to the experimental observations
for YBCO with $x=0.93$ (optimum) or $x=0.65$ (slightly under-doped) 
for $\w\sim1000{\rm cm}^{-1}$
 \cite{Drew04}.
On the other hand,
the result for $\w=0.04$ resembles
the observation for heavily under-doped
non-superconducting sample ($x=0.4$).
We guess from this fact that
the electronic states in heavily under-doped 
systems are qualitatively reproduced,
although the experimental value of $\gamma_{\rm H}^\ast$
is much larger.

%%%%%%%%%%%%%%%%%%%%%%%%%%%%%
\subsection{Predictions for Electron-Doped Systems}
%%%%%%%%%%%%%%%%%%%%%%%%%%%%
DC transport phenomena under magnetic field
in electron-doped systems (e.g., NCCO) also shows
striking NFL behaviors which originate
from the CVC
 \cite{Kontani-Hall,Kontani-S}.
Surprisingly,
both $R_{\rm H}$ and $S$ in a under-doped NCCO 
is negative, and its absolute value increase
as $T$ decreases.
Their behavior looks approximately
symmetrical to those in hole-doped systems.
Contrary to these experimental facts,
the RTA predicts the positive Hall coefficient
because it has a hole-like FS whose shape 
is similar to YBCO.
This discrepancy is naturally solved
if one take the CVC into account,
since $\d\theta^J/\d k_\parallel$ becomes positive
around the cold spot of NCCO
whose location is different from that of YBCO;
see fig.\ref{fig:FS} (a)
 \cite{Kontani-Hall}.

%%%%%%%%%%%%%%%%%%%%%%%%%%%%%%%%%%%%%%%%%%%%%%%%%%%%%
\begin{figure}
%\vspace{10mm}
\begin{center}
\epsfig{file=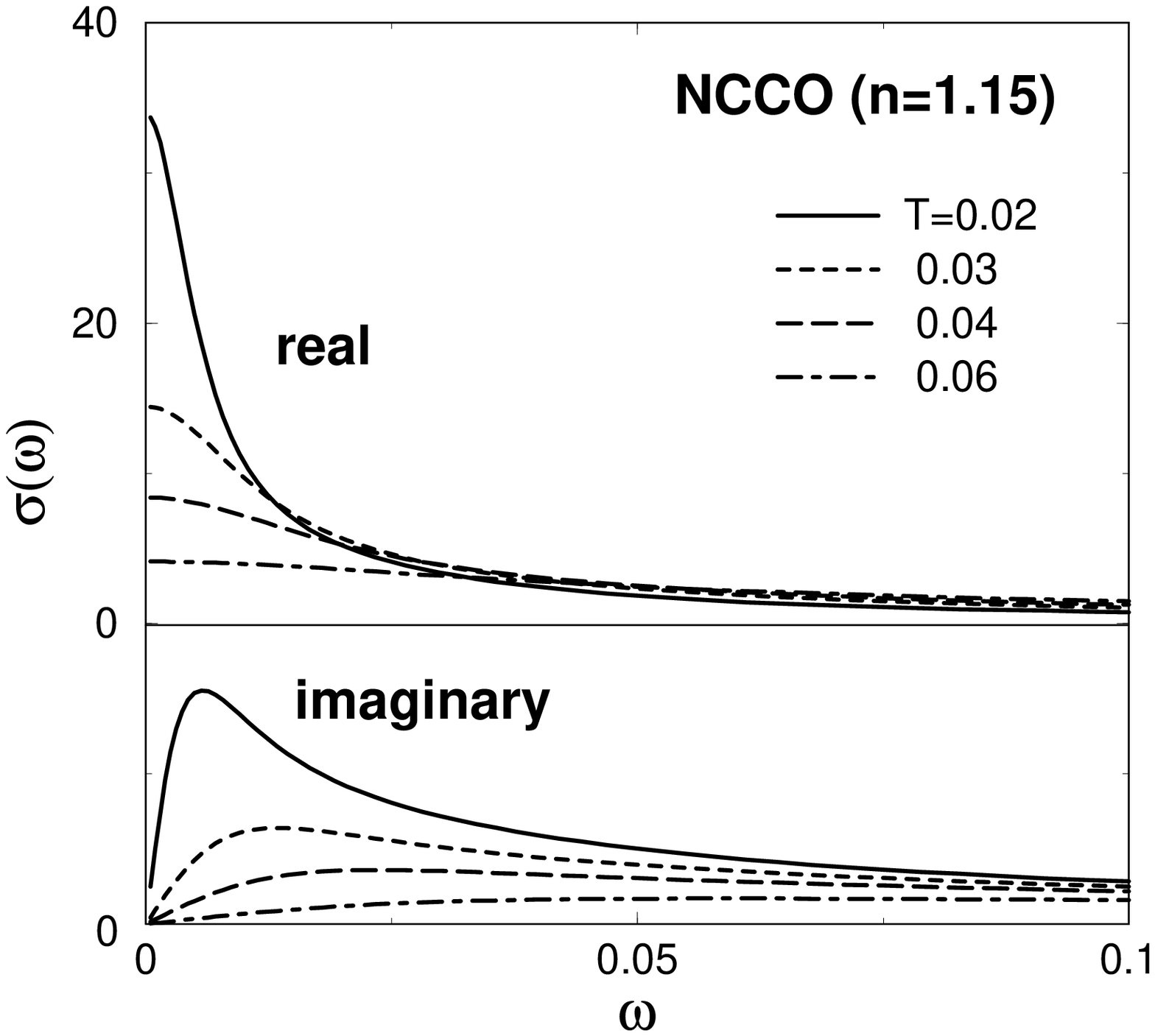,width=7cm}
\epsfig{file=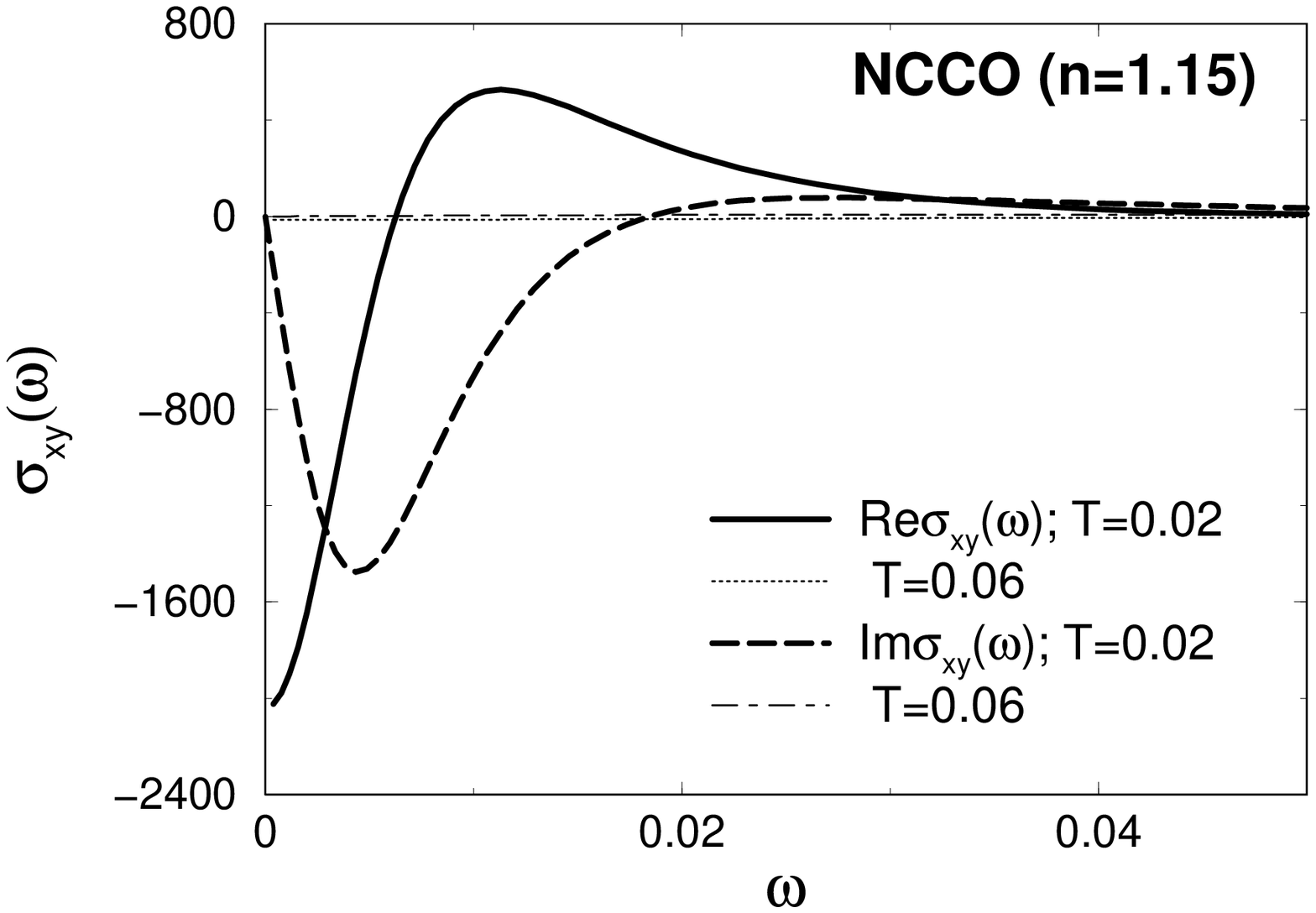,width=7cm}
\epsfig{file=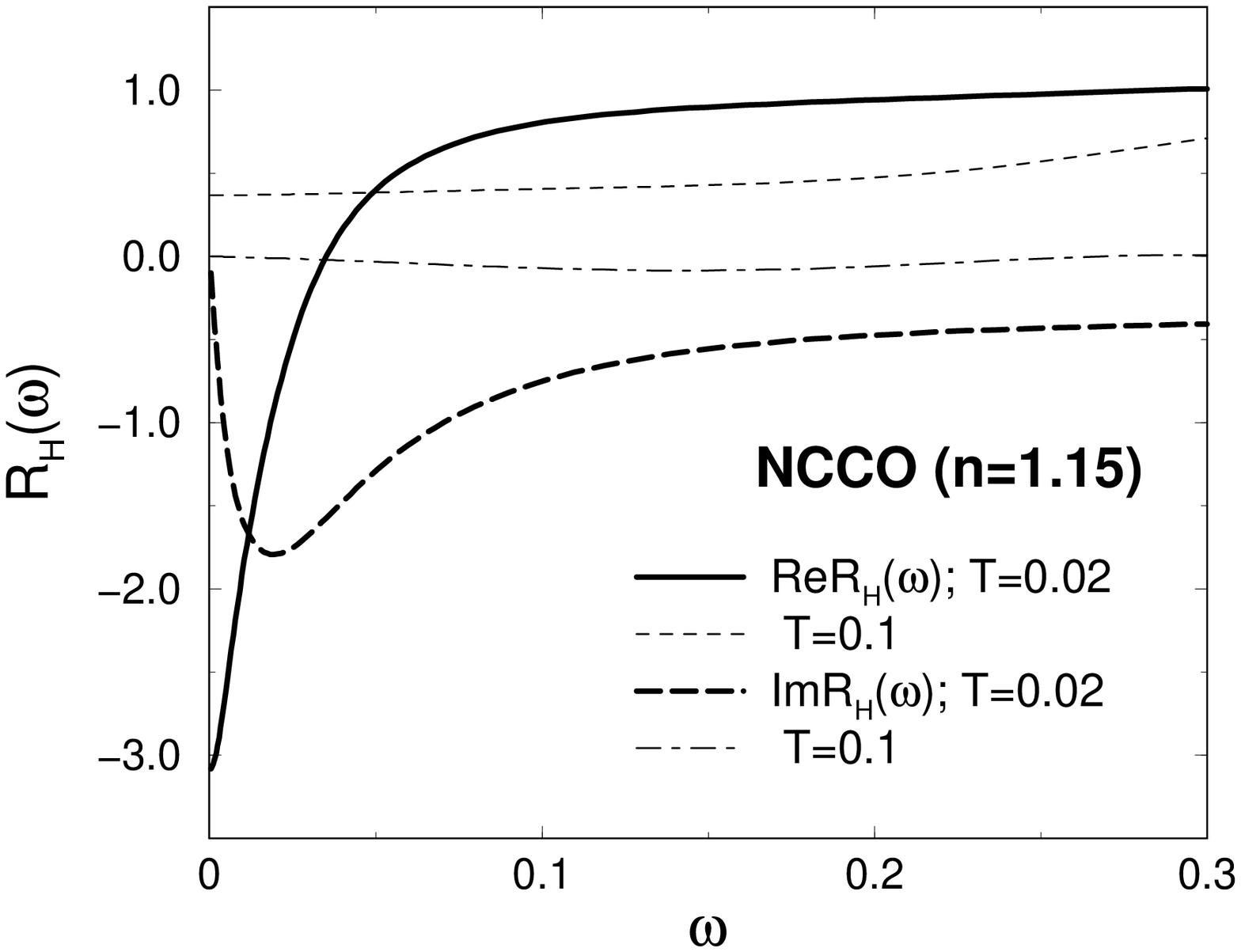,width=7cm}
\end{center}
%\vspace{10mm}
\caption{
Obtained $\sigma(\w)$, $\sigma_{xy}(\w)$ 
and $R_{\rm H}(\w)$ for NCCO
by the CVC-FLEX approximation.
}
  \label{fig:N}
\end{figure}
%%%%%%%%%%%%%%%%%%%%%%%%%%%%%%%%%%%%%%%%%%%%%%%%%%%%%

%To the best of my knowledge, optical Hall conductivity in
%electron-doped systems has not been measured
%experimentally.
Quite recently, optical Hall conductivity in
electron-doped systems has been observed by Zimmers et al
 \cite{Zimmers}.
Here, we analyze the $\w$-dependences
of $\s_{\mu\nu}(\w)$ in electron-doped systems
based on the conserving approximation.
Figure \ref{fig:N} shows $\s(\w)$, $\s_{xy}(\w)$
and $R_{\rm H}(\w)$ obtained by the CVC-FLEX approximation.
Both $\s_{xy}(\w)$ and $R_{\rm H}(\w)$
for NCCO are similar to those for LSCO given in Figs.
\ref{fig:L-sxy} and \ref{fig:RH},
except their signs.
We stress that Im$R_{\rm H}(\w)$ is as large as Re$R_{\rm H}(\w)$
for finite $\w$, which means that the simple
ED-form of $\s_{xy}(\w)$ is violated.
We predict that the signs of Re$R_{\rm H}(\w)$ and Re$\s_{xy}(\w)$
change from negative to positive with $\w$.
Thus, the CVC in NCCO plays important roles.
In future, measurements of $\s_{xy}(\w)$ in NCCO are 
highly anticipated.

We found that an accurate numerical calculation
(Pade approximation) for NCCO is much difficult 
than that for LSCO and YBCO.
By this reason, we could not obtain reliable
results for $0.04\simge T \simge0.08$.
It is a future important problem to improve
the stability of the Pade approximation
in case of NCCO.

%%%%%%%%%%%%%%%%%%%%%%%%%%%%%%%
\section{Summary and Future problems}
%%%%%%%%%%%%%%%%%%%%%%%%%%%%%%%

% Summary of the numerical study
In the present work,
we have calculated the optical conductivities
$\s(\w)$ and $\s_{xy}(\w)$ for HTSC's
by the CVC-FLEX approximation.
%The optical Hall angle and Hall coefficient
%are given by $\theta_{\rm H}(\w)\equiv \s_{xy}(\w)
%/\s(\w)$ and $R_{\rm H}(\w)\equiv \s_{xy}(\w)
%/(\s(\w))^2$, respectively.
Experimentally observed
anomalous behaviors for $\s(\w)$, $\s_{xy}(\w)$,
$\theta_{\rm H}(\w)$ and $R_{\rm H}(\w)$
are well reproduced {\it for enough wide range of 
frequencies and temperatures},
without assuming any fitting parameters
 \cite{Letter}.
Especially, (I) $\s(\w)$ given by the CVC-FLEX approximation
follows the ED-form shown in eq. (\ref{eqn:ED1})
with the relaxation time in eq. (\ref{eqn:AV}), 
whereas $\s_{xy}(\w)$ strongly deviates 
from the ED-form, eq. (\ref{eqn:ED2}), because the $\w$-dependence
is much exaggerated due to the CVC in nearly AF Fermi liquids.
By this reason, (II) Im$R_{\rm H}(\w) \sim {\rm Re}R_{\rm H}(\w)$
is realized even when for $\w\ll\gamma$, as shown
in Fig. \ref{fig:RH}.
Moreover, (III) $\theta_{\rm H}(\w)$ follows a simple Drude form
given in eq.(\ref{eqn:D-HA}) for $\w\simle0.2$, 
as shown in Figs. \ref{fig:L-IHA}, \ref{fig:L-IHA2}
and \ref{fig:Y-IHA}.
They are consistent with the characteristic
experimental results for HTSC's reported by Drew et al.
 \cite{Drew96,Drew00,Drew04}.
These anomalous AC transport phenomena 
cannot be reproduced by previous 
theoretical works based on the RTA,
even if one assume extremely anisotropic $\tau_\k$.
Instead, they are naturally explained by taking the CVC into account
in accordance with the Ward identity.

In the present study,
we have pointed out the important role of the 
back-flow in the optical conductivities for the first time.
The enhancement of ${\rm Im}\s_{xy}(\w)/i\w|_{\w=0}$
due to the CVC is not same as the enhancement of $\s_{xy}(0)$;
The former is more prominent than the latter
as explained in eqs.(\ref{eqn:beki6})-(\ref{eqn:beki9}).
This fact leads to the breakdown of the extended Drude-form
at very low frequencies.
The back-flow decreases monotonically with $\w$
as one approaches the collisionless region ($\w\simge\gamma$).
This fact gives an approximate Drude-form of
the Hall angle in eq.(\ref{eqn:D-HA}) for $\w\simle0.2$, 
nonetheless of the fact that $\s(\w)$ deviates from a simple 
Drude-form (instead it follows a ED-form) due to 
the $\w$-dependence of $\gamma(\w)$.
Note that the back-flow in the collisionless region
is given by the real part of ${\cal T}_{22}$
 \cite{Eliashberg,Okabe,Jujo}.
This is an important future problem for us 
to find a simple physical explanation for this numerical result.

We stress that  
both AC and DC anomalous transport phenomena
in HTSC's are explained {\it in a unified way}
based on the Fermi liquid theory,
if one take the CVC to satisfy the conservation laws.
As for the DC Hall coefficient,
one frequently attribute the enhancement of $R_{\rm H}$ to 
the small area of the cold spot (Fermi arc) 
observed by ARPES in under-doped compounds.
However, this idea contradicts the fact that
the $R_{\rm H}$ decreases in the pseudo-gap
region while the Fermi arc shrinks further.
In the same way, anomalous behaviors of 
$\Delta\rho/\rho$, $S$ and $\nu$
in the pseudo-gap region cannot be understood
within the scheme of the RTA.
Such contradictions are satisfactorily solved
by the CVC-FLEX approximation, by taking the
superconducting fluctuations induced by the
AF fluctuations 
 \cite{Kontani-N}.
We stress that
the natural extension of this DC transport theory
to AC transport phenomena, 
with taking the same diagrams for the CVC,
succeeds in explaining the optical Hall effect
observed in HTSC's.
This fact means that the qualitative dynamical electronic 
properties of HTSC's, from the over-doped to the slightly
under-doped systems, are well understood 
in terms of the Fermi liquid theory with strong AF fluctuations.

There remain many important issues 
for the Future study.
For example, one can study various AC-transport
coefficients other than $\s_{\mu\nu}$
based on the CVC-FLEX approximation,
using the similar method developed in the present study.
Study of the role of the CVC at finite frequencies
for $\Delta\rho/\rho$, $S$ and $\nu$
would be very interesting, although experimental
observation would be difficult at the present stage.
In addition, we are planning to study
the optical conductivities in the pseudo-gap region
based on the FLEX+T-matrix approximation,
which ascribes the pseudo-gap phenomena in HTSC's 
to the strong superconducting fluctuations
 \cite{Yamada-rev}.
As we mentioned in the previous section,
reference \cite{Drew02} reports that
the far-IR ($\w=20\sim250{\rm cm}^{-1}$)
Hall angle in YBa$_2$Cu$_3$O$_7$ 
deviates from the Drude-form in eq.(\ref{eqn:D-HA}),
although it is well satisfied for $\w\sim 1000{\rm cm}^{-1}$.
We would like to find out whether
such an anomaly in far-IR Hall angle
could be understood as the pseudo-gap effect
using the AF+SC fluctuation theory.

%%%%%%%%%%%%%%%%%%%%%
\acknowledgements
The author is grateful to H.D. Drew and A. Zimmers
for fruitful discussions.

\appendix
%%%%%%%%%%%%%%%%%%%%%%%%%%%%%%%
\section{Physical Meaning of the Back-Flow}
%%%%%%%%%%%%%%%%%%%%%%%%%%%%%%%
% VC in Fermi liquid, back-flow
% Landau
Throughout the present work,
we have stressed the importance of the back-flow
for both DC and AC transport phenomena.
Here, we would like to depict the 
physical aspect of the back-flow in nearly AF Fermi liquid
based on the phenomenological Landau-Fermi liquid theory
 \cite{Nozieres}.
According to Landau,
the energy of the quasiparticle ${\tilde \e}_{\k\s}$ 
is expressed as
\begin{eqnarray}
{\tilde \e}_{\k\s}= {\e}_{\k\s} + \sum_{\k'\s'} 
 f_{\k\s,\k'\s'}\delta n_{\k'\s'} + O((\delta n)^2),
 \label{eqn:LFLT}
\end{eqnarray}
where ${\e}_{\k\s}= \k^2/2m^\ast$,
$n_{\k\s}$ is the distribution function of quasiparticles,
$f_{\k\s,\k'\s'}$ is the Landau function, and
$\delta n_{\k\s}= n_{\k\s} - n_{\k\s}^{T=0}$.
Equation (\ref{eqn:LFLT}) means that
the energy of quasiparticles are changed  
when the quasiparticle excitation exists.
%through $f_{\k\s,\k'\s'}$. 
By this reason,
once we add a quasiparticle at $\k$ outside the FS, 
the Fermi sphere is deformed to minimize the total energy
unless $f_{\k\s,\k'\s'}=0$.
As a result, the Fermi sphere has a finite momentum,
which is the physical meaning of the back-flow.
Thus, the existence of the back-flow is assured by the 
most essential relation in the Fermi liquid, eq.(\ref{eqn:LFLT}).
Apparently, the back-flow would be indispensable
in strongly correlated Fermi liquids, like in HTSC.

The importance of the back-flow has been 
understood very well in a spherical system,
where $f_{\k\s,\k'\s'}$ can be expanded by
Legendre polynomials, $P_l({\hat \k}\cdot{\hat \k'})$.
Landau first studied the back-flow in the 
collisionless region $\w\gg\gamma$, 
where the lifetime of an quasiparticles 
is longer than the period of the outer field.
Based on the Kubo formula,
Yamada and Yosida analyzed the opposite region 
$\w\ll\gamma$ in order to study the role of the
back-flow on the DC conductivity.
They rigorously proved that the conductivity diverges
even at finite temperatures if no Umklapp scattering
process exists.
In contrast, the RTA always gives finite conductivity 
$\s\propto\gamma^{-1}$ at $T\ne0$ even in the absence 
of the Umklapp process, reflecting the 
the violation of the momentum conservation laws.

In contrast, importance of the back-flow
{\it in anisotropic systems with strong correlations}
has not been recognized until recently.
As explained in \S III,
we found that total current ${\bf J}_\k$ becomes
quite different from the quasiparticle velocity
${\bf v}_\k$ due to the CVC 
when the AF fluctuations with $\q\approx {\bf Q}$ are strong
 \cite{Kontani-Hall}.
This is the origin of various anomalous transport phenomena
in HTSC's.
This unexpected behavior of ${\bf J}_\k$
comes from the fact that
${\cal T}_{22}(\k-\k')$
in the Bethe-Salpeter equation (\ref{eqn:JJJ}),
which corresponds to the Landau function for $\w\ll\gamma$,
takes large values only for $\k-\k'\approx{\bf Q}$.
In this case, according to eq.(\ref{eqn:LFLT}),
a quasiparticle added at $\k'$ strongly modifies 
${\tilde \e}_\k$ only when $\k\approx \k'+{\bf Q}$,
which makes the induced current (back-flow)
proportional to ${\bf v}_{\k}$.
The induced current is not parallel to the 
source velocity ${\bf v}_{\k'}$,
in contrast to the case of spherical systems.
The schematic behavior of ${\bf J}_\k$
in HTSC's is shown in fig.\ref{fig:FS} (b).
${\bf J}_\k$ at the hot spot takes enhanced values
because $\a_\k\simle 1$ in eq.(\ref{eqn:Jy}),
which is interpreted as the ``resonance'' 
between ${\bf v}_\k$ and ${\bf v}_{\k'}$.

In the present paper, we studied the 
optical conductivity and Hall conductivity
by taking the $\w$-dependence of the CVC into account appropriately, 
which has not been performed in previous studies.
We find that $\s_{xy}(\w)$ shows a striking $\w$-dependence
when the AF fluctuations are strong, which cannot be
expressed by an ED-form.
Such a non-Fermi liquid-like behavior comes from 
the prominent $\w$-dependence of the CVC,
which was detected in the present study for the first time.
As shown in \S III,
the total current at finite $\w$ is given by
${\bf J}_\k(\w)= {\bf J}_k+i\w{\bf J}_k^{(1)}+ O(\w^2)$,
where ${\bf J}_k^{(1)}$ is real and its $\k$-dependence 
is much larger than the first term.
This strong $\w$-dependence of the total current
gives rich variety of spectrum in optical conductivities.

%%%%%%%%%%%%%%%%%%%%%%%%%%%%%%%
\section{Comments on Previous Theoretical Studies}
%%%%%%%%%%%%%%%%%%%%%%%%%%%%%%%

Anomalous DC transport phenomena in HTSC's,
as represented by the enhancement of the Hall coefficient,
have been frequently ascribed to the reduction of the
effective carrier number within the RTA.
For example, Ref. \cite{Pines-Hall}
%Stojkovi{\'c} and Pines
proposed the highly anisotropic $\tau_\k$ model
based on a spin fluctuation theory;
$\tau_{\rm h}\propto T^{-1}$ for hot electrons
whose density is $n_{\rm h}$,
$\tau_{\rm c}\propto T^{-2}$ for cold electrons
whose density is $n_{\rm c}(=n-n_{\rm h})$.
They assume that $n_{\rm c} \ll n_{\rm h}$ and 
$\tau_{\rm c}/\tau_{\rm h}\sim 100$ at lower temperatures.
Their model cannot give a comprehensive explanation
for anomalous DC transport phenomena in HTSC's,
while the CVC-FLEX approximation can give it.

Here, we examine the optical conductivities 
within the RTA based on a simplified anisotropic 
$\gamma_\k$ model as follows:
\begin{eqnarray}
\s(\w)&\propto& \frac{n_{\rm c}}{2\gamma_{\rm c}-i\w}
 + \frac{n_{\rm h}}{2\gamma_{\rm h}-i\w},
 \\
\s_{xy}(\w)&\propto& \frac{n_{\rm c}}{(2\gamma_{\rm c}-i\w)^2}
 + \frac{n_{\rm h}}{(2\gamma_{\rm h}-i\w)^2},
\end{eqnarray}
where $\tau_{\rm h,c}= 1/2\gamma_{\rm h,c}$.
The Hall coefficient is highly enhanced 
in proportion to $1/en_{\rm c}(\propto T^{-1})$ 
when $n_{\rm c}/\gamma_{\rm c}^2 \gg n_{\rm h}/\gamma_{\rm h}^2$.

In the case of $n_{\rm c} \ll n_{\rm h}$ and
$\gamma_{\rm c} \ll \gamma_{\rm h}$,
frequencies $\w_{xx}$, $\w_{xy}$ and $\w_{\rm RH}$
which give the maximum Im$\s(\w)$, Im$\s_{xy}(\w)$
and Im$R_{\rm H}(\w)$ respectively, are given by
\begin{eqnarray}
\w_{xx} &=& 2\gamma_{\rm c}, 
 \\
\w_{xy} &=& (2/\sqrt{3})\gamma_{\rm c}=1.16\gamma_{\rm c},
 \\
\w_{\rm RH} &=& \frac{2(\gamma_{\rm c}n_{\rm h}
 + \gamma_{\rm h}n_{\rm c})}{\sqrt{3}n_{\rm h}} ,
\end{eqnarray}
where $\w_{\rm RH}$ is approximately given by 
\begin{eqnarray}
\w_{\rm RH} &\approx& 1.16 n_{\rm c}\gamma_{\rm h}/n_{\rm h}
 \gg \gamma_{\rm c}
 \ \ \ \ \ \ {\rm for} \ 
 {n_{\rm c}}/{\gamma_{\rm c}} \gg {n_{\rm h}}/{\gamma_{\rm h}} ,
  \nonumber \\
   &\approx& 2.31 \gamma_{\rm c}
 \ \ \ \ \ \ \ \ \ \ \ \ \ \ \ \ \ \ \ \ \ {\rm for} \ 
  {n_{\rm c}}/{\gamma_{\rm c}} = {n_{\rm h}}/{\gamma_{\rm h}} ,
  \nonumber \\
   &\approx& 1.16\gamma_{\rm c}
 \ \ \ \ \ \ \ \ \ \ \ \ \ \ \ \ \ \ \ \ \ {\rm for} \ 
  {n_{\rm c}}/{\gamma_{\rm c}} \ll {n_{\rm h}}/{\gamma_{\rm h}} .
  \nonumber 
\end{eqnarray}
Because ${n_{\rm c}}/{\gamma_{\rm c}} (\propto T^{-3})$
will be larger than 
${n_{\rm h}}/{\gamma_{\rm h}} (\propto T^{-1})$
at lower temperatures,
the relation $\w_{\rm RH} \simge \w_{xx}$ is expected
in this model.
This result is inconsistent with experimental fact
$\w_{\rm RH} \ll \w_{xx}$, as mentioned in \S I.

In a similar way,
d-density wave (DDW) model
\cite{DDW} have been proposed
to explain the enhancement of of the Hall coefficient:
$R_{\rm H}$ increases below the 
d-density wave transition temperature,
inversely proportional to the area of the ``Fermi arc''.
However, it is hopeless to reproduce the
characteristic experimental behavior of 
$R_{\rm H}(\w)$ in this model.

We also comment on the 2D Luttinger liquid model
with two kinds of relaxation times
($\tau_{\rm tr}\propto T^{-1}$, $\tau_{\rm H}\propto T^{-2}$)
proposed by Anderson
 \cite{Anderson}.
In this model, DC conductivities are given by
$\s\propto \tau_{\rm tr}^{-1}$ and
$\s_{xy}\propto (\tau_{\rm tr}\tau_{\rm H})^{-1}$, respectively.
Their natural extensions to the optical conductivities
are given as $\s(\w)\propto (\tau_{\rm tr}^{-1}-i\w)^{-1}$ and
$\s_{xy}(\w)\propto (\tau_{\rm tr}/\tau_{\rm H})\s^2(\w)$
 \cite{Romero}.
This result directly means that Im$R_{\rm H}(\w)\equiv0$, 
which apparently contradicts experiments.
In addition, the Hall angle in this model is 
$\theta_{\rm H}(\w)\propto (\tau_{\rm tr}/\tau_{\rm H})
\cdot (\tau_{\rm tr}^{-1}-i\w)^{-1}$;
the temperature dependences of coefficients
are different from the experimental ones 
given in eq.(\ref{eqn:D-HA}).
Note that another functional form of $\s_{\mu\nu}(\w)$
which predict finite Im$R_{\rm H}(\w)$
was proposed in ref. 
 \cite{Drew96},
although its theoretical verification is uncertain.

In summary,
a comprehensive understanding for the optical 
Hall coefficient in HTSC's cannot be obtained 
by previous theoretical works based on the RTA,
or by the 2D Luttinger liquid theory.
The transport theory based on the Fermi liquid theory
presented in the present paper, where the CVC
is correctly taken into account,
can explain various experimental anomalies
at the same time.

%%%%%%%%%%%%%%%%%%%%
% references
%%%%%%%%%%%%%%%%%%%%

%\end{multicols}

\end{document}